\documentclass[journal,twoside]{IEEEtran}
\usepackage{cite}

\ifCLASSINFOpdf
  \usepackage[pdftex]{graphicx}
   \graphicspath{}
\else
   \usepackage[dvips]{graphicx}
   \graphicspath{}
   \DeclareGraphicsExtensions{.eps}
\fi
\usepackage[cmex10]{amsmath}
\usepackage{amssymb}

\usepackage[caption=false,font=footnotesize]{subfig}
\usepackage{url}

\usepackage[latin1]{inputenc} 
\usepackage[squaren]{SIunits}
\usepackage{bm}
\usepackage{multirow}
\usepackage{color}
\usepackage[usenames,dvipsnames]{xcolor}
\usepackage{algorithm}
\usepackage{algorithmic}
\usepackage{abbrevs}
\usepackage{paralist}

\begin{document}
\setlength{\interdisplaylinepenalty}{2500}
\title{An extended target tracking model with multiple random matrices and unified kinematics}
%
%
%

\author{Karl~Granstr\"om
\thanks{Karl~Granstr\"om is with the Department
of Electrical Engineering, Division of Automatic Control, Link\"oping University, Link\"oping, SE-581~83, Sweden, e-mail: \texttt{karl@isy.liu.se}.}
}
\maketitle

\begin{abstract}

This paper presents a model for tracking of extended targets, where each target is represented by a given number of elliptic subobjects. A gamma Gaussian inverse Wishart implementation is derived, and necessary approximations are suggested to alleviate the data association complexity. A simulation study shows the merits of the model compared to previous work on the topic.

\end{abstract}

\begin{IEEEkeywords}
Target tracking, extended target, group target, measurement rate, random matrix, gamma distribution, Gaussian distribution, inverse Wishart distribution.
\end{IEEEkeywords}

%
\IEEEpeerreviewmaketitle

%
%
%
%



\section{Introduction}
Target tracking can be defined as the processing of a sequence of measurements obtained from a target in order to maintain an estimate of the target's current state. In this context a point target is defined as a target which is assumed to give rise to at most one measurement per time step. With modern and more accurate sensors the target may occupy multiple resolution cells of the sensor, thus potentially giving rise to more than one measurement per time step. An extended target is defined as a target that potentially gives rise to more than one measurement per time step. Examples of extended target tracking include vehicle tracking using automotive radar and pedestrian tracking using laser range sensors. Closely related to extended target is group target, defined as a cluster of point targets which cannot be tracked individually, but has to be treated as a single object.

In point target tracking the estimated state typically corresponds to the targets position and its kinematics (velocity, heading, etc). In extended target tracking the multiple measurements make it possible to estimate also the target's extension in the measurement domain, \iep to estimate the shape, the size and the orientation of the target. To estimate the target's extension requires a measurement model that relates the multiple measurements to the states that govern the extension. 

Spatial distribution models in extended target tracking appeared in \cite{GilholmGMS:2005,GilholmS:2005}. Under this model each extended target measurement is a random sample from a probability distribution that is dependent on the extended target state. A number of different extended target models have been presented, where the targets are modeled as, \egp, sticks \cite{GilholmS:2005,BaumFH:2012,Boersetal:06a}, circles \cite{PetrovMGA:2011}, ellipses \cite{Koch:2008,BaumNH:2010,GranstromLO:2011,ZhuHL:2011,ReuterD:2011,DegermanWS:2011,LanRL:2012,ReuterWD:2012}, rectangles \cite{GranstromLO:2011}, or general shapes \cite{LundquistGO:11,BaumH:2011,LanRL:2014,LanRL:2012irregular}.

\begin{figure}[t]
\centering
\subfloat{\includegraphics[width=0.40\columnwidth]{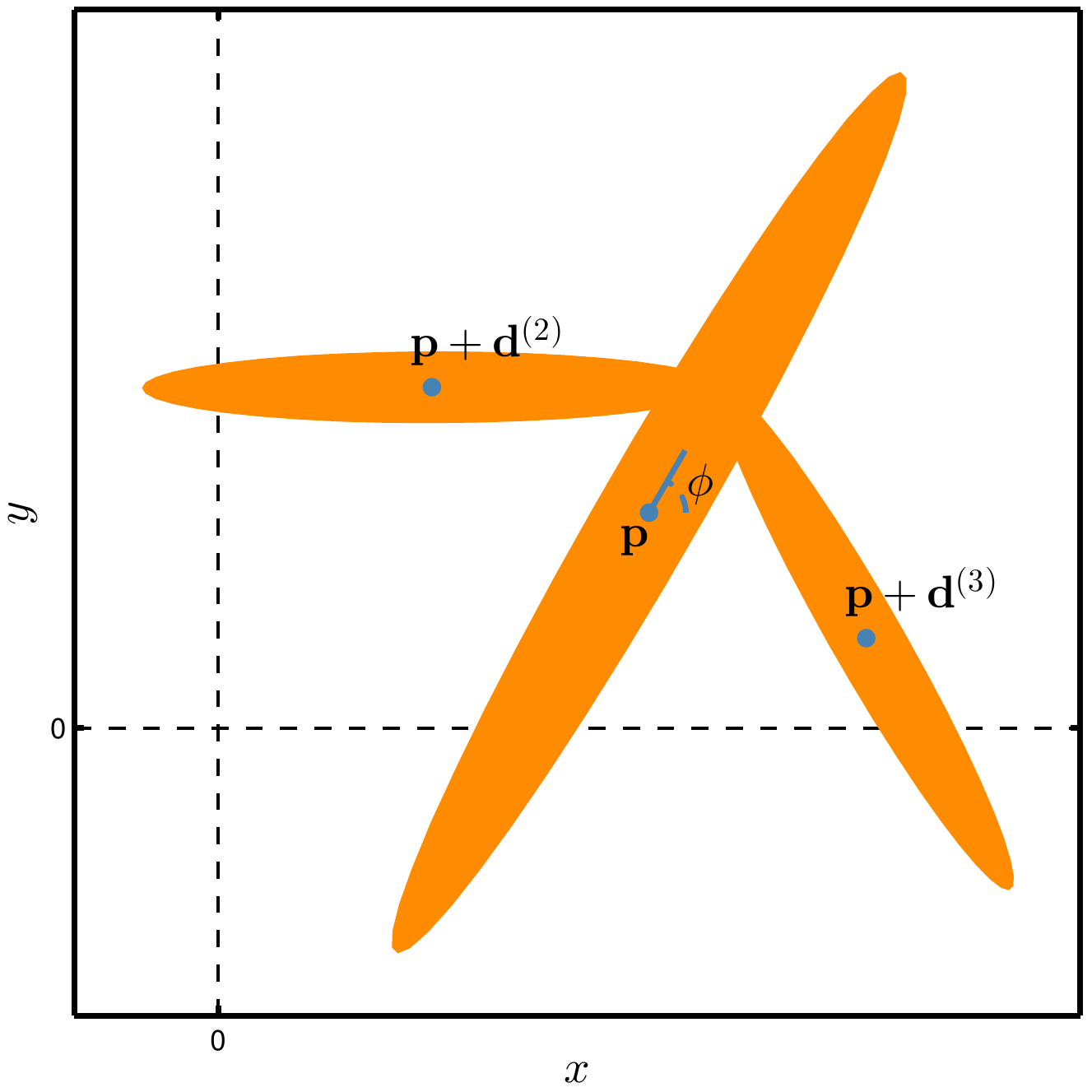}}%
\hfil
\subfloat{\includegraphics[width=0.40\columnwidth]{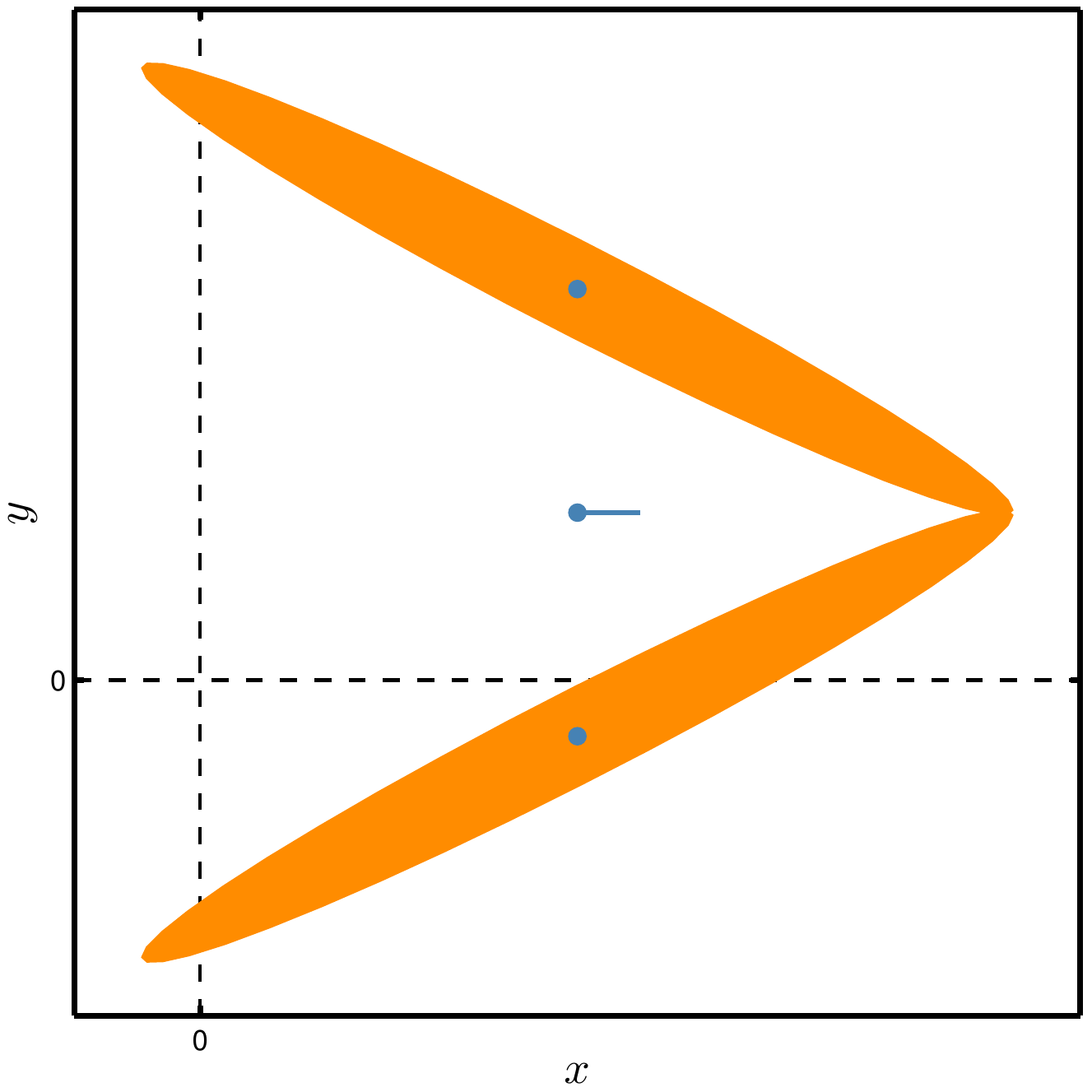}}%
\caption{Examples in 2D of extended/group targets that are represented by elliptic subobjects. Neither one of the examples has a shape that can be described by a single ellipse. Left: The overall position of the extended target is $\mathbf{p}$, and coincides with one of the subobject's position. The positions of remaining subobjects are given by offsets $\mathbf{d}$ from the overall position. Right: The overall position does not coincide with either subobject's position.}
\label{fig:multiple_ellipse_example_target}
\end{figure}

In this paper we consider state estimation for extended targets whose extensions cannot be approximated by a simple geometric shape such as an ellipse or a rectangle.
The extended target is modeled as a collection of elliptical subobjects, see \figurename~\ref{fig:multiple_ellipse_example_target}, and the positions and extensions of the subobjects are Gaussian inverse Wishart distributed. The scope of the paper is limited by the assumptions that
\begin{inparaenum}[\itshape a\upshape)]
	\item there is exactly one target present;
	\item there are no clutter measurements; and
	\item the number of subobjects is constant and known.
\end{inparaenum}
To handle multiple targets and clutter, the presented work can be integrated into a multiple target framework, \egp an extended target \phd/\cphd filter \cite{mahler_FUSION_2009_extTarg,GranstromLO:2010,GranstromLO:2012,GranstromO:2012a,LundquistGO:2013,SwainC:2010,SwainC:2012}. Estimating the number of subobjects is left for future work.


\begin{table}
\caption{Notation}
\label{tab:notation}
\vspace{-5mm}
\rule[0pt]{\columnwidth}{1pt}
$\bullet \ $ $\mathbb{R}^{n}$ is the set of real column vectors of length $n$, $\mathbb{S}_{++}^{n}$ is the set of symmetric positive definite $n\times n$ matrices, $\mathbb{S}_{+}^{n}$ is the set of symmetric positive semi-definite $n\times n$ matrices, and $\mathbb{N}$ is the set of non-negative integers.	

$\bullet \ $ $\Id$ is a $d\times d$ eye matrix, and $\mathbf{0}_{d\times e}$ is a $d\times e$ all-zero matrix.

$\bullet \ $ $|\cdot|$ is absolute value, $\left\|\cdot\right\|_{2}$ is Euclidean norm, and $\left\|\cdot\right\|_{F}$ is Frobenius norm.

$\bullet \ $ $\mathcal{PS}\left(n;\ \gamma\right)$ denotes a Poisson probability mass function (pmf) defined of the integer $n\in\mathbb{N}$ with rate parameter $\gamma>0$,
		\begin{align*}
			\mathcal{PS}\left(n;\ \gamma\right) = & {\gamma^{n}e^{-\gamma}}{\left(n!\right)^{-1}}.
		\end{align*}	

$\bullet \ $ $\Gammapdf{\gamma}{\alpha}{\beta}$ denotes a gamma probability density function (pdf) defined over the scalar $\gamma >0$ with scalar shape parameter $\alpha > 0$ and scalar inverse scale parameter $\beta > 0$,
\begin{align*}
	\Gammapdf{\gamma}{\alpha}{\beta} = {\beta^{\alpha}}{\Gamma(\alpha)^{-1}}\gamma^{\alpha-1}e^{-\beta\gamma},
\end{align*}
where $\Gamma(\cdot)$ is the gamma function.

$\bullet \ $ $\Npdfbig{\sx}{\GaussMean}{P}$ denotes a multi-variate Gaussian pdf defined over the vector $\sx\in\mathbb{R}^{n_x}$ with mean vector $\GaussMean\in\mathbb{R}^{n_x}$, and covariance matrix $P\in\mathbb{S}_{+}^{n_x}$,
	\begin{align*}
	\Npdfbig{\sx}{\GaussMean}{P} = \frac{\exp\left({-\frac{1}{2}\left(\sx-\GaussMean\right)^{\tp}P^{-1}\left(\sx-\GaussMean\right)}\right)}{\left(2\pi\right)^{\frac{n_x}{2}}\det(P)^{\frac{1}{2}}}.
	\end{align*}
	where $\det (\cdot)$ is the matrix determinant function.

$\bullet \ $ $\IWishpdf{\ext}{v}{V}$ denotes an inverse Wishart pdf defined over the matrix $\ext\in\mathbb{S}_{++}^{d}$ with scalar degrees of freedom $v>2d$ and parameter matrix $V\in\mathbb{S}_{++}^{d}$, \cite[Definition 3.4.1]{GuptaN:2000}
	\begin{align*}
	\IWishpdf{\ext}{v}{V} = \frac{2^{-\frac{v-d-1}{2}}\det(V)^{\frac{v-d-1}{2}}}{\Gamma_{d}\left(\frac{v-d-1}{2}\right)\det(\ext)^{\frac{v}{2}}}\etr\left(-\frac{1}{2}\ext^{-1}V\right),
	\end{align*}
	where $\etr(\cdot) = \exp \left(\tr(\cdot)\right)$ is exponential of the matrix trace, and $\Gamma_{d}\left(\cdot\right)$ is the multivariate gamma function. The multivariate gamma function can be expressed as a product of ordinary gamma functions, see \cite[Theorem 1.4.1]{GuptaN:2000}.

$\bullet \ $ $\Wishpdf{\ext}{w}{W}$ denotes a Wishart pdf defined over the matrix $\ext\in\mathbb{S}_{++}^{d}$ with scalar degrees of freedom $w\geq d$ and parameter matrix $W\in\mathbb{S}_{++}^{d}$, \cite[Definition 3.2.1]{GuptaN:2000}
	\begin{align*}
		\Wishpdf{\ext}{w}{W} = \frac{2^{-\frac{wd}{2}}\det(\ext)^{\frac{w-d-1}{2}}}{\Gamma_{d}\left(\frac{w}{2}\right)\det(W)^{\frac{n}{2}}}\etr\left(-\frac{1}{2}W^{-1}\ext\right).
	\end{align*}
\rule[0pt]{\columnwidth}{1pt}
\end{table}


\section{Previous work and paper contributions}

\subsection{Overview of random matrix framework}
Notation is given in Table~\ref{tab:notation}. In the random matrix extended target model, originally proposed by Koch in \cite{Koch:2008}, the extended target state is the combination of a kinematic state vector $\sx_{k}$ and an extension matrix $\ext_{k}$. The vector $\sx_{k}$ represents the target's position and kinematics, and the matrix $\ext_{k}$ represents the target's size and shape, \iep its spatial extension. The matrix $\ext_{k}$ is modeled as being symmetric and positive definite, which implies that the target shape is approximated by an ellipse. The ellipse shape may seem limiting, however the model is applicable to many real scenarios, e.g. pedestrian tracking \cite{GranstromO:2012a}.

In \cite{Koch:2008} the target state, the target generated measurements, and the transition density, are modeled as
\begin{subequations}
\begin{align}
p\left(\sx_{k},\ext_{k}|\setZ^{k}\right) = & p\left(\sx_{k}|\ext_{k},\setZ^{k}\right)p\left(\ext_{k}|\setZ^{k}\right) \\
= & \Npdfbig{\sx_{k}}{m_{k|k}}{P_{k|k}\otimes\ext_{k}} \nonumber \\
& \times \IWishpdf{\ext_{k}}{v_{k|k}}{V_{k|k}}, \label{eq:Koch_GIW_model}\\
p\left(\sz_{k}|\sx_{k},\ext_{k}\right) = & \Npdfbig{\sz_{k}}{H_{k}\sx_{k}}{\ext_{k}}. \label{eq:Linear_Gaussian_Measurements}\\
p\left(\sx_{k+1},\ext_{k+1}|\sx_{k},\ext_{k}\right) = & p\left(\sx_{k+1}|\ext_{k+1},\sx_{k}\right)p\left(\ext_{k+1}|\ext_{k}\right), \label{eq:Koch_transition_density}
\end{align}%
\label{eq:Koch_random_matrix_model}%
\end{subequations}%
In this model the kinematic state $\sx_{k}$ consists of a spatial state component $\mathbf{r}_{k}$  (target position), and derivatives of $\mathbf{r}_{k}$ (typically velocity and acceleration) \cite{Koch:2008}. Non-linear dynamics, such as turn-rate, are not included in the kinematic vector. The measurement update is linear \cite{Koch:2008}, a derivation of the predicted likelihood can be found in \cite[Appendix A]{GranstromO:2012a}. A linear Gaussian transition density is used for the kinematic state, and for the extension a simple heuristic is used in which the expected value is kept constant and the variance is increased \cite{Koch:2008}. The extension transition density $p\left(\ext_{k+1}|\ext_{k}\right)$ in \eqref{eq:Koch_transition_density} assumes independence of the kinematic state $\sx_{k}$, which does not account for, e.g., rotations during a turning maneuver \cite{GranstromO:2014}.

The random matrix model \eqref{eq:Koch_random_matrix_model} was modified in \cite{FeldmannF:2008,FeldmannFK:2011}, where the target state, the target generated measurements, and the transition density, are modeled as
\begin{subequations}
\begin{align}
p\left(\sx_{k},\ext_{k}|\setZ^{k}\right)  \approx &  p\left(\sx_{k}|\setZ^{k}\right)p\left(\ext_{k}|\setZ^{k}\right) \label{eq:Feldmann_random_matrix_state} \\
= & \Npdfbig{\sx_{k}}{m_{k|k}}{P_{k|k}} \nonumber \\
& \times\IWishpdf{\ext_{k}}{v_{k|k}}{V_{k|k}},\\
p\left(\sz_{k}|\sx_{k},\ext_{k}\right) = & \Npdfbig{\sz_{k}}{H_{k}\sx_{k}}{z\ext_{k}+R}, \label{eq:Feldmann_measurement_model} \\
p\left(\sx_{k+1},\ext_{k+1}|\sx_{k},\ext_{k}\right) = & p\left(\sx_{k+1}|\sx_{k}\right)p\left(\ext_{k+1}|\ext_{k}\right), \label{eq:Feldmann_transition_density}
\end{align}%
\label{eq:Feldmann_random_matrix_model}%
\end{subequations}%
where $z$ is a scaling factor and $R$ is measurement noise. Note the assumed independence between the kinematic state $\sx_{k}$ and $\ext_{k}$ in \eqref{eq:Feldmann_random_matrix_state}, an assumption that cannot be fully theoretically justified\footnote{Conditioned on a set of measurements $\setZ$ the kinematic state $\sx$ and extension state $\ext$ are necessarily dependent.}. Further the measurement update is no longer linear and must be approximated, see \cite{FeldmannFK:2011} for details. However, there are considerable practical advantages to the model \cite{FeldmannFK:2011}.

This model allows for a more general class of kinematic state vectors $\sx_{k}$, e.g. including non-linear dynamics such as heading and turn-rate, and the Gaussian covariance is no longer intertwined with the extension matrix. This measurement model is better when the size of the extension and the size of the sensor noise are within the same order of magnitude \cite{FeldmannFK:2011}. The assumed independence between $\sx_{k}$ and $\ext_{k}$ is alleviated in practice by the measurement update which provides for the necessary interdependence between kinematics and extension estimation \cite{FeldmannFK:2011}. An alternative measurement update for the measurement model \eqref{eq:Feldmann_measurement_model}, based on variational Bayes approximation, is given in \cite{Orguner:2012}.

The kinematics transition density $p\left(\sx_{k+1}|\sx_{k}\right)$ in \eqref{eq:Feldmann_transition_density} is assumed independent of the extension. This neglects factors such as wind resistance, which can be modeled as a function of the extension $\ext_{k}$, however the assumption is necessary to retain the functional form \eqref{eq:Feldmann_random_matrix_state} in a Bayesian recursion. A linear Gaussian transition density is used for the kinematic state, and a heuristic transition similar to the one in \cite{Koch:2008} is used for the extension.

An alternative to the heuristic extension predictions from \cite{Koch:2008,FeldmannFK:2011} is to use a Wishart transition density \cite{Koch:2008}, see also \cite{LianHLYZ:2010,LanRL:2012,GranstromO:2014}. In \cite{LanRL:2012} transformations of the extension are allowed via parameter matrices $A_{k}$,
\begin{align}
p(\ext_{k+1}|\ext_{k})=&\Wishpdf{\ext_{k+1}}{\delta_{k}}{A_{k}\ext_{k}A_{k}^{\tp}}. \label{eq:LanRongLi_prediction}
\end{align}
The parameter matrices correspond to, \egp, rotation matrices. This is generalized in \cite{GranstromO:2014} to allow for transformation matrices $M(\sx_{k})$ that are functions of the kinematic state,
\begin{align}
p(\ext_{k+1}|\sx_{k},\ext_{k}) = & \Wishpdf{\ext_{k+1}}{n_{k}}{\frac{M({\sx_{k}})\ext_{k}(M({\sx_{k}}))^{\tp}}{n_{k}}}, \label{eq:GranstromOrguner_prediction}
\end{align}
which means that the rotation angle can be coupled to, e.g., the turn-rate and estimated online. The transition density \eqref{eq:GranstromOrguner_prediction} relaxes the assumption (made in \eqref{eq:Koch_transition_density}, \eqref{eq:Feldmann_transition_density}, and \eqref{eq:LanRongLi_prediction}) that the extension's time evolution is independent of the kinematic state. A comparison of the models \eqref{eq:Koch_transition_density}, \eqref{eq:Feldmann_transition_density}, \eqref{eq:LanRongLi_prediction} and \eqref{eq:GranstromOrguner_prediction} is given in \cite{GranstromO:2014}, where \eqref{eq:GranstromOrguner_prediction} is shown to give lower errors at lower computational complexity.

In addition to the transition density \eqref{eq:LanRongLi_prediction}, a measurement model is also suggested in \cite{LanRL:2012},
\begin{align}
p\left(\sz_{k}|\sx_{k},\ext_{k}\right) = & \Npdfbig{\sz_{k}}{H_{k}\sx_{k}}{B_{k}\ext_{k}B_{k}^{\tp}} \label{eq:LanRL_measurement_model}
\end{align}
where $B_{k}$ is a parameter matrix. Under the assumption $\ext_{k}\approx\hat{\ext}_{k|k-1}=\E[\ext_{k}|\setZ^{k-1}]$  the model \eqref{eq:LanRL_measurement_model} incorporates \eqref{eq:Feldmann_measurement_model} approximately when $B_{k} = (z\hat{\ext}_{k|k-1}+R)^{1/2}\hat{\ext}_{k|k-1}^{-1/2}$.

The random matrix model has been integrated into the Probabilistic Multi-Hypothesis Tracking (\textsc{pmht}) framework \cite{StreitL:1993}, see \cite{WienekeK:2010,WienekeD:2011,WienekeK:2012}. The model has also been used in \phd- and \cphd-filters for multiple extended target tracking in clutter, see \cite{GranstromO:2012a,LundquistGO:2013}.

In \cite{LanRL:2014} a single extended target model is given where the extended target is a combination of multiple subobjects with kinematic state vectors $\sx_{k}^{(i)}$ and extension matrices $\ext_{k}^{(i)}$. Each subobject is modeled using \eqref{eq:Koch_GIW_model},
\begin{align}
\Npdfbig{\sx_{k}^{(i)}}{m_{k|k}^{(i)}}{P_{k|k}^{(i)}\otimes\ext_{k}^{(i)}}\IWishpdf{\ext_{k}^{(i)}}{v_{k|k}^{(i)}}{V_{k|k}^{(i)}}. \label{eq:LanRongLiStateModel}
\end{align}
Using multiple instances of a simpler shape alleviates the limitations posed by the implied elliptic target shape\footnote{As the number of ellipses grows, their combination can form nearly any given shape.}, and also retains, on a subobject level, the simplicity of the random matrix model \cite{Koch:2008,FeldmannFK:2011}.


\subsection{Paper contributions}
The major contributions in this paper are:

$\bullet$ A new state space model where all subobjects's positions are modeled as fully dependent using a single state vector, there are fully unified kinematics, and where the measurement rate and extension of each subobject are individually modeled.

$\bullet$ A derivation of the prediction update, the measurement update, and the predicted likelihood.

$\bullet$ A computationally effcient gamma Gaussian inverse Wishart implementation, including an initialization method that does not rely on any a priori information about the target.

$\bullet$ A simple and effective method that minimizes the number of association events that have to be considered without relying on a predicted target estimate. The method is capable of handling partial occlusion of the extended target, i.e. one or a few of the subobjects are hidden from sensor view.

$\bullet$ The proposed extended target model is validated on simulated data from realistic scenarios, and the results are compared to previous work on the topic. 


\section{Proposed multiple ellipse model}
\label{sec:new_multi_ellipse_model}

\subsection{Extended target state}
The extended target is made up of a combination of $N_{s,k}$ $d$-dimensional subobjects, where $N_{s,k}$ is known. Each subobject $i$ is described by a position $\mathbf{p}_{k}^{(i)}\in\mathbb{R}^{d}$, a measurement rate $\gamma_{k}^{(i)}>0$ and an extension state $\ext_{k}^{(i)}\in\mathbb{S}_{++}^{d}$, where sub-index $k$ refers to discrete time step $t_k$. The measurement rate governs how many measurements the subobject generates per time step, and the extension describes the size and the shape of the subobject. Because extended targets in most cases can be assumed to be rigid bodies the subobjects have unified dynamics, by which we mean that all subobjects move forward with the same velocity and the same heading, turn with the same turn-rate, etc. The unified dynamics are denoted $\mathbf{c}_{k}\in\mathbb{R}^{n_c}$, where $\mathbf{c}_{k}$ includes parameters for, \egp, velocity, acceleration, heading and turn-rate. Note that $\mathbf{c}_{k}$, in addition to parameters for unified dynamics, also may include parameters for individual subobject dynamics. This is useful for group tracking, where the individual targets in the group may shift their positions within the group.

The positions $\mathbf{p}_{k}^{(i)}$ of the subobjects are $\mathbf{p}_{k}^{(1)} = \mathbf{p}_{k}$ and $\mathbf{p}_{k}^{(i)} = \mathbf{p}_{k} + \mathbf{d}_{k}^{(i)}$, $i=2,\ldots,N_{s,k}$,
\iep the positions of subobjects $i=2,\ldots,N_{s,k}$ are offset by vectors $\mathbf{d}_{k}^{(i)}\in\mathbb{R}^{d}$ from the first subobject's position $\mathbf{p}_{k}\in\mathbb{R}^{d}$. The first subobject is also referred to as the main subobject, and the position $\mathbf{p}_{k}$ is referred to as the overall position. The unified kinematics are defined w.r.t. the overall position. For linear dynamics it does not matter which subobject is denoted the first subobject. However, this is important for non-linear dynamics, \egp a turning maneuver that causes the extended target to rotate.

The positions and dynamics of all subobjects are jointly described by a kinematic state $\sx_{k}\in\mathbb{R}^{n_x}$,
\begin{align}
\sx_{k} = \begin{bmatrix} \mathbf{p}_{k}^{\tp} & \mathbf{c}_{k}^{\tp} & \left(\mathbf{d}_{k}^{(2)}\right)^{\tp} & \ldots &  \left(\mathbf{d}_{k}^{(N_{s,k})}\right)^{\tp} \end{bmatrix}^{\tp}. \label{eq:kinematic_state_vector}
\end{align}
Note that the position of one of the subobjects must coincide with the overall position, because if $\mathbf{p}_{k}^{(i)} = \mathbf{p}_{k} + \mathbf{d}_{k}^{(i)}$ for all $i$ then $\mathbf{p}_{k}$ is not observable.

For brevity the measurement rates, kinematic state and extension states are abbreviated as follows
\begin{subequations}
\begin{align}
\xi_{k} = & \left(\gamma_{k}^{(1)},\ldots,\gamma_{k}^{(N_{s,k})},\sx_{k},\ext_{k}^{(1)},\ldots,\ext_{k}^{(N_{s,k})}\right) \\
= & \left(\setGamma_{k},\sx_{k},\setX_{k}\right)
\end{align}%
\end{subequations}%
where $\xi_{k}$ is referred to as the extended target state. Let $\setZ_{k}$ be a set of target generated measurements $\setZ_{k} = \{\sz_{k}^{(j)}\}_{j=1}^{n_{z,k}}$, $\sz_{k}^{(j)}\in\mathbb{R}^{d}, \ \forall j$,
and let $\setZ^{k}$ be a sequence of measurement sets from time $t_{0}$ to time $t_{k}$. The distribution of the extended target state $\xi_{k}$, conditioned on the history of measurement sets, is
\begin{align}
p\left(\xi_{k}\left|\setZ^{k}\right.\right)= & p\left(\setGamma_{k}\left|\sx_{k},\setX_{k},\setZ^{k}\right.\right) p\left(\sx_{k}\left|\setX_{k},\setZ^{k}\right.\right) p\left(\setX_{k}\left|\setZ^{k}\right.\right),
\end{align}%
The following assumptions are made about the state $\xi_{k}$.
\begin{assumption}
The measurement rates are independent of the kinematic vector and the random matrices.\hfill$\square$
\end{assumption}
\begin{remark}
In reality the measurement rate is often dependent on the distance to the sensor (i.e. on the position) and on the size of the target (i.e. the extension). However, the distribution of the measurement rate, conditioned on the kinematic and extension states, is unknown in many applications. The variance of the estimated measurement rate is sufficient to model the variations over time \cite{GranstromO:2012c}, and the assumption ensures practical computational tractability. 
\hfill$\square$\end{remark}
\begin{assumption}
The kinematic vector is approximated as independent of the random matrices.\hfill$\square$
\end{assumption}
\begin{remark}
This assumption is analogous to \eqref{eq:Feldmann_random_matrix_state}, and similarly it neglects dependence between the subobjects positions and extension states -- an assumption that cannot be fully theoretically justified. However, the assumption is necessary to enable the subobject positions and unified kinematics to be modeled as a single state vector. Just as in \cite{FeldmannFK:2011,GranstromO:2014}, the time update and measurement update that are derived in this paper will provide for the necessary practical interdependency between the kinematic state estimate and the extension state estimate. 
\hfill$\square$\end{remark}
\begin{assumption}
The measurement rates are independent of each other. The same holds for the random matrices.\hfill$\square$
\end{assumption}
\begin{remark}
For symmetric targets, \egp airplanes, the subobjects that correspond to the wings are not fully independent since the wings are symmetric. Modeling how random matrices, or measurement rates, depend on each other is difficult in a general case, and the assumption simplifies further analysis greatly and ensures practical computational tractability.  An important topic for future work is to consider how symmetry can be used to relax this assumption. For the random matrices, the same is assumed in \cite{LanRL:2014}, see \eqref{eq:LanRongLiStateModel}. 
\hfill$\square$\end{remark}

This gives the following extended target state distribution,
\begin{align}
p \left(\xi_{k} \left| \setZ^{k}\right.\right) = & p\left(\sx_{k}\left|\setZ^{k}\right.\right) \prod_{i=1}^{N_{s,k}} p\left(\gamma_{k}^{(i)} \left| \setZ^{k}\right.\right) p\left(\ext_{k}^{(i)} \left| \setZ^{k} \right.\right) . \label{eq:multiple_ellipse_ET_distribution}%
\end{align}%

Because of the many uncertainties involved the extended target state distribution can also be represented by a distribution mixture. In this case the distribution is
\begin{align}
p \left(\xi_{k} \left| \setZ^{k}\right.\right) = \sum_{\ell=1}^{J} w^{(\ell)} p^{(\ell)} \left(\xi_{k} \left| \setZ^{k}\right.\right),
\end{align}
where $J$ is the number of components in the mixture, the weights $w^{(\ell)}$ sum to unity, and each $p^{(\ell)} \left(\cdot\right)$ are of the form \eqref{eq:multiple_ellipse_ET_distribution}. The weights can be interpreted as probabilities that the $\ell$:th mixture component $p^{(\ell)} \left(\cdot\right)$ is the true distribution.

\subsection{Prediction}
For brevity and increased readability, in this section we drop sub-index $_k$ and write sub-index $_{k+1}$ as sub-index $_{+}$, \iep we write $\xi$ and $\xi_{+}$ instead of $\xi_{k}$ and $\xi_{k+1}$. The state transition density $p\left(\xi_{+}|\xi_{}\right)$ describes the time evolution of the extended target state from time $t_{}$ to time $t_{+}$. The transition density decomposes as follows
\begin{align}
p\left(\xi_{+}|\xi_{}\right) = & p\left( \setGamma_{+} \left|\sx_{+},\setX_{+},\setGamma_{} \right. \right) p\left(\sx_{+} \left|\setX_{+}, \setGamma_{},\sx_{} \right.\right) \nonumber \\
& \times p\left(\setX_{+} \left|\setGamma_{},\sx_{},\setX_{} \right.\right),
\end{align}
where we have used Bayes rule and Markov-property assumptions. We now make the following assumptions:
\begin{assumption}
The measurement rates can be predicted independently of the kinematic vector and the random matrices. \hfill$\square$
\end{assumption}
\begin{remark}
This assumption is also made in \cite{GranstromO:2012c}. In reality the measurement rates typically depend on both the size of the target, and its distance from the sensor (\iep depends on the position), and this assumption neglects such dependencies. However, constructing a general model for these dependencies seems difficult, and the assumption simplifies further analysis significantly. Furthermore, the variances of the estimated measurement rates are sufficient to capture the variations over time \cite{GranstromO:2012c}.
\hfill$\square$\end{remark}
\begin{assumption}
The kinematic vector and the random matrices can be predicted independently of the prior measurement rates $\setGamma_{}$.\hfill$\square$
\end{assumption}
\begin{remark}
This assumption can be justified in most -- if not all -- practical cases, since neither the kinematics nor the size and shape of a target evolve differently depending on how many sensor detections the target generates.
\hfill$\square$\end{remark}
\begin{assumption}
The kinematic vector can be predicted independently of the random matrices.\hfill$\square$
\end{assumption}
\begin{remark}
This assumption is analogous to \eqref{eq:Feldmann_transition_density}, and similarly it neglects aspects such as wind resistance, which can be modeled as dependent on the target's size (\iep dependent on the random matrices). In this work the assumption is necessary for the predicted state distribution to be of the same functional form as the posterior state distribution, which is a typical requirement in Bayesian estimation. Both the measurement update and the prediction update that are used in this work provide for interdependency between the predicted kinematic vector and the random matrices. For single ellipse targets it is shown in \cite{GranstromO:2014} that estimation performance is not negatively affected by the assumption.
\hfill$\square$\end{remark}
\begin{assumption}
Each measurement rate can be predicted independently of the other measurement rates. The same holds for the random matrices.\hfill$\square$
\end{assumption}
\begin{remark}
In practice the measurement rates and extensions evolve over time dependent on the position and kinematics. Because the rates and extensions are parts of the same object (i.e. subobjects), the evolution over time should be dependent. Under this assumption this dependence is not modeled, however the assumption is necessary for the predicted state distribution to be of the same functional form as the posterior. For the measurement rates the same motion model is used for all estimated rates, and the estimated variances are sufficient to model the time variations. For the extension states, the kinematic state is used in the prediction, which introduces sufficient dependence.
\hfill$\square$\end{remark}
This gives the following transition density for $\xi_{}$,
\begin{align}
p \left(\xi_{+}|\xi_{}\right) = & p\left(\sx_{+} \left| \sx_{} \right.\right) \prod_{i=1}^{N_{s,k}} p\left( \gamma_{+}^{(i)} \left| \gamma_{}^{(i)} \right. \right) p\left( \ext_{+}^{(i)} \left| \sx_{}, \ext_{}^{(i)} \right. \right) . \label{eq:multiple_ellipse_transition_density}
\end{align}%
With a posterior \eqref{eq:multiple_ellipse_ET_distribution} and a transition density \eqref{eq:multiple_ellipse_transition_density} the Bayes predicted distribution is
\begin{subequations}
\begin{align}
p & \left(\xi_{+}\left|\setZ^{k}\right.\right) = \int p(\xi_{+}|\xi) p\left(\xi|\setZ^{k}\right) \diff \xi_{} \\
= & \left(\prod_{i=1}^{N_{s,k}}\int p\left( \gamma_{+}^{(i)} \left| \gamma_{}^{(i)} \right. \right) p\left(\gamma_{}^{(i)}\left|\setZ^{k}\right.\right) \diff\gamma_{}^{(i)}\right) \\
& \times \int \left( \prod_{i=1}^{N_{s,k}}\int p\left(\ext_{+}^{(i)} \left| \sx_{},\ext_{}^{(i)} \right.\right) p\left(\ext_{}^{(i)} \left| \setZ^{k}\right.\right) \diff\ext_{}^{(i)} \right) \nonumber \\
& \hspace{10mm}  \times p\left(\sx_{+} \left| \sx_{} \right.\right) p\left(\sx_{}\left|\setZ^{k}\right.\right) \diff\sx_{} \nonumber \\
= & \left(\prod_{i=1}^{N_{s,k}} p\left( \gamma_{+}^{(i)}\left|\setZ^{k}\right.\right) \right) \\
& \times \int \left( \prod_{i=1}^{N_{s,k}} p\left(\ext_{+}^{(i)} \left| \sx_{} , \setZ^{k} \right.\right) \right) p\left(\sx_{+} \left| \sx_{} \right.\right) p\left(\sx_{}\left|\setZ^{k}\right.\right)  \diff\sx_{} \nonumber
\end{align}
\end{subequations}
We want the predicted distribution to be of the same functional form as the posterior \eqref{eq:multiple_ellipse_ET_distribution}, however in general the following equality does not hold,
\begin{align}
\int & \left( \prod_{i=1}^{N_{s,k}} p\left(\ext_{+}^{(i)} \left| \sx_{} , \setZ^{k} \right.\right) \right) p\left(\sx_{+} \left| \sx_{} \right.\right) p\left(\sx_{}\left|\setZ^{k}\right.\right)  \diff\sx_{} \nonumber \\
= & p\left(\sx_{+}\left|\setZ^{k}\right.\right) \left( \prod_{i=1}^{N_{s,k}} p\left(\ext_{+}^{(i)} \left| \setZ^{k} \right.\right) \right).
\end{align}
Therefore, to obtain a predicted distribution of the same functional form as the posterior \eqref{eq:multiple_ellipse_ET_distribution}, following the discussion in \cite{GranstromO:2014} we solve independent integrals instead,
\begin{subequations}
\begin{align}
p\left(\sx_{+}\left|\setZ^{k}\right.\right) = & \int p\left(\sx_{+} \left| \sx_{} \right.\right) p\left(\sx_{}\left|\setZ^{k}\right.\right)  \diff\sx_{}, \label{eq:kinematics_prediction_integral} \\
p\left(\ext_{+}^{(i)} \left| \setZ^{k} \right.\right) = & \int p\left(\ext_{+}^{(i)} \left| \sx_{} , \setZ^{k} \right. \right) p\left(\sx_{}\left|\setZ^{k}\right.\right)  \diff\sx_{}. \label{eq:extension_prediction_integral}
\end{align}%
\end{subequations}%
With a mixture distribution, the predicted mixture is the mixture of the predicted components,
\begin{subequations}
\begin{align}
p\left(\xi_{+}\left|\setZ^{k}\right.\right) = & \int p(\xi_{+}|\xi) \sum_{\ell=1}^{J} w^{(\ell)} p^{(\ell)} \left(\xi \left| \setZ^{k}\right.\right) \diff \xi_{} \\
= & \sum_{\ell=1}^{J} w^{(\ell)} p^{(\ell)}\left(\xi_{+}\left|\setZ^{k}\right.\right).
\end{align}
\end{subequations}

\subsection{Correction}
\label{sec:multi_ellipse_correction}
Let $\theta$ denote a possible measurement-to-subobject association event, and let $\setTheta$ denote the set of all possible association events.  For measurement generation, we assume the following:
\begin{assumption}
The subobjects generate measurements independently of each other.
For each subobject, the generated measurements are independent.
Each measurement is generated by exactly one subobject.
Measurement origin is unknown. \hfill$\square$
\end{assumption}
\begin{remark}
These assumptions are analogous to multiple target tracking, where it is typically assumed that each target generates measurements independently of the other targets, that the target generated measurements are independent, that each measurement is generated by exactly one target, and that measurement origin is unknown, see \egp \cite{BarShalomWT:2011}.
\hfill$\square$\end{remark}

Under an association event $\theta$ the measurement set $\setZ_{k}$ can be partitioned into $N_{s,k}$ (possibly empty) subsets,
\begin{align}
\setZ_{k} = & \bigcup_{i=1}^{N_{s,k}} \setZ_{k}^{(\theta,i)}, & \setZ_{k}^{(\theta,i)} = & \left\{\sz_{k}^{(\theta,i,j)}\right\}_{j=1}^{n_{z,k}^{(\theta,i)}},
\end{align}
where the $i$th subset $\setZ_{k}^{(\theta,i)}$ was generated by the $i$th subobject. Conditioned on $\theta$ the measurement likelihood is
\begin{subequations}
\begin{align}
p\left(\setZ_{k}\left|\xi_{k},\theta\right.\right) = & \prod_{i=1}^{N_{s,k}} p\left(\setZ_{k}^{(\theta,i)}\left|\gamma_{k}^{(i)},\sx_{k},\ext_{k}^{(i)}\right.\right).
\end{align}%
\end{subequations}%
If the $i$th subset is empty (\iep $n_{z,k}^{(\theta,i)}=0$) the subset likelihood is simply the likelihood of an empty set of measurements,
\begin{align}
& p\left(\setZ_{k}^{(\theta,i)}\left|\gamma_{k}^{(i)},\sx_{k},\ext_{k}^{(i)}\right.\right) = P\left(n_{z,k}^{(\theta,i)}=0\left| \gamma_{k}^{(i)}\right.\right).
\end{align}%
If $n_{z,k}^{(\theta,i)}>0$ the subobject likelihood is
\begin{subequations}
\begin{align}
p&  \left(\setZ_{k}^{(\theta,i)}\left|\gamma_{k}^{(i)},\sx_{k},\ext_{k}^{(i)}\right.\right) \nonumber \\
= & n_{z,k}^{(\theta,i)}! P\left(n_{z,k}^{(\theta,i)}\left| \gamma_{k}^{(i)}\right.\right) \prod_{j=1}^{n_{z,k}^{(\theta,i)}} p\left(\sz_{k}^{(\theta,i,j)}\left|\sx_{k},\ext_{k}^{(i)}\right.\right).
\end{align}%
\end{subequations}%
Let the predicted mixture distribution be
\begin{align}
p \left(\xi_{k} \left| \setZ^{k-1}\right.\right) = \sum_{\ell=1}^{J} w^{(\ell)} p^{(\ell)} \left(\xi_{k} \left| \setZ^{k-1}\right.\right).
\end{align}
By the total probability theorem the density $p\left(\xi_{k}|\setZ^{k}\right)$ is 
\begin{align}
p\left(\xi_{k}|\setZ^{k}\right) = & \sum_{\theta\in\setTheta} p\left(\xi_{k}|\setZ^{k},\theta\right)P\left(\theta|\setZ^{k}\right),
\label{eq:Bayes_updated_all_assoc_events}
\end{align}
where $p\left(\xi_{k}|\setZ^{k},\theta\right)$ is the Bayes updated distribution for the association event $\theta$, and $P\left(\theta|\setZ^{k}\right)$ is the probability of the association event $\theta$. As noted in \cite[Assumption 3]{LanRL:2014}, without any prior information the association events can be assumed to be equally likely, \iep $P\left(\theta|\setZ^{k-1}\right)= |\setTheta|^{-1}$. In this case we have
\begin{subequations}
\begin{align}
P\left(\theta|\setZ^{k}\right) = & \frac{p\left(\setZ_{k}|\theta,\setZ^{k-1}\right)P\left(\theta|\setZ^{k-1}\right)}{\sum_{\theta'\in\setTheta}p\left(\setZ_{k}|\theta',\setZ^{k-1}\right)P\left(\theta'|\setZ^{k-1}\right)} \\
= & \frac{\sum_{\ell=1}^{J} w^{(\ell)} p^{(\ell)}\left(\setZ_{k}|\theta,\setZ^{k-1}\right)}{\sum_{\theta'\in\setTheta} \sum_{\ell'=1}^{J} w^{(\ell')} p^{(\ell')}\left(\setZ_{k}|\theta',\setZ^{k-1}\right)},
\end{align}%
\label{eq:prob_assoc_event}%
\end{subequations}%
where we have again used the total probability theorem in the second equality. For the association event $\theta$ the Bayes updated distribution is
\begin{subequations}
\begin{align}
p\left(\xi_{k}\left|\setZ^{k},\theta\right.\right) = & \frac{p\left(\setZ_{k}\left|\xi_{k},\theta\right.\right) p\left(\xi_{k}\left|\setZ^{k-1}\right.\right)}{\int p\left(\setZ_{k}\left|\xi_{k},\theta\right.\right) p\left(\xi_{k}\left|\setZ^{k-1}\right.\right) \diff \xi_{k}} \\
= & \frac{ \sum_{\ell=1}^{J} w^{(\ell)} p^{(\ell)}\left(\setZ_{k}\left|\theta,\setZ^{k-1}\right.\right) p^{(\ell)} \left(\xi_{k} \left| \setZ^{k},\theta\right.\right)}{\sum_{\ell=1}^{J} w^{(\ell)} p^{(\ell)}\left(\setZ_{k}\left|\theta,\setZ^{k-1}\right.\right) }.
\end{align}%
\label{eq:Bayes_updated_single_assoc_event}%
\end{subequations}%
Combining \eqref{eq:Bayes_updated_all_assoc_events}, \eqref{eq:prob_assoc_event} and \eqref{eq:Bayes_updated_single_assoc_event} gives the posterior distribution
\begin{align}
p\left(\xi_{k}|\setZ^{k}\right) = & \sum_{\theta\in\setTheta} \sum_{\ell=1}^{J} w^{(\ell)}\left(\theta\right) p^{(\ell)} \left(\xi_{k} \left| \setZ^{k},\theta\right.\right), \\
w^{(\ell)}\left(\theta\right) = & \frac{w^{(\ell)} p^{(\ell)}\left(\setZ_{k}|\theta,\setZ^{k-1}\right)}{\sum_{\theta'\in\setTheta} \sum_{\ell'=1}^{J} w^{(\ell')} p^{(\ell')}\left(\setZ_{k}|\theta',\setZ^{k-1}\right)},
\end{align}
where, following the assumption that the subobjects generate measurements independently, for the predicted likelihood we have
\begin{subequations}
\begin{align}
& p^{(\ell)}\left(\setZ_{k}|\theta,\setZ^{k-1}\right) = \prod_{i=1}^{N_{s,k}} p^{(\ell)}\left(\left.\setZ_{k}^{(\theta,i)}\right|\setZ^{k-1}\right). \label{eq:meas_lik_prod} \\
& p\left(\setZ_{k}|\setZ^{k-1}\right) = \frac{1}{|\setTheta|} \sum_{\theta\in\setTheta} \sum_{\ell=1}^{J} w^{(\ell)} p^{(\ell)}\left(\setZ_{k}|\theta,\setZ^{k-1}\right),
\end{align}
\end{subequations}
The predicted likelihood $p\left(\setZ_{k}|\setZ^{k-1}\right)$ is useful in a multiple target tracking scenario, \egp if the presented extended target model is used in an implementation of an extended target \phd or \cphd filter \cite{mahler_FUSION_2009_extTarg,LundquistGO:2013}.


\section{A Gamma Gaussian inverse Wishart implementation}
\label{sec:GGIW_implementation}
In this section we give a gamma Gaussian inverse Wishart implementation of the multiple random matrix extended target model outlined above. To handle different types of motion $M_{k}$ different motion models are used.

\subsection{Extended target state distribution}
In Gilholm \etal's extended target model \cite{GilholmGMS:2005,GilholmS:2005} the number of measurements that each target generates is Poisson distributed with a parameter $\varsigma\left(\xi\right)$ that is a function of the extended target state. In practice this means that the expected value of the number of measurements generated by a target with state $\xi$ is $\varsigma\left(\xi\right)$. Here the Poisson model is adopted for each subobject and the parameters are given by the measurement rates, \iep $\varsigma^{(i)}\left(\xi_{k}\right) = \gamma_{k}^{(i)}, \ \forall i$.
The gamma distribution is the conjugate prior for the Poisson distribution's parameter, and the subobjects' measurement rates are modeled as gamma distributed.
Following the random matrix model \cite{Koch:2008,FeldmannFK:2011} the subobjects' random matrices are modeled as inverse Wishart distributed.
The kinematic vector is modeled as Gaussian distributed.
The extended target state distribution is
\begin{subequations}
\begin{align}
p\left(\xi_{k} \left| \setZ^{k}\right.\right) = & \Npdfbig{\sx_{k}}{m_{k|k}}{P_{k|k}} \prod_{i=1}^{N_{s,k}} \bigg( \Gammapdf{\gamma_{k}^{(i)}}{\alpha_{k|k}^{(i)}}{\beta_{k|k}^{(i)}} \big. \nonumber \\
& \times \big. \IWishpdf{\ext_{k}^{(i)}}{v_{k|k}^{(i)}}{V_{k|k}^{(i)}} \bigg) \label{eq:GGIW_extended_target_state} \\
= & \GGIWpdf{\xi_{k}}{\zeta_{k|k}},
\end{align}
\end{subequations}
where $\GGIWpdf{\cdot}{\cdot}$ is introduced for brevity and $\zeta_{k|k}$ denotes all of the involved parameters. As noted above, the extended target state distribution is represented by a distribution mixture
\begin{align}
p\left(\xi_{k} \left| \setZ^{k}\right.\right) = & \sum_{\ell=1}^{J_{k|k}} w_{k|k}^{(\ell)} \GGIWpdf{\xi_{k}}{\zeta_{k|k}^{(\ell)}}, \label{eq:mixtureGGIW_extended_target_state}
\end{align}
where $\sum_{\ell}w_{k|k}^{(\ell)}=1$.

\begin{figure}[!t]
\centerline{
\subfloat{\includegraphics[width=0.25\columnwidth]{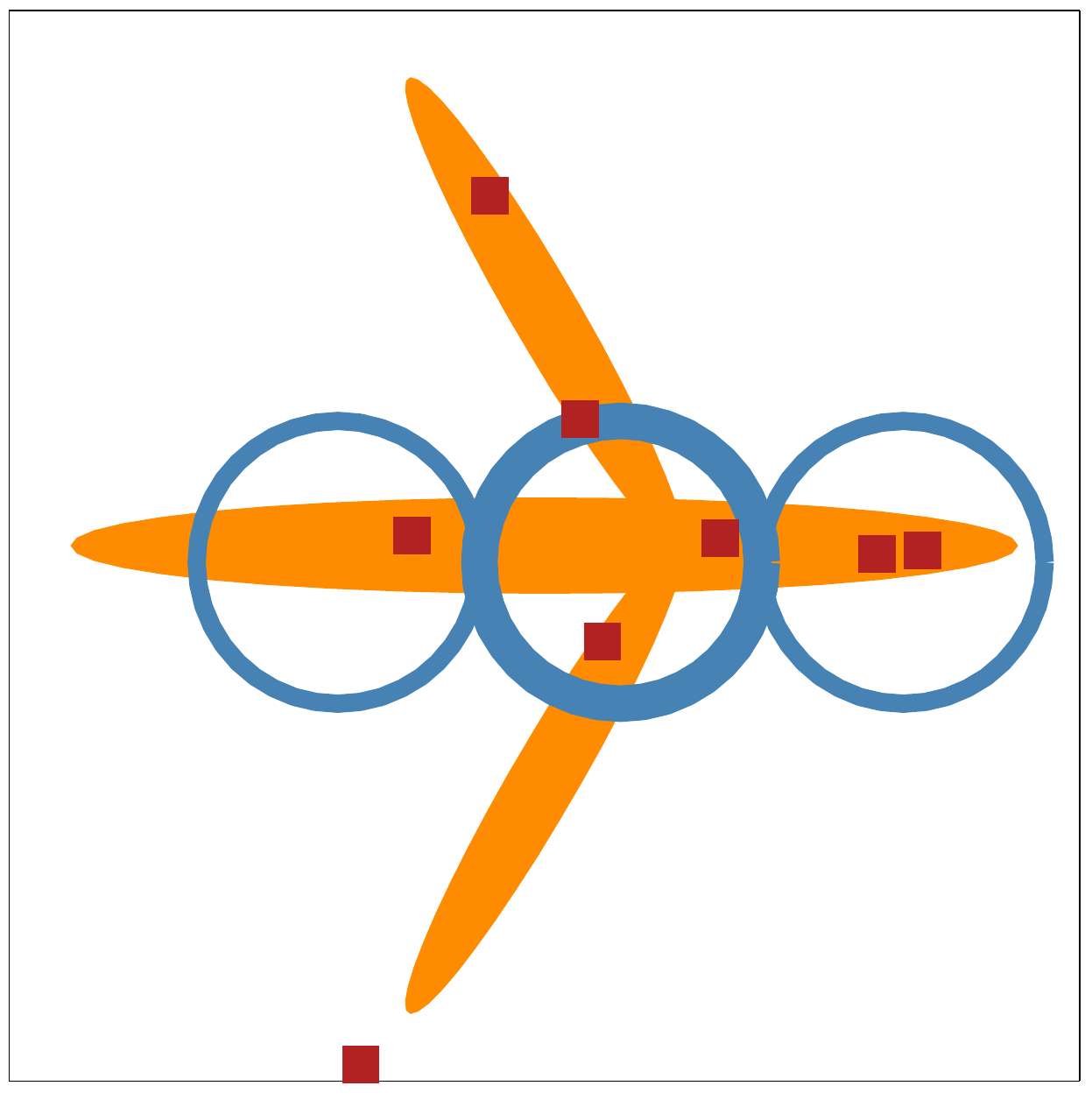} \label{fig:Initialization_Example_1}}
\subfloat{\includegraphics[width=0.25\columnwidth]{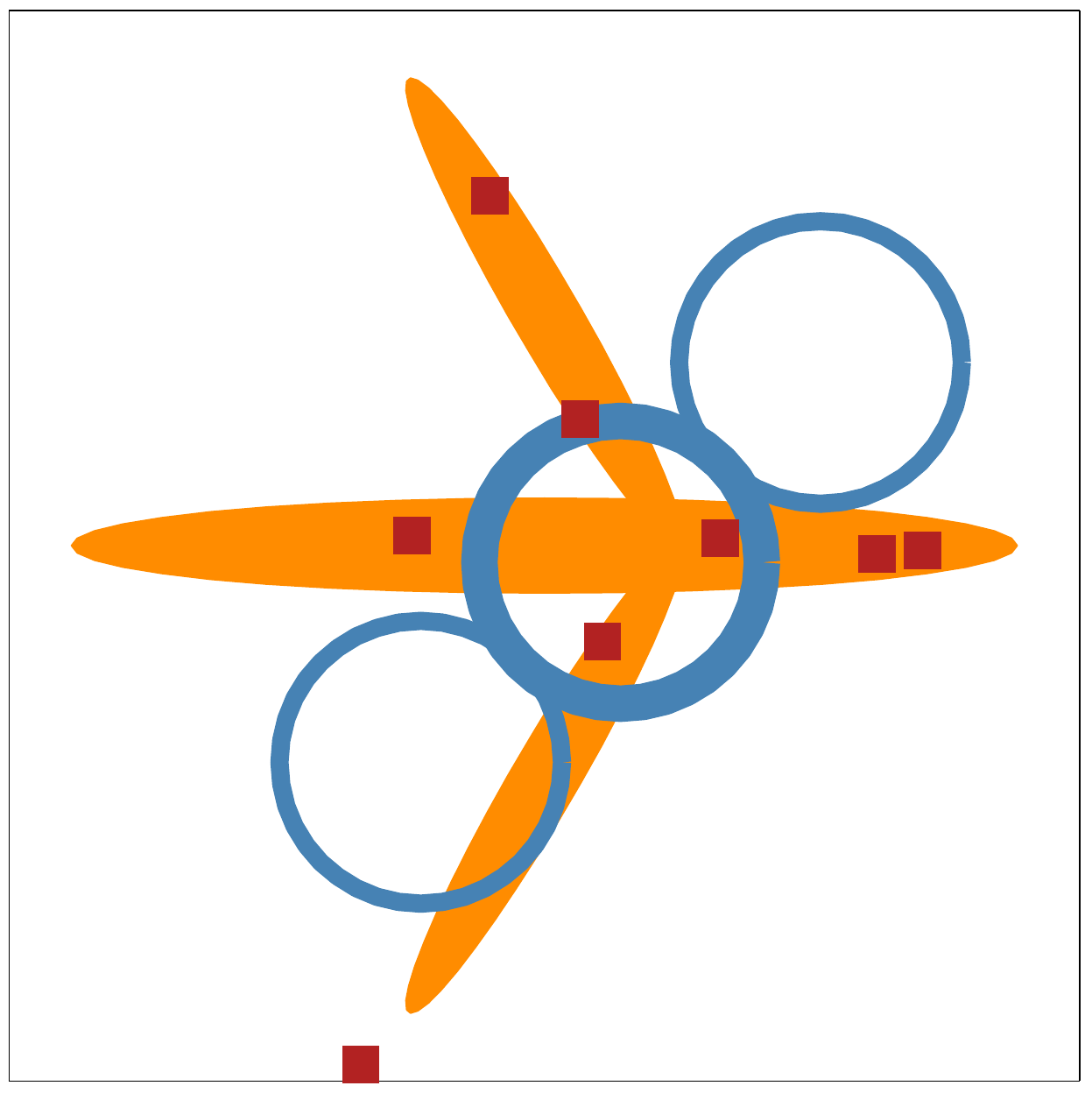} \label{fig:Initialization_Example_2}}
\subfloat{\includegraphics[width=0.25\columnwidth]{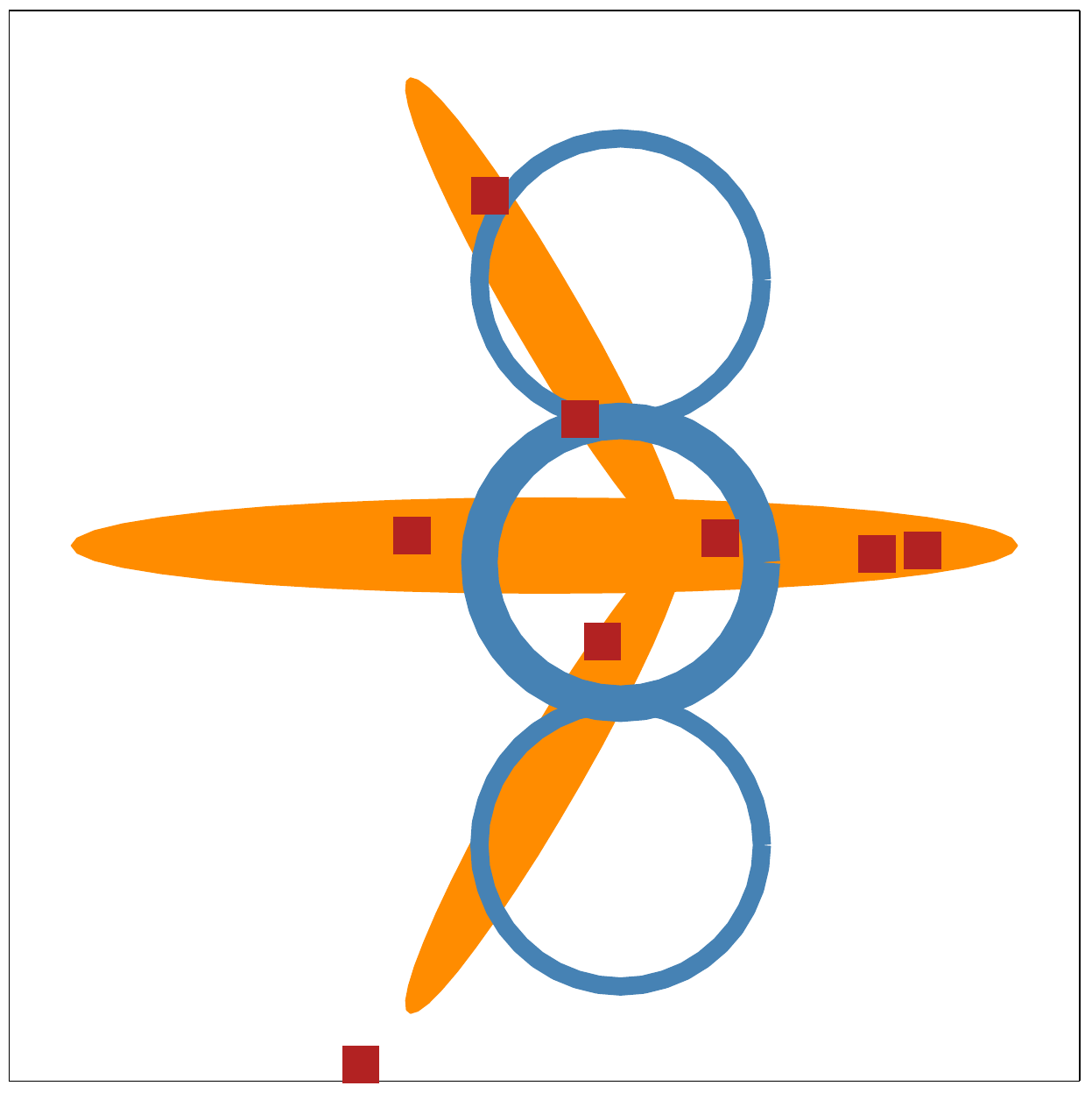} \label{fig:Initialization_Example_3}}
\subfloat{\includegraphics[width=0.25\columnwidth]{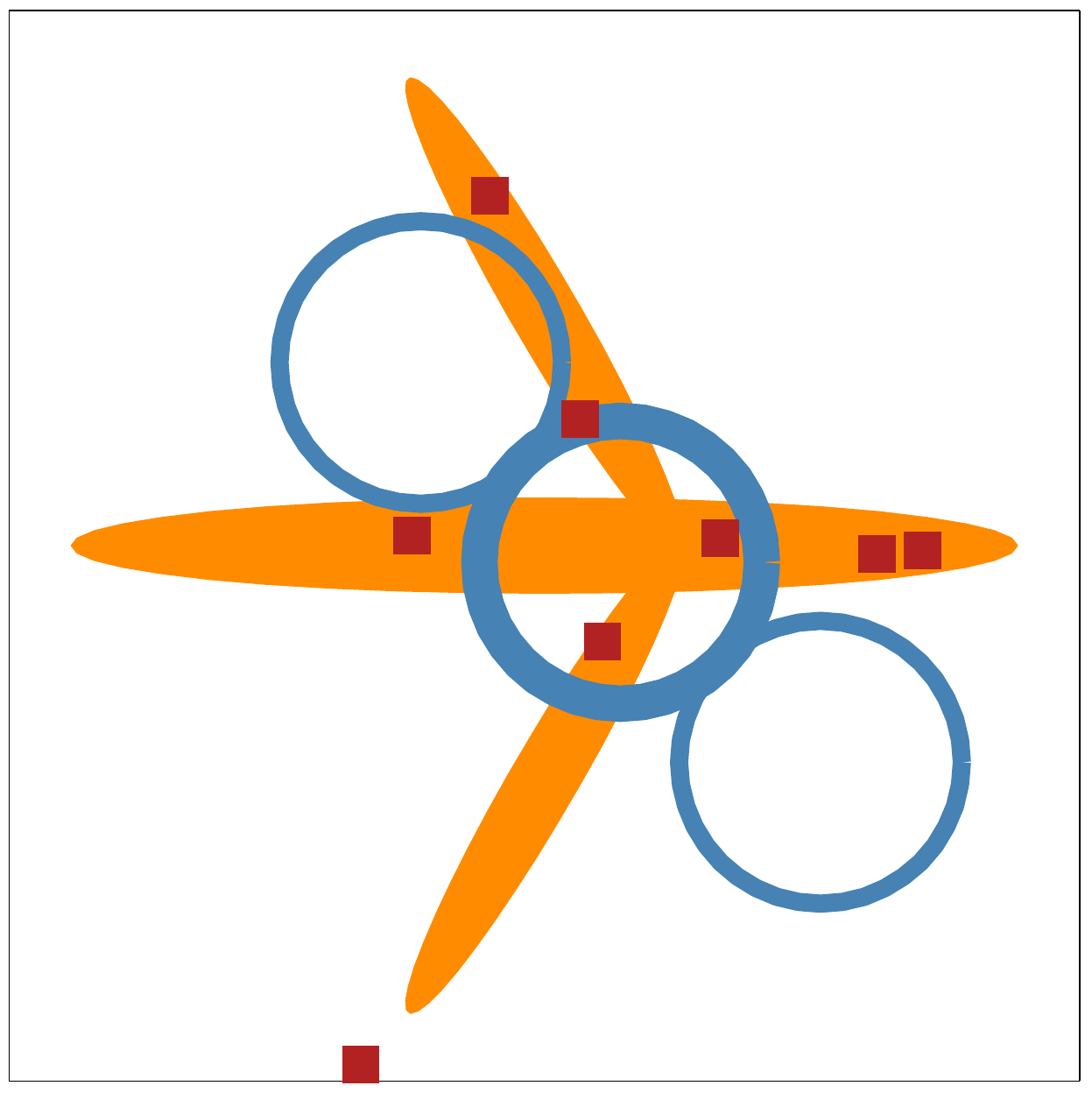} \label{fig:Initialization_Example_4}}
}
\caption{Initialization example. True underlying extended target (orange area), measurements (red squares), and initialized estimates (blue ellipses, initialized main subobject shown by thicker line).}
\label{fig:Initialization_Example}
\end{figure}

When a new target appears the parameters $\zeta^{(\ell)}$ of the estimate must be initialized. Table~\ref{tab:multi_ellipse_initilization} gives a simple algorithm where this is performed using the first set of measurements. The algorithm initializes $N_{p}$ hypotheses in each motion mode.
A simple initialization example is given in Figure~\ref{fig:Initialization_Example_1}. In this example there is a single motion mode, and $N_{p}=4$ hypotheses are generated using only $8$ measurements.

\begin{table}[tb]
\caption{Multiple ellipse parameter initialization}
\label{tab:multi_ellipse_initilization}
\vspace{-0.5cm}
\rule[0pt]{\columnwidth}{1pt}
\vspace{-0.3cm}
\begin{algorithmic}[1]
\STATE \textbf{Input:} Set of measurements $\setZ=\{\sz_{i}\}_{i=1}^{n}$. Desired number of initial hypotheses $N_{p}$. Initial kinematics $\mathbf{c}_{0}$ and initial covariance $P_{0}$. Initial mean $e$ and variance $v$ for measurement rates.
\STATE Define $\sz_{c} = \frac{1}{n}\sum_{i=1}^{n}\sz_{i} $, $r_{z} = \frac{1}{2}\max_{i} \left\| \sz_{i}-\sz_{c} \right\|_{2}$. Set $\ell=0$.
\FOR{$p=1,\ldots,N_{p}$}
	\FOR{$m=1,\ldots,M_{k}$}
		\STATE Set $\ell = \ell+1$
		\STATE $\gamma$: $\alpha_{0}^{(\ell,i)} = \frac{e^2}{v}$, $\beta_{0}^{(\ell,i)} = \frac{e}{v}$.
		\STATE $\sx$: $P_{0}^{(\ell)}=P_{0}$, $\mathbf{p}_{0}^{(\ell)}=\sz_{c}$, $\mathbf{c}_{0}^{(\ell)}=\mathbf{c}_{0}$,
			\begin{align*}
				\mathbf{d}_{0}^{(\ell,i)} = r_{z}\begin{bmatrix} \cos\left(\frac{2\pi(i-2)}{N_{s}-1}+\frac{2\pi(p-1)}{N_{s}N_{p}}\right) \\ \sin\left(\frac{2\pi(i-2)}{N_{s}-1}+\frac{2\pi(p-1)}{N_{s}N_{p}}\right) \end{bmatrix}.
			\end{align*}
		\STATE $\ext$: $v_{0}^{(\ell,i)} = 2d+5$, $V_{0}^{(\ell,i)} = \left(\frac{r_z}{4}\right)^{2}\Id\left(v_{0}^{(\ell,i)}-2d-2\right)$.
	\ENDFOR
\ENDFOR
\STATE \textbf{Output:} $p\left(\xi_{0}\right) = \sum_{\ell=1}^{J_{0}} w_{0}^{(\ell)} \GGIWpdf{\xi_{0}}{\zeta_{0}^{(\ell)}}$ where $w_{0}^{(\ell)} = \frac{1}{J_{0}}$.
\end{algorithmic}
\vspace{-0.1cm}
\rule[0pt]{\columnwidth}{1pt}
\end{table}

\subsection{Prediction}
With a posterior distribution of the form \eqref{eq:mixtureGGIW_extended_target_state} the predicted distribution is 
\begin{align}
p\left(\xi_{k+1} \left| \setZ^{k}\right.\right) = & \sum_{m=1}^{M_{k}}\sum_{\ell=1}^{J_{k|k}} \pi_{m,m'(\ell)} w_{k|k}^{(\ell)} \GGIWpdf{\xi_{k}}{\zeta_{k+1|k}^{(m,\ell)}}, \label{eq:predicted_GGIW_distribution}%
\end{align}%
where $\pi_{m,m'(\ell)}$ is the probability of a transition to the current mode $m$ from the previous mode $m'(\ell)$ that component $\ell$ was in.

\subsubsection{Measurement rates}
For the measurement rates the exponential forgetting prediction from \cite{GranstromO:2012c} is used. For the $m$th motion model the parameters are predicted as
\begin{align}
\alpha_{k+1|k}^{(m,\ell,i)} & = \frac{\alpha_{k|k}^{(\ell,i)}}{\eta_k^{(m)}}, \qquad \beta_{k+1|k}^{(m,\ell,i)} =  \frac{\beta_{k|k}^{(\ell,i)}}{\eta_k^{(m)}},
\end{align}%
which corresponds to keeping the expected value of $\gamma_{k}^{(i)}$ constant, while increasing the variance with a factor $\eta_{k}^{(m)}$ \cite{GranstromO:2012c}. This prediction has an effective window length of $w_{e} =\frac{\eta_{k}^{(m)}}{\eta_{k}^{(m)}-1}$, where $\frac{1}{\eta_k^{(m)}} < 1$ is the forgetting factor.

\subsubsection{Kinematic state}
For the $m$th motion model the kinematic state transition density is modeled as
\begin{align}
p(\sx_{k+1}|\sx_k) = & \Npdfbig{\sx_{k+1}}{f^{(m)}(\sx_k)}{Q_{k+1}^{(m)}},
\end{align}
where $f^{(m)}(\cdot):\mathbb{R}^{n_x}\rightarrow\mathbb{R}^{n_x}$ is a state transition function, and $Q_{k+1}^{(m)}$ is the process noise covariance for the kinematic state. The transition function can be partitioned into $N_{s,k}$ parts,
\begin{align}
f^{(m)}(\sx_k) = & \begin{bmatrix} f_{\mathbf{p},\mathbf{c}}^{(m)}\left(\mathbf{p}_{k},\mathbf{c}_{k}\right)^{\tp} & \ldots & f_{\mathbf{d}}^{(m)}\left(\mathbf{d}_{k}^{(i)},\mathbf{c}_{k}\right)^{\tp} & \ldots \end{bmatrix}^{\tp},
\end{align}
where $f_{\mathbf{p},\mathbf{c}}^{(m)}(\cdot)$ describes the time evolution of the overall position and the kinematics, and $f_{\mathbf{d}}^{(m)}(\cdot)$ describes the time evolution of the subobject offsets. The functions $f_{\mathbf{p},\mathbf{c}}^{(m)}(\cdot)$ and $f_{\mathbf{d}}^{(m)}(\cdot)$ are generally nonlinear, see \cite{RongLiJ:2003} for a thorough overview of state transition functions.

In case $f^{(m)}\left(\sx_{k}\right)$ is a linear function, the solution to \eqref{eq:kinematics_prediction_integral}
is given by the Kalman filter prediction \cite{Kalman:1960}. If $f^{(m)}(\sx_{k})$ is non-linear it is straightforward to solve \eqref{eq:kinematics_prediction_integral} approximately. Using the extended Kalman filter prediction formulas, see \egp \cite{Jazwinski:1970}, the predicted mean $m_{k+1|k}^{(m,\ell)}$ and covariance $P_{k+1|k}^{(m,\ell)}$ are
\begin{subequations}
\begin{align}
m_{k+1|k}^{(m,\ell)} & = f^{(m)}(\mkk^{(\ell)}),\\
P_{k+1|k}^{(m,\ell)} & = F_{k|k}^{(m,\ell)}P_{k|k}^{(\ell)}\left(F_{k|k}^{(m,\ell)}\right)^{\tp}+Q_{k+1}^{(m)}
\end{align}%
\end{subequations}%
where $F_{k|k}^{(m,\ell)}=\left.\nabla_{\sx} f^{(m)}(\sx)\right|_{\sx=\mkk^{(\ell)}}$ is the gradient of $f^{(m)}(\cdot)$ evaluated at the mean $\mkk^{(\ell)}$.

\subsubsection{Random matrices}
For the $m$th motion model we use the transition density suggested in \cite{GranstromO:2014},
\begin{align}
p&(\ext_{k+1}^{(i)}|\sx_{k},\ext_{k}^{(i)}) \\
& = \Wishpdf{\ext_{k+1}^{(i)}}{n_{k+1}^{(m)}}{\left(n_{k+1}^{(m)}\right)^{-1}M_{\sx_{k}}^{(m)}\ext_{k}^{(i)}\left(M^{(m)}_{\sx_{k}}\right)^{\tp}}, \nonumber
\end{align}
where $n_{k+1}^{(m)}>d-1$ is a scalar design parameter and the matrix transformation $M_{\sx_k}^{(m)}\triangleq M^{(m)}\left(\sx_k\right):\mathbb{R}^{n_x}\rightarrow \mathbb{R}^{d\times d}$ is a non-singular matrix valued function of the kinematic state. The extension state's time evolution is modeled as being dependent on the kinematic state mainly because it allows for the modeling of rotation of extended targets, however in general the only requirement is that the output is a non-singular $d\times d$ matrix \cite{GranstromO:2014}.
Details on how the parameters $v_{k+1|k}^{(m,\ell,i)}$ and $V_{k+1|k}^{(m,\ell,i)}$ of the solutions to \eqref{eq:extension_prediction_integral}
are computed are given in \cite{GranstromO:2014}.

\subsection{Generation of association events}
\label{sec:subset_association_events}
The correction step, see Section~\ref{sec:multi_ellipse_correction}, involves a summation over the set $\setTheta$ of all possible measurement-to-subobject association events. For $n_{z,k}$ measurements and $N_{s,k}$ subobjects there are $(N_{s,k})^{n_{z,k}}$ possible measurement-to-subobject association events. For example, if $N_{s,k}=3$ and $n_{z,k}=5$ there are $243$ possible association events, and if $n_{z,k}=10$ there are $59049$ possible association events. Due to the quickly increasing size of the full set of association events approximations are necessary to achieve tractable computational complexity.

Using different methods to simplify the data association problem is common in target tracking. Popular methods for multiple point target tracking include probabilistic data association (\textsc{pda}), and multiple hypothesis tracking (\textsc{mht}), see e.g. \cite{BarShalomWT:2011}. Data clustering methods are used in extended target \phd/\cphd filters to reduce the number of measurement partitions that are considered, see e.g. \cite{GranstromLO:2010,GranstromLO:2012,GranstromO:2012a,LundquistGO:2013}. In \cite{LanRL:2014} it is proposed to reduce the number of measurement-to-subobject association events by using a combination of $k$-means clustering, see e.g. \cite{Bishop:2006,HastieTF:2009}, and gating with a suitable pseudo-likelihood.


In this paper a subset $\bar{\setTheta}\subseteq\setTheta$ of association events is computed using a method that is based on the Expectation Maximization algorithm \cite{DempsterLR:1977} for Gaussian Mixtures (\emgm), see \egp \cite[Chapter 9]{Bishop:2006}. \emgm is favored over $k$-means clustering because it has been shown that the $k$-means clustering algorithm often can give unfavorable results for elliptically shaped extended targets, see \cite{GranstromO:2012a}. 
First \emgm is used to partition the current set of measurements into $N_{c}$ clusters. To accommodate the possibility that one or more subobjects does not generate any measurements \emgm is used for $N_{c}\in[1, \ 2,\ldots,N_{s,k}]$. Because the solution to \emgm has many local optimums, for each $N_{c}$ the algorithm is given several different initializations. Note that care is taken to ensure that the set of partitions returned by \emgm only contains unique partitions.

The next step is to use the clusters to obtain measurement-to-subobject associations. Given a partition of the measurement set with $N_{c}$ clusters, and an estimate with $N_{s,k}$ subobjects, there are ${N_{s,k}!}/{(N_{s,k}-N_{c})!}$
possible cluster-to-subobject associations. A cluster-to-subobject association defines a measurement-to-subobject association event $\theta$ because each measurement is associated to a cluster, which in turn is associated to a subobject. Let $C(N_{c})$ denote the number of unique partitions with $N_{c}$ clusters obtained using \emgm. Then the number of measurement-to-subobject association events that has to be considered is
\begin{align}
\left|\bar{\setTheta}\right| = \sum_{N_c=1}^{N_{s,k}} C(N_{c}) \frac{N_{s,k}!}{\left(N_{s,k}-N_{c}\right)!}.
\end{align}

\subsection{Correction}
Each object generates a Poisson distributed number of measurements and the measurement models are linear Gaussian,
\begin{subequations}
\begin{align}
P\left(n_{z,k}^{(i)}\left| \gamma_{k}^{(i)}\right.\right) = & \mathcal{PS}\left(n_{z,k}^{(i)};\ \gamma_{k}^{(i)}\right) \\
p\left(\sz_{k}^{(i,j)}\left|\sx_{k},\ext_{k}^{(i)}\right.\right) = & \Npdfbig{\sz_{k}^{(i,j)}}{H_{k}^{(i)}\sx_{k}}{\ext_{k}^{(i)}}. \label{eq:LinGaussMeasurementModel}
\end{align}
\end{subequations}
For the kinematic state \eqref{eq:kinematic_state_vector} the models $H_{k}^{(i)}$ are
\begin{subequations}
\begin{align}
H_{k}^{(1)} = & \begin{bmatrix} \Id & \mathbf{0}_{d\times n_c} & \mathbf{0}_{d\times (N_{s,k}-1)d} \end{bmatrix},\\
H_{k}^{(i)} = & \begin{bmatrix} \Id & \mathbf{0}_{d\times n_c} & \mathbf{0}_{d\times (i-2)d} & \Id & \mathbf{0}_{d\times (N_{s,k}-i)d} \end{bmatrix},
\end{align}%
\end{subequations}%
for $i=2,\ldots,N_{s,k}$.
For an association event $\theta\in\bar{\setTheta}$ the centroid measurement and scatter matrix are defined as follows,
\begin{subequations}
\begin{align}
\bar{\sz}_{k}^{(\theta,i)} & = \frac{1}{n_{z,k}^{(\theta,i)}} \sum_{j=1}^{n_{z,k}^{(\theta,i)}} \sz_{k}^{(\theta,i,j)},\\
Z_{k}^{(\theta,i)} & = \sum_{j=1}^{n_{z,k}^{(\theta,i)}} \left(\sz_{k}^{(\theta,i,j)}-\bar{\sz}_{k}^{(\theta,i)}\right) \left(\sz_{k}^{(\theta,i,j)}-\bar{\sz}_{k}^{(\theta,i)}\right)^{\tp}.
\end{align}
\end{subequations}
The same measurement model is used for all motion models. With a predicted distribution 
\begin{align}
p\left(\xi_{k} \left| \setZ^{k-1}\right.\right) = & \sum_{\ell=1}^{J_{k|k-1}} w_{k|k-1}^{(\ell)} \GGIWpdf{\xi_{k}}{\zeta_{k|k-1}^{(\ell)}}
\end{align}
the corrected distribution is 
\begin{align}
p\left(\xi_{k} \left| \setZ^{k}\right.\right) = & \sum_{\theta\in\bar{\setTheta}}\sum_{\ell=1}^{J_{k|k-1}}w_{k|k}^{(\theta,\ell)}\GGIWpdf{\xi_{k}}{\zeta_{k|k}^{(\theta,\ell)}}
\end{align}
In the sections that follow the parameters $\zeta_{k|k}^{(\theta,\ell)}$ of the corrected distribution are given. The derivation of the measurement update is given in the Appendix. 


\subsubsection{Measurement rates}
The corrected parameters are
\begin{align}
\alpha_{k|k}^{(\theta,\ell,i)} & = \alpha_{k|k-1}^{(\ell,i)} + n_{z,k}^{(\theta,i)}, \qquad \beta_{k|k}^{(\theta,\ell,i)} = \beta_{k|k-1}^{(\ell,i)} + 1.
\end{align}

\subsubsection{Kinematic state}
The corrected parameters are
\begin{subequations}
\begin{align}
m_{k|k}^{(\theta,\ell)} & = m_{k|k-1}^{(\ell)} + K_{k}^{(\theta,\ell)}\left(\bar{\sz}_{k}^{(\theta)}-\mathbb{H}_{k}m_{k|k-1}^{(\ell)}\right),\\
P_{k|k}^{(\theta,\ell)} & = P_{k|k-1}^{(\ell)} + K_{k}^{(\theta,\ell)}\mathbb{H}_{k}P_{k|k-1}^{(\ell)},\\
\bar{\sz}_{k}^{(\theta)} & = \begin{bmatrix} \left(\bar{\sz}_{k}^{(\theta,1)}\right)^{\tp} & \cdots & \left(\bar{\sz}_{k}^{(\theta,N_{s,k})}\right)^{\tp}  \end{bmatrix}^{\tp}, \\
\mathbb{H}_{k} & = \begin{bmatrix} \left(H_{k}^{(1)}\right)^{\tp} & \cdots & \left(H_{k}^{(N_{s,k})}\right)^{\tp} \end{bmatrix}^{\tp},\\
K_{k}^{(\theta,\ell)} & = P_{k|k-1}^{(\ell)}\mathbb{H}_{k}^{\tp}\left(S_{k}^{(\theta,\ell)}\right)^{-1},\\
S_{k}^{(\theta,\ell)} & = \mathbb{H}_{k}P_{k|k-1}^{(\ell)}\mathbb{H}_{k}^{\tp} + \hat{\mathbb{X}}_{k|k-1}^{(\theta,\ell)},\\
\hat{\mathbb{X}}_{k|k-1}^{(\theta,\ell)} & = \blkdiag\left( \frac{\hat{\ext}_{k|k-1}^{(\ell,1)}}{n_{z,k}^{(\theta,1)}} , \ldots , \frac{\hat{\ext}_{k|k-1}^{(\ell,N_{s,k})}}{n_{z,k}^{(\theta,N_{s,k})}} \right),\\
\hat{\ext}_{k|k-1}^{(\ell,i)} & = \frac{V_{k|k-1}^{(\ell,i)}}{v_{k|k-1}^{(\ell,i)}-2d-2}.
\end{align}
\end{subequations}

\subsubsection{Random matrices}
The corrected parameters are
\begin{subequations}
\begin{align}
v_{k|k}^{(\theta,\ell,i)} = & v_{k|k-1}^{(\ell,i)} + n_{z,k}^{(\theta,i)},\\
V_{k|k}^{(\theta,\ell,i)} = & V_{k|k-1}^{(\ell,i)} + Z_{k}^{(\theta,i)} + N_{k|k-1}^{(\theta,\ell,i)},\\
N_{k|k-1}^{(\theta,\ell,i)} = & \left(\hat{\ext}_{k|k-1}^{(\ell,i)}\right)^{\frac{1}{2}} \left(S_{k}^{(\theta,\ell,i)}\right)^{-\frac{1}{2}} \varepsilon_{k|k-1}^{(\theta,\ell,i)} \nonumber \\
& \times \left(\varepsilon_{k|k-1}^{(\theta,\ell,i)}\right)^{\tp} \left(S_{k}^{(\theta,\ell,i)}\right)^{-\frac{\tp}{2}} \left(\hat{\ext}_{k|k-1}^{(\ell,i)}\right)^{\frac{\tp}{2}},\\
\varepsilon_{k|k-1}^{(\theta,\ell,i)} = & \bar{\sz}_{k}^{(\theta,i)}-H_{k}^{(i)}m_{k|k-1}^{(\ell)},\\
S_{k}^{(\theta,\ell,i)} = & H_{k}^{(i)}P_{k|k-1}^{(\ell)}\left(H_{k}^{(i)}\right)^{\tp} + \frac{\hat{\ext}_{k|k-1}^{(\ell,i)}}{n_{z,k}^{(\theta,i)}},
\end{align}
\end{subequations}
where the matrix square-roots are computed using, \egp, Cholesky factorization.

\subsubsection{Weights}
The weights are computed as
\begin{subequations}
\begin{align}
w_{k|k}^{(\theta,\ell)} = & \frac{w_{k|k-1}^{(\ell)} \prod_{i=1}^{N_{s,k}}\mathcal{L}_{k}^{(\theta,\ell,i)}}{\sum_{\theta'\in\bar{\setTheta}} \sum_{\ell'=1}^{J_{k|k-1}} w_{k|k-1}^{(\ell')} \prod_{i'=1}^{N_{s,k}}\mathcal{L}_{k}^{(\theta',\ell',i')}},\\
\mathcal{L}_{k}^{(\theta,\ell,i)} = & \frac{\Gamma\left(\alpha_{k|k}^{(\theta,\ell,i)}\right)}{\Gamma\left(\alpha_{k|k-1}^{(\ell,i)}\right)} \frac{\left(\beta_{k|k-1}^{(\ell,i)}\right)^{\alpha_{k|k-1}^{(\ell,i)}}}{\left(\beta_{k|k}^{(\theta,\ell,i)}\right)^{\alpha_{k|k}^{(\theta,\ell,i)}}} \nonumber \\
& \times \frac{ \left(n_{z,k}^{(\theta,i)} \pi^{n_{z,k}^{(\theta,i)}} \right)^{-\frac{d}{2}} 2^{-\frac{n_{z,k}^{(\theta,i)} (d-1)}{2}} }{\left|\left(\hat{\ext}_{k|k-1}^{(\ell,i)}\right)^{-\frac{1}{2}}S_{k}^{(\theta,\ell,i)} \left(\hat{\ext}_{k|k-1}^{(\ell,i)}\right)^{-\frac{\tp}{2}} \right|^{\frac{1}{2}}} \nonumber \\
& \times  \frac{\Gamma_{d}\left(\frac{v_{k|k}^{(\theta,\ell,i)}-d-1}{2}\right)}{\Gamma_{d}\left(\frac{v_{k|k-1}^{(\ell,i)}-d-1}{2}\right)} \frac{\left|V_{k|k-1}^{(\ell,i)} \right|^{\frac{v_{k|k-1}^{(\ell,i)}-d-1}{2}}}{ \left|V_{k|k}^{(\theta,\ell,i)} \right|^{\frac{v_{k|k}^{(\theta,\ell,i)}-d-1}{2}} }.
\end{align}
\end{subequations}

\subsubsection{Predicted likelihoods}
The predicted likelihoods are
\begin{subequations}
\begin{align}
& p^{(\ell)}\left(\setZ_{k}|\theta,\setZ^{k-1}\right) = \prod_{i=1}^{N_{s,k}} \mathcal{L}_{k}^{(\theta,\ell,i)} \\
& p\left(\setZ_{k}|\setZ^{k-1}\right) = \frac{1}{|\bar{\setTheta}|} \sum_{\theta\in\bar{\setTheta}} \sum_{\ell=1}^{J_{k|k-1}} w^{(\ell)}_{k|k-1} \prod_{i=1}^{N_{s,k}} \mathcal{L}_{k}^{(\theta,\ell,i)}.
\end{align}
\end{subequations}

\begin{table}[tb]
\caption{Change main subobject}
\label{tab:ChangeMainSubobject}
\vspace{-0.5cm}
\rule[0pt]{\columnwidth}{1pt}
\vspace{-0.3cm}
\begin{algorithmic}[1]
\STATE \textbf{Input:} Component $\GGIWpdf{\xi_{k}}{\zeta_{k|k}}$.
\STATE Center: $\hat{\mathbf{p}}_{k|k}^{(c)} = \sum_{i=1}^{N_{s,k}}\hat{\mathbf{p}}_{k|k}^{(i)}$, where $\hat{\mathbf{p}}_{k|k}^{(i)} = \E\left[\left.\mathbf{p}_{k}^{(i)}\right|\setZ^{k}\right]$.
\STATE Distances to center: $\delta_{k|k}^{(i)} = \left\| \hat{\mathbf{p}}_{k|k}^{(c)} - \hat{\mathbf{p}}_{k|k}^{(i)} \right\|_{2}$.
\STATE Closest to center: $j = \argmin{i} \delta_{k|k}^{(i)}$
\IF{ $j \neq 1$ }
	\STATE Change main subobject from $1$ to $j$.
	\STATE $\tilde{m}_{k|k} = A_{1,j}m_{k|k}$, $\tilde{P}_{k|k} = A_{1,j}P_{k|k}A_{1,j}^{\tp}$, where $A_{1,j}$ is a permutation matrix that changes place between subobjects $1$ and $j$.
	
	\STATE $\left(\tilde{\alpha}_{k|k}^{(1)},\tilde{\beta}_{k|k}^{(1)},\tilde{v}_{k|k}^{(1)},\tilde{V}_{k|k}^{(1)}\right) = \left({\alpha}_{k|k}^{(j)},{\beta}_{k|k}^{(j)},{v}_{k|k}^{(j)},{V}_{k|k}^{(j)}\right)$, 
	
	$\left(\tilde{\alpha}_{k|k}^{(j)},\tilde{\beta}_{k|k}^{(j)},\tilde{v}_{k|k}^{(j)},\tilde{V}_{k|k}^{(j)}\right) = \left({\alpha}_{k|k}^{(1)},{\beta}_{k|k}^{(1)},{v}_{k|k}^{(1)},{V}_{k|k}^{(1)}\right)$, and
	
	$\left(\tilde{\alpha}_{k|k}^{(i)},\tilde{\beta}_{k|k}^{(i)},\tilde{v}_{k|k}^{(i)},\tilde{V}_{k|k}^{(i)}\right) = \left({\alpha}_{k|k}^{(i)},{\beta}_{k|k}^{(i)},{v}_{k|k}^{(i)},{V}_{k|k}^{(i)}\right)$ for $i\neq 1,j$.

	\STATE $\tilde{\zeta}_{k|k} = \left( \left\{\tilde{\alpha}_{k|k}^{(i)},\tilde{\beta}_{k|k}^{(i)} \right\}_{i=1}^{N_{s,k}},\tilde{m}_{k|k},\tilde{P}_{k|k}, \left\{\tilde{v}_{k|k}^{(i)},\tilde{V}_{k|k}^{(i)} \right\}_{i=1}^{N_{s,k}} \right) $.
\ELSE
	\STATE No change of main subobject, $\tilde{\zeta}_{k|k} = \zeta_{k|k}$
\ENDIF
\STATE \textbf{Output:} Component $\GGIWpdf{\xi_{k}}{\tilde{\zeta}_{k|k}}$.
\end{algorithmic}
\vspace{-0.1cm}
\rule[0pt]{\columnwidth}{1pt}
\end{table}

\subsection{Mixture reduction}
With $J_{k|k}$ components, $M_{k}$ motion models and $|\bar{\setTheta}|$ association events there are $J_{k+1|k+1} = |\bar{\setTheta}| M_{k} J_{k|k}$ components after one iteration of prediction and correction. Mixture reduction is used in each iteration after the correction step to keep the number of components at a tractable level.
Hypotheses with weights lower than a threshold $\tau$ are pruned and the weights are re-normalized. Merging is then performed on the mixture, where we have used a combination of the gamma mixture merging from \cite{GranstromO:2012c} and the Gaussian inverse Wishart merging from \cite{GranstromO:2012e}. Note that when multiple motion models are used, merging is only performed within the same motion modes, and not across the motion modes.

\subsection{Change of main subobject}
When the motion model $f^{(m)}(\sx_k)$ is linear it does not matter which subobject is defined as the main one. However with non-linear motion, \egp a coordinated turn, it is typically most intuitive to define the main subobject as the subobject closest to the center of the overall extended target. Here center is defined as the mean of the subobjects' positions.

It can happen that an estimate hypothesis $p^{(\ell)}(\cdot)$ initialized by the method in Table~\ref{tab:multi_ellipse_initilization}, after a couple of prediction-correction-iterations, converges to a configuration where the main subobject is not the one closest to the center. In this case another subobject can be defined as the main one, an operation that corresponds to a simple re-ordering of the measurement rates and random matrices, and a simple linear transformation of the kinematic vector. Note that any non-linear dynamics could be re-defined too, however empirically we have found that it is sufficient to keep the kinematics $\mathbf{c}_{k}$ constant during the change of main subobject. The algorithm that is used is given in Table~\ref{tab:ChangeMainSubobject}.

An important special case is when there are two subobjects, since in this case both subobjects are equally close to the center. In this case the main subobject is arbitrarily assigned upon initialization, and the change of main subobject is never utilized in the filter recursion.


\section{Comparison to alternative models}
In this section we discuss and compare the proposed model to other extended target models that are available in the litterature. Models for specific geometric shapes, \egp sticks, circles, ellipse and rectangles can be found in \cite{GilholmS:2005,BaumFH:2012,Boersetal:06a,PetrovMGA:2011,Koch:2008,BaumNH:2010,GranstromLO:2011,ZhuHL:2011,ReuterD:2011,DegermanWS:2011,LanRL:2012,ReuterWD:2012}. Because these models consider specific shapes we do not compare to them further. Three models capable of handling general and irregular shapes can be found in \cite{LundquistGO:11,BaumH:2011,LanRL:2014}. The model in \cite{LundquistGO:11} considers measurements that are spread along the outline of the target's shape (e.g. laser range measurements), and in this paper we consider measurement that are spread across the target's surface. The model in \cite{LundquistGO:11} is thus not applicable to the scenarios considered here. In the next section we present simulation results that compare the proposed model to the two models presented in \cite{BaumH:2011,LanRL:2014}, and in the remainder of this section we elaborate on the theoretical similarities and differences between the proposed model and the ones from \cite{BaumH:2011,LanRL:2014}.

\subsection{Star-Convex model \cite{BaumH:2011}}
The Star-Convex Random Hypersurface Model \cite{BaumH:2011}, denoted M1, parametrizes the boundary of the target shape as a radial function. The radial function is parameterized using Fourier series, and the Fourier coefficients are estimated. With more coefficients, the radial function has more degrees of freedom and increasingly complex shapes can be described. In a sense, this is analogous to how a larger number of subobjects can describe a more complex shape. Model M1 does not decompose the extended target into subobjects, and thus does not need to solve the measurement-to-subobject association problem.

\subsection{Multi-Ellipse model \cite{LanRL:2014}}
The model in \cite{LanRL:2014}, denoted M2, models the target using multiple elliptic subobjects, see \eqref{eq:LanRongLiStateModel}. The proposed model is very similar to model M2 as they both extend the random matrix framework \cite{Koch:2008,FeldmannFK:2011} to model the subobjects. The two models also have the following differences:

\subsubsection{Measurement rates}
The proposed model includes a model of the number of detections per subobject per timestep and estimates the measurement rates for each subobject, which M2 does not.

\subsubsection{Main subobject}
The proposed model defines one of the subobjects as the main subobject, around which remaining subobjects are located. 
In comparison, M2 does not define one of the subobjects as the main one.

\subsubsection{Position covariance and unified kinematics}
M2 is based on the random matrix model \eqref{eq:Koch_random_matrix_model}, see \eqref{eq:LanRongLiStateModel}, while the proposed model is based on the random matrix model \eqref{eq:Feldmann_random_matrix_model}. This difference is fundamental, because it is what allows the positions and kinematics of all subobjects to the modeled as a single random vector. Due to the form of the Gaussian covariances in \eqref{eq:Koch_random_matrix_model} ($P\otimes X$), under this model the state vectors of the subobjects cannot easily be treated as a single random vector. Modeling with a single random vector improves the overall modeling in the following ways:
\begin{enumerate}
	\item The proposed model estimates unified kinematics $\mathbf{c}_{k}$ (\iep a single velocity, a single turn-rate etc for the extended target as a whole) for all subobjects', and if necessary individual subobject kinematics can be included in $\mathbf{c}_{k}$. In comparison, M2 estimates individual velocities and accelerations for the subobjects. A multiple motion model framework is used in M2 where there are some common kinematics via the process noise parameters, however unified kinematics are not estimated.
	\item The proposed model maintains a full covariance matrix for the subobjects' positions and the kinematics, i.e. the dependencies between the subobjects' positions and the kinematics are modeled and estimated. In comparison, M2 does not model the dependencies between the subobjects' kinematic states $\sx_{k}^{(i)}$, cf \eqref{eq:LanRongLiStateModel}.
\end{enumerate}

\begin{figure*}[htbp]
\centerline{
\subfloat{\includegraphics[width=0.33\textwidth]{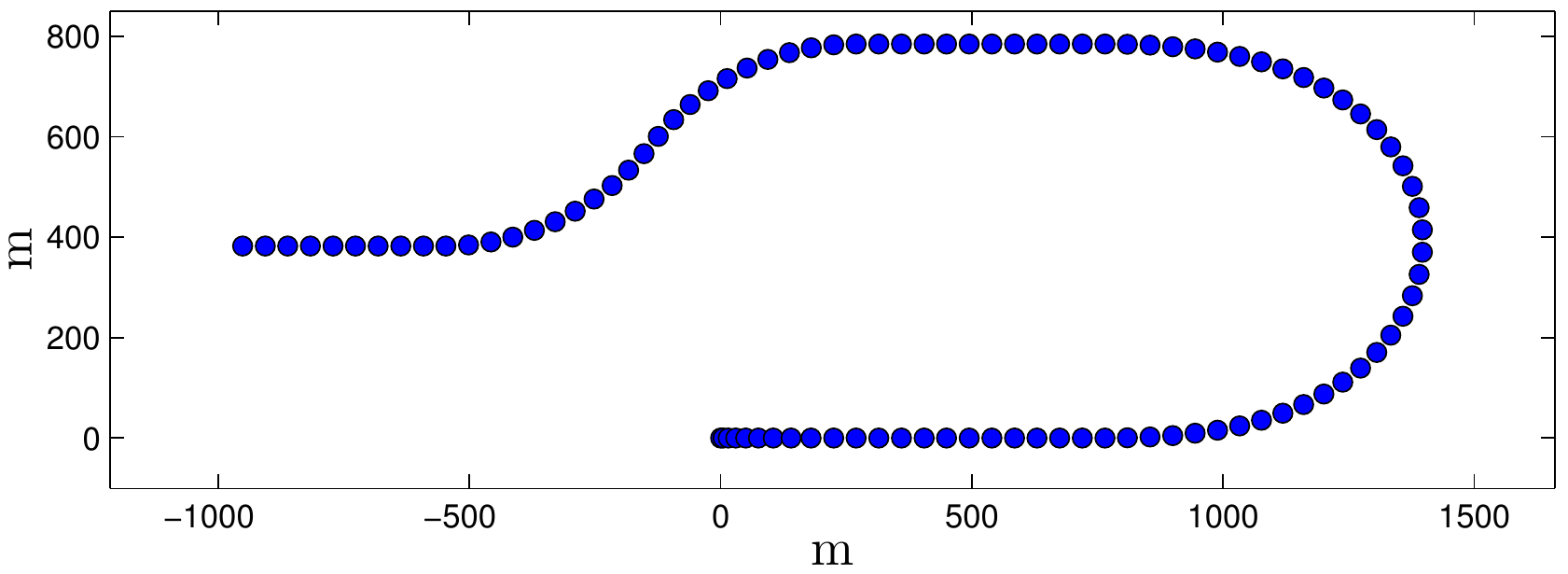} \label{fig:multi_ellipse_true_tracks_xy}}
\hfill
\subfloat{\includegraphics[width=0.33\textwidth]{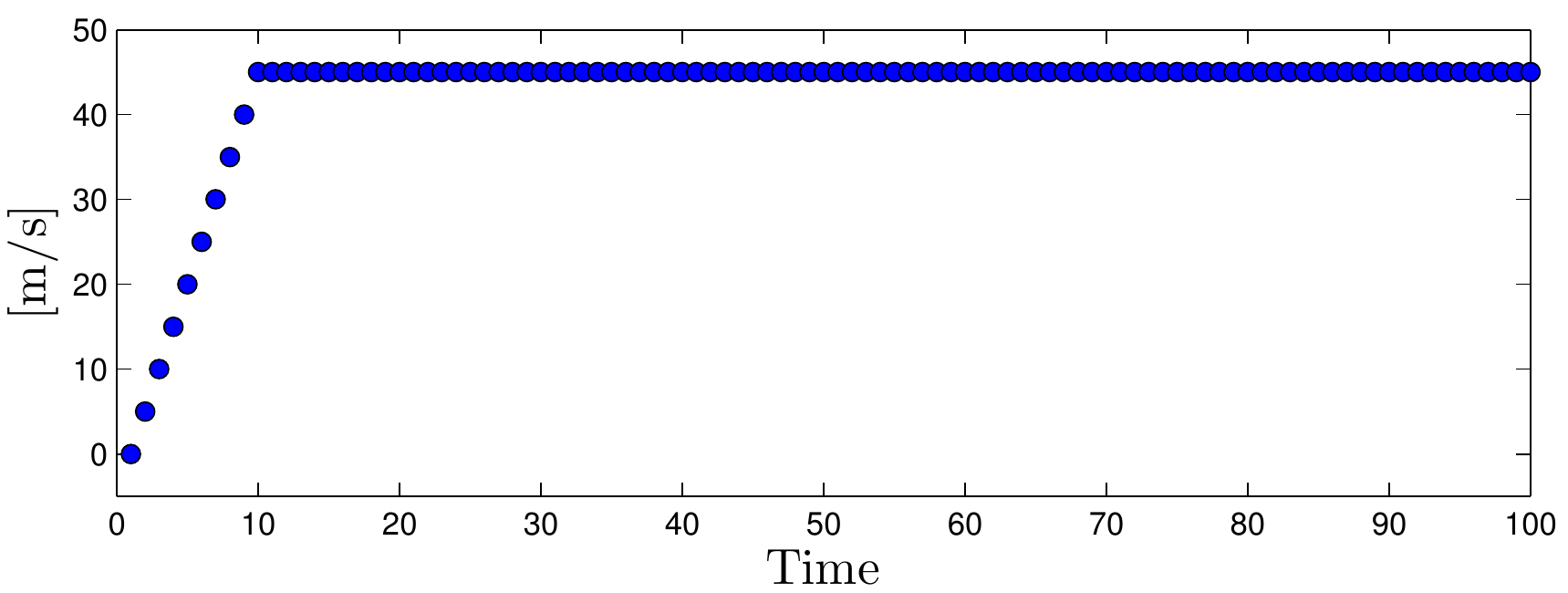} \label{fig:multi_ellipse_true_tracks_v}}
\hfill
\subfloat{\includegraphics[width=0.33\textwidth]{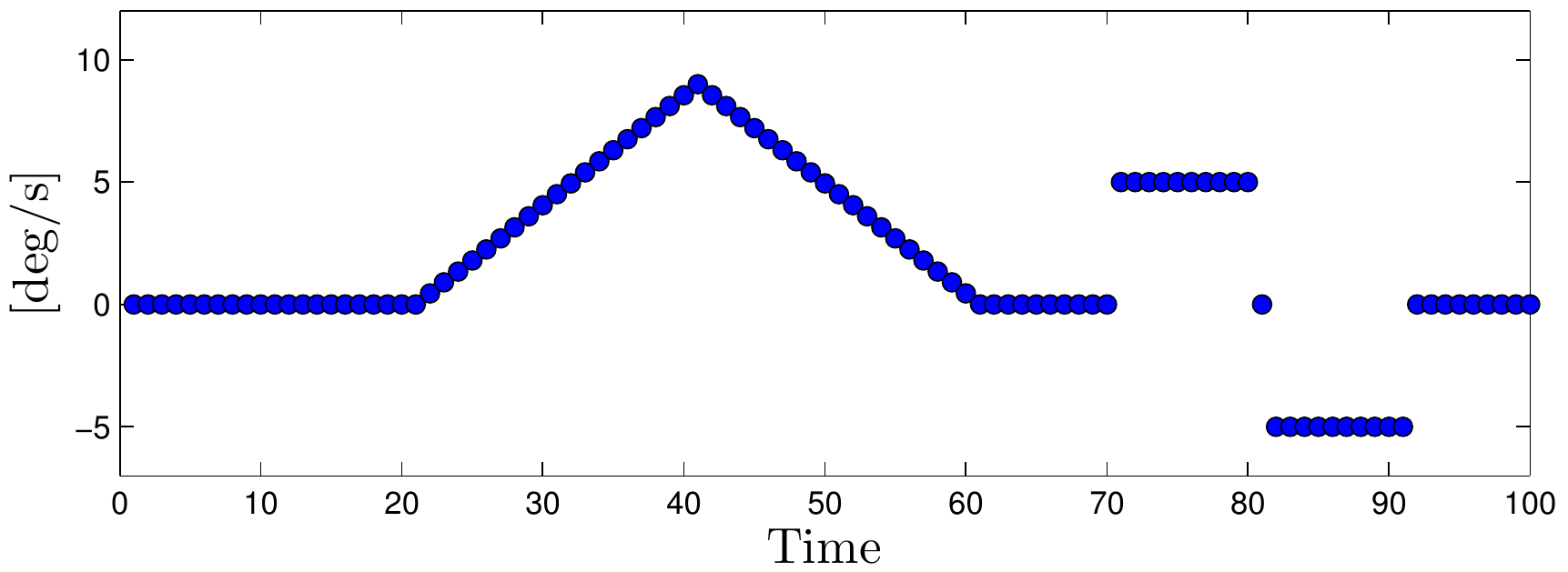} \label{fig:multi_ellipse_true_tracks_omega}}
}
\caption{True target trajectory, initial position is origin. Left: $x,y$-position. Middle: velocity. Right: turn-rate.}
\label{fig:multi_ellipse_true_tracks}
\end{figure*}

\subsubsection{Kinematics-extension-dependence}
The proposed model does not model the dependence between the position and the extension of a subobject, which M2 does. Further, under an assumed linear Gaussian measurement model the proposed model's measurement update is approximate, while M2's is exact. Both these differences follow from the proposed model being based on the random matrix model \eqref{eq:Feldmann_random_matrix_model} and M2 being based on the random matrix model \eqref{eq:Koch_random_matrix_model}.

With regards to the dependence not being modeled, the measurement update and the prediction update provides for the necessary interdependence between the kinematic state estimate and the extension state estimate. This is analogous to how the measurement update \cite{FeldmannFK:2011} and the prediction update \cite{GranstromO:2014} provide for interdependence in case the target is modeled using a single ellipse. In simulation comparisons for single elliptic targets, models based on \cite{FeldmannFK:2011} outperform models based on \cite{Koch:2008}, see \cite{GranstromO:2014}.

We conclude the comparison by noting that different models should not be judged and compared only on their theoretical properties, but also on their practical properties. In the next section we present a simulation study that compares the practical performance of the proposed model and models M1 and M2.


\section{Simulation results}
\label{sec:simulation_results}

\subsection{Target extraction and performance evaluation}
\label{sec:performance_eval}
To extract a target estimate from a mixture \eqref{eq:mixtureGGIW_extended_target_state}, merging is first performed, this time across the motion modes. Expected values of the measurement rates, positions and extension matrices are then computed w.r.t. the component with the highest weight $w_{k|k}^{(\ell)}$. Both the predicted estimate $\hat{\xi}_{k|k-1}$ and the filtered estimate $\hat{\xi}_{k|k}$ is compared to the true target state $\xi_{k}$. The following error metrics are used for the measurement rates, subobject positions, and random matrices,
\begin{subequations}
\begin{align}
d^{\gamma}_{k|k} = & \sum_{i=1}^{N_{s,k}} \left| \gamma_{k}^{(i)} - \hat{\gamma}_{k|k}^{(\pi(i))} \right|, \ \hat{\gamma}_{k|k}^{(i)}=\E\left[\left.\gamma_{k}^{(i)}\right|\setZ^{k}\right]\\
d^{\mathbf{p}}_{k|k} = & \sum_{i=1}^{N_{s,k}} \left\| \mathbf{p}_{k}^{(i)} - \hat{\mathbf{p}}_{k|k}^{(\pi(i))}\right\|_{2}, \ \hat{\mathbf{p}}_{k|k}^{(i)} = \E\left[ \left. \mathbf{p}_{k}^{(i)} \right| \setZ^{k} \right] \\
d^{\ext}_{k|k} = & \sum_{i=1}^{N_{s,k}} \left\| \ext_{k}^{(i)} - \hat{\ext}_{k|k}^{(\pi(i))} \right\|_{F}, \hat{\ext}_{k|k}^{(i)}=\E\left[\left.\ext_{k}^{(i)}\right|\setZ^{k}\right]
\end{align}
\end{subequations}
A subobject-to-subobject association $\pi(i)$ is obtained by minimizing $d^{\mathbf{p}}_{k|k}$. Because $\gamma_{k}^{(i)}$, $\mathbf{p}_{k}^{(i)}$ and $\ext_{k}^{(i)}$ all have different units we refrain from computing an overall metric for the extended target state $\xi_{k}$.

\subsection{True tracks and setup}
We simulate both stationary and moving targets. In a comparison of shape estimation, stationary targets are used because we wish to emphasize the shape estimation, not the motion estimation. For the moving target, the target trajectory that was simulated is shown in \figurename~\ref{fig:multi_ellipse_true_tracks}; the true position is shown in \figurename~\ref{fig:multi_ellipse_true_tracks_xy} and the corresponding speed and turn-rate is shown in \figurename~\ref{fig:multi_ellipse_true_tracks_v} and \figurename~\ref{fig:multi_ellipse_true_tracks_omega}.

Three different $d=2$ dimensional extended target shapes were simulated. The first is T-shaped, with measurements generated by uniformly sampling across the shape and adding Gaussian noise with covariance $\mathbf{R}=\mathbf{I}_{2}$. The measurement rate was $4\gamma_{0}$. The other two shapes consist of three and two subobjects, respectively. The shape of the targets are consistent with the examples given in \figurename~\ref{fig:multiple_ellipse_example_target}, \iep the shape resembles that of an airplane and of the letter V, respectively. For these two shapes, the measurement model \eqref{eq:Linear_Gaussian_Measurements} was used. For the plane-like target, for the subobject that corresponds to the fuselage the measurement rate was $2\gamma_{0}$, and the extension matrix was $\ext=\diag{[10^2 \ , \ 2^2]}$. For the subobjects that correspond to the wings the measurement rates were $\gamma_{0}$, and the extension matrices were $\ext=\diag{[5^2 \ , \ 1^2]}$. For the V-shaped target, the subobjects both had measurement rates $\gamma_{0}$ and extension matrices $\ext=\diag{[20^2\ , \ 1^2]}$. The scenarios were simulated for different values of $\gamma_{0}$: $2$, $5$ and $20$.

For the presented filter, a constant turn-rate (\ctmod) motion model $f_{\mathbf{p},\mathbf{c}}^{(m)}(\cdot)$ with polar velocity, see \cite[Eq. 75]{RongLiJ:2003}, was used for the overall position and kinematics. In this case the unified kinematics are given by $\mathbf{c}_{k} = \begin{bmatrix} \cv_{k} & \phi_{k} & \omega_{k} \end{bmatrix}^{\tp}$, where $\cv_{k}$ is the speed, $\phi_{k}$ is the heading and $\omega_{k}$ is the turn-rate. For this type of motion model the time evolution of the subobject offsets is
\begin{align}
\mathbf{d}_{k+1}^{(i)} = f_{\mathbf{d}}^{(m)}\left(\mathbf{d}_{k}^{(i)},\mathbf{c}_{k}\right) = R \left(T\omega_{k}\right) \mathbf{d}_{k}^{(i)}
\end{align}
where $T$ is the sampling time and $R \left(\cdot\right)$ is a rotation matrix. The matrix transformation function is also a rotation matrix, $M^{(m)}\left(\sx_k\right) = R \left(T\omega_{k}\right)$.
Two \ctmod models were implemented, one with small process noise corresponding to non-maneuver, and one with larger process noise corresponding to maneuver. The transition probabilities were set to $95\%$ probability to stay in the same mode, and $5\%$ probability for mode switch.

\begin{figure*}[!t]
\centerline{
\subfloat{\includegraphics[width=0.125\textwidth]{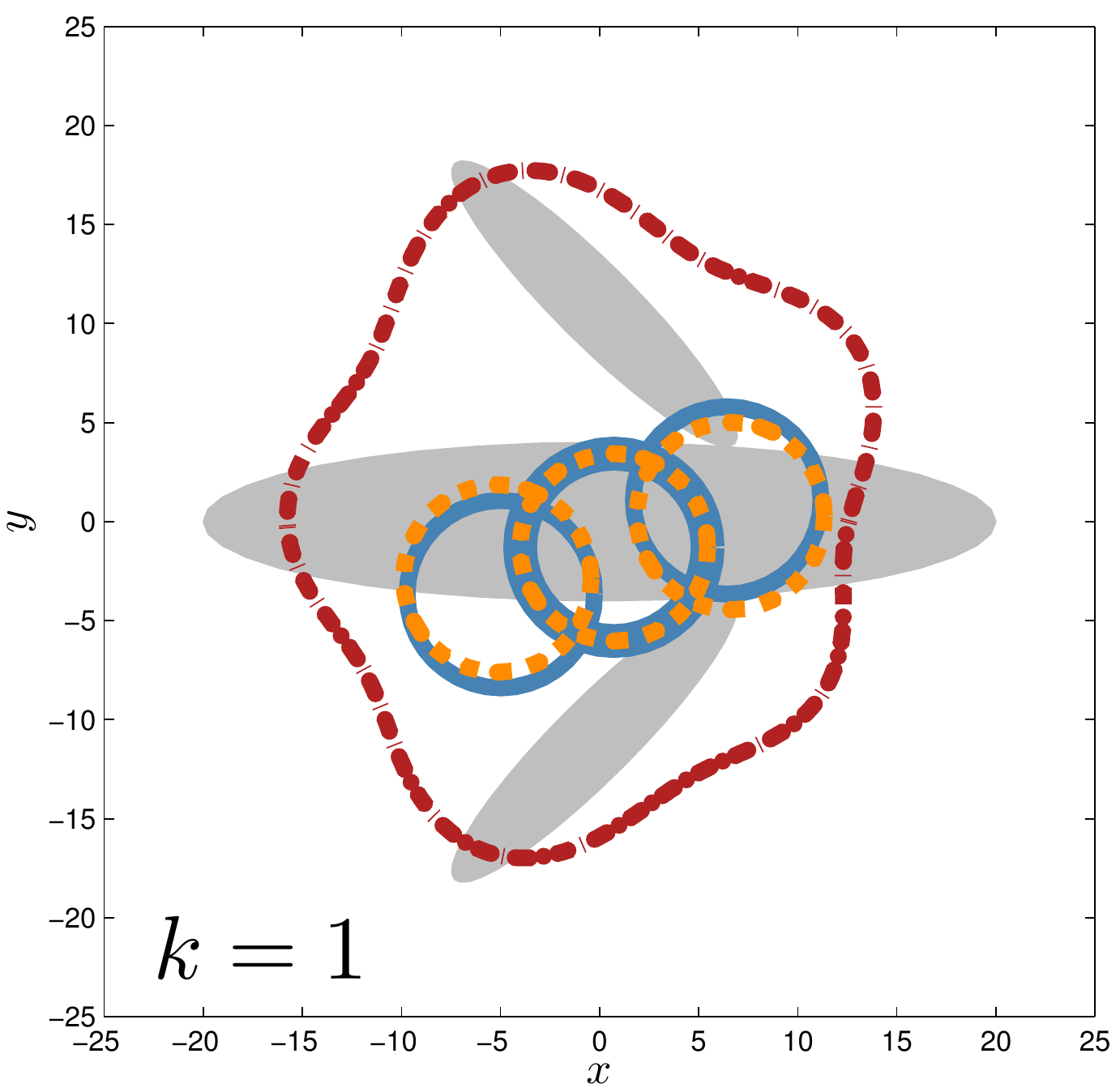} \label{fig:multi_ellipse_shape_comparison_gam_2_ex_1}}
\hfill
\subfloat{\includegraphics[width=0.125\textwidth]{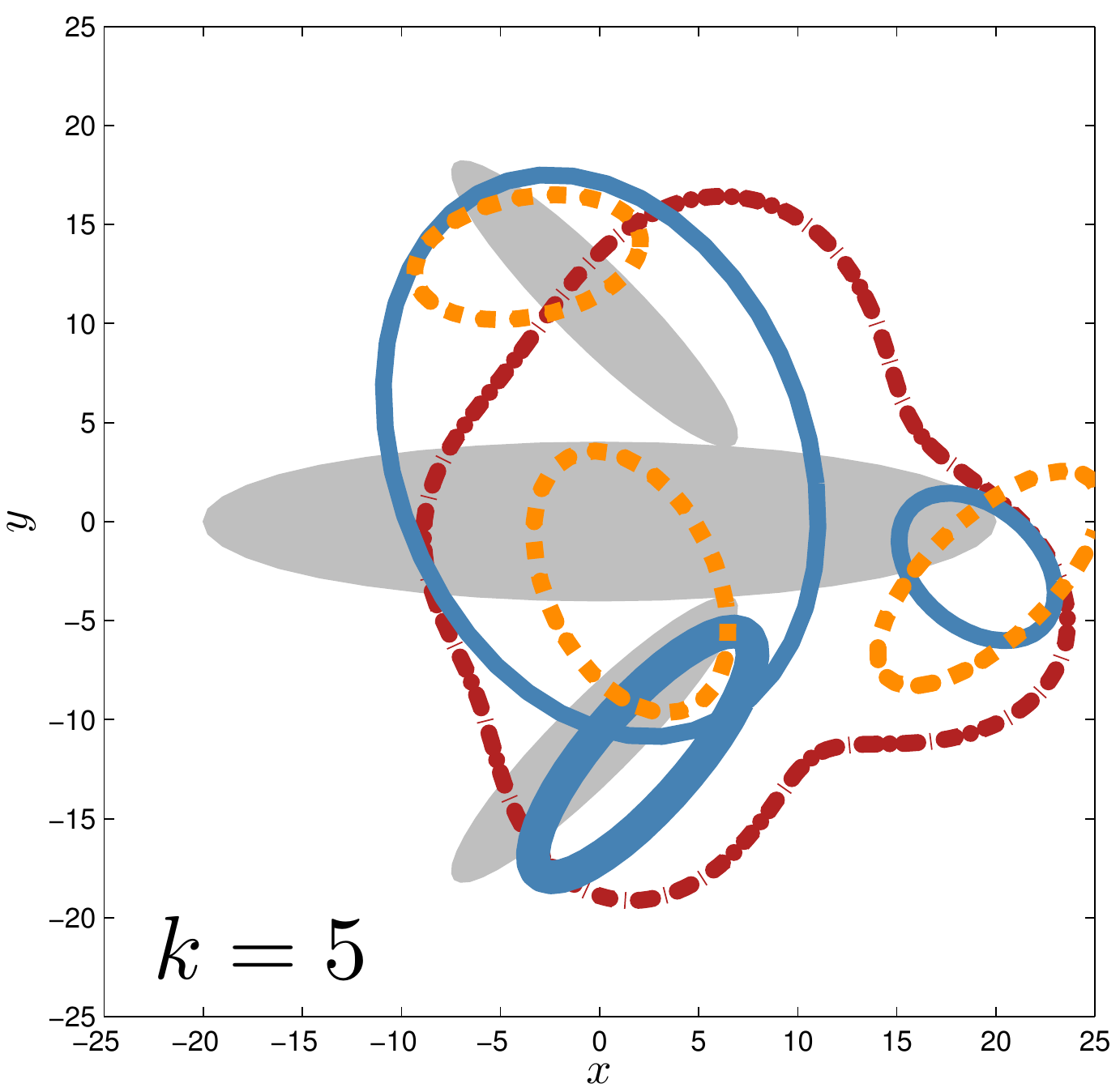} \label{fig:multi_ellipse_shape_comparison_gam_2_ex_2}}
\hfill
\subfloat{\includegraphics[width=0.125\textwidth]{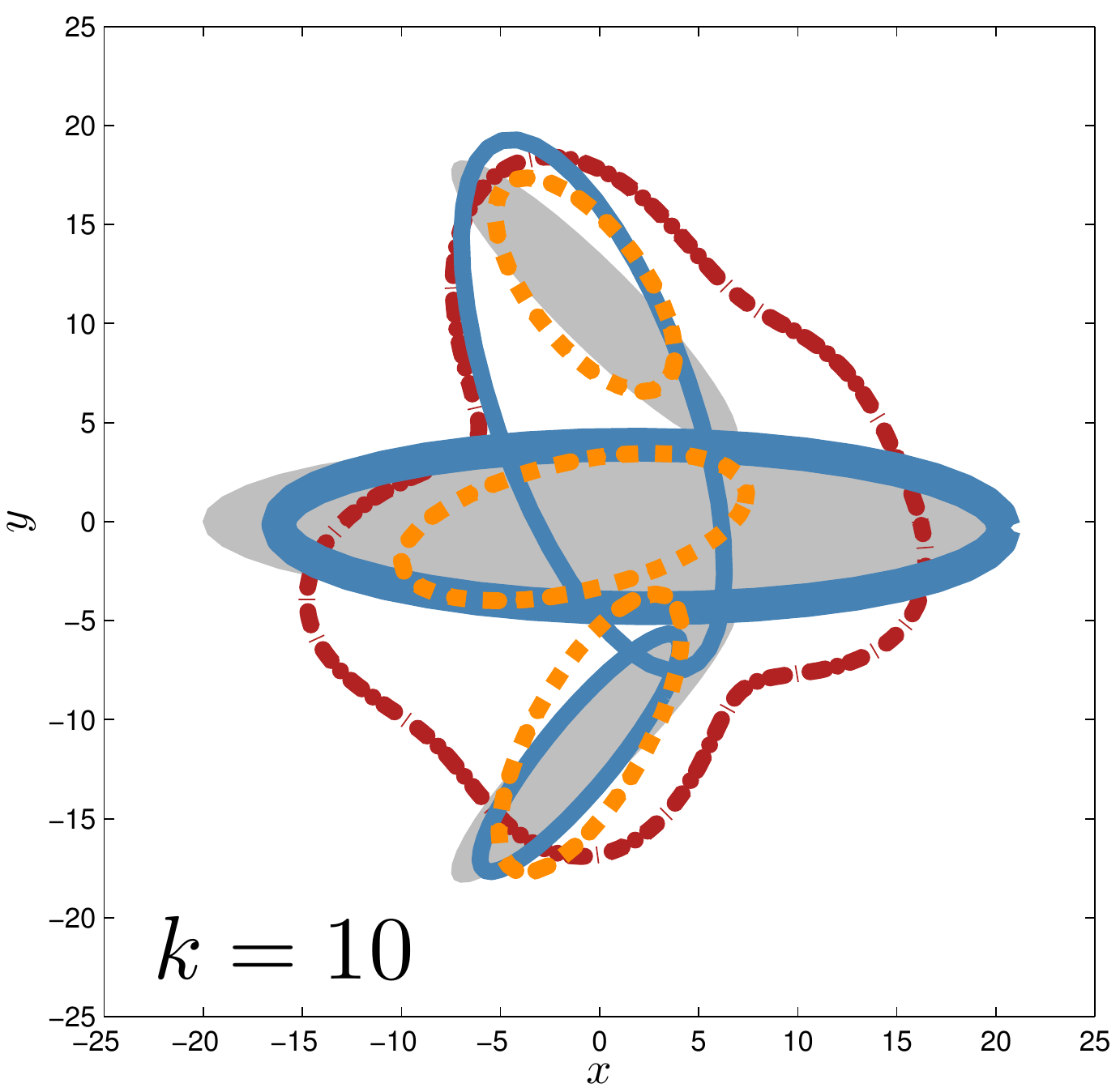} \label{fig:multi_ellipse_shape_comparison_gam_2_ex_3}}
\hfill
\subfloat{\includegraphics[width=0.125\textwidth]{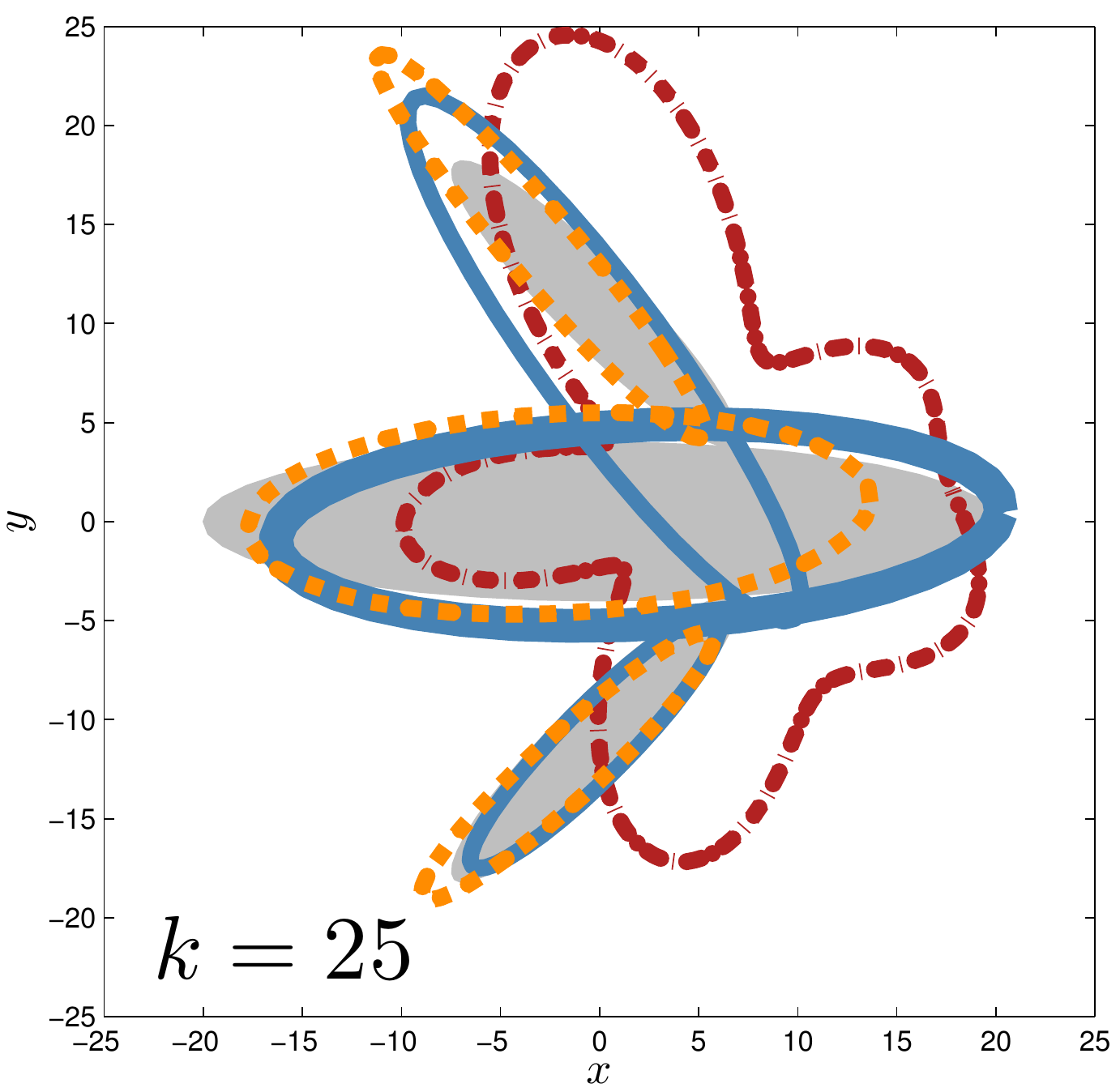} \label{fig:multi_ellipse_shape_comparison_gam_2_ex_4}}
\hfill
\subfloat{\includegraphics[width=0.125\textwidth]{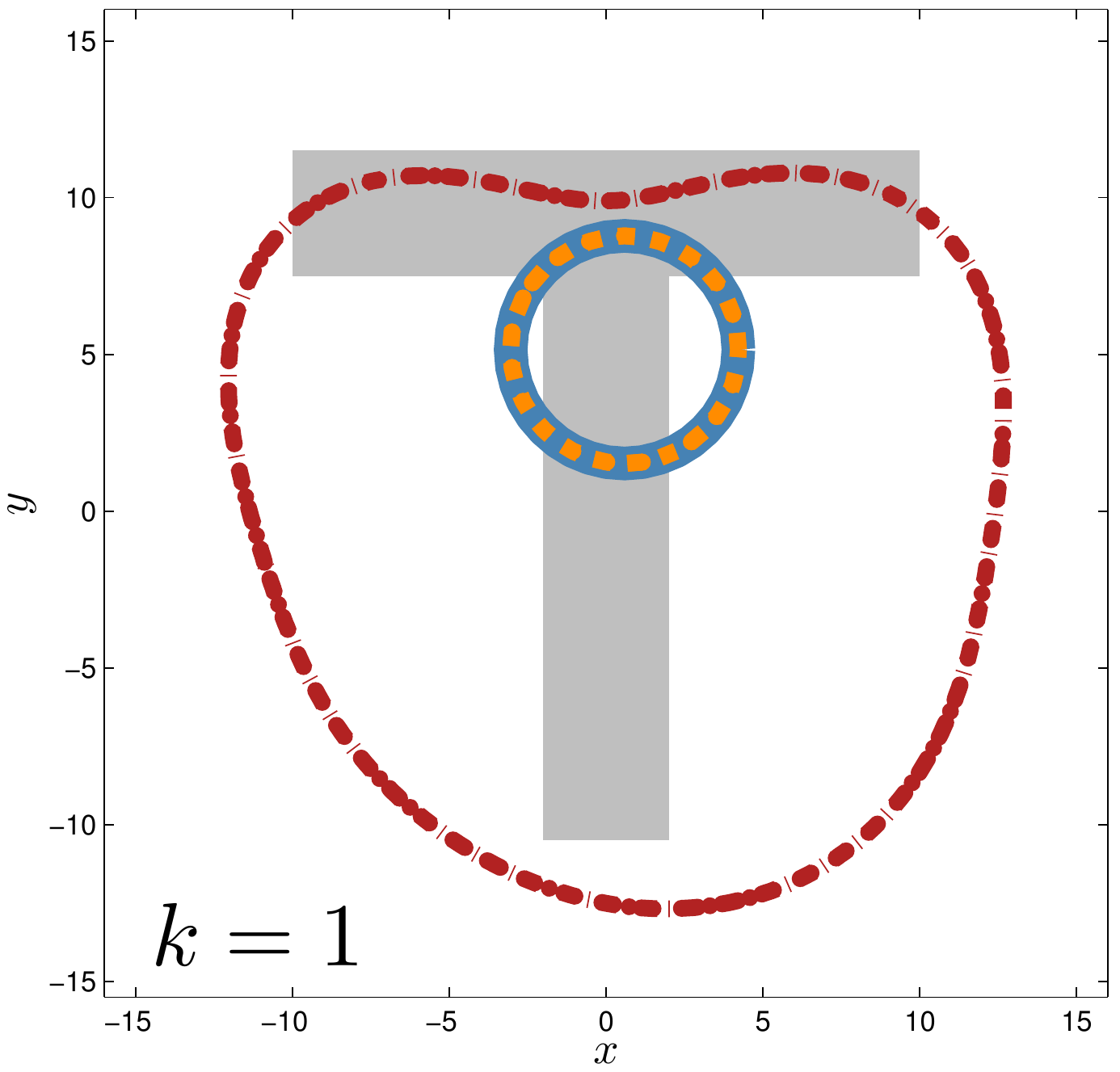} \label{fig:multi_ellipse_shape_comparison_Tshape_gam_2_ex_1}}
\hfill
\subfloat{\includegraphics[width=0.125\textwidth]{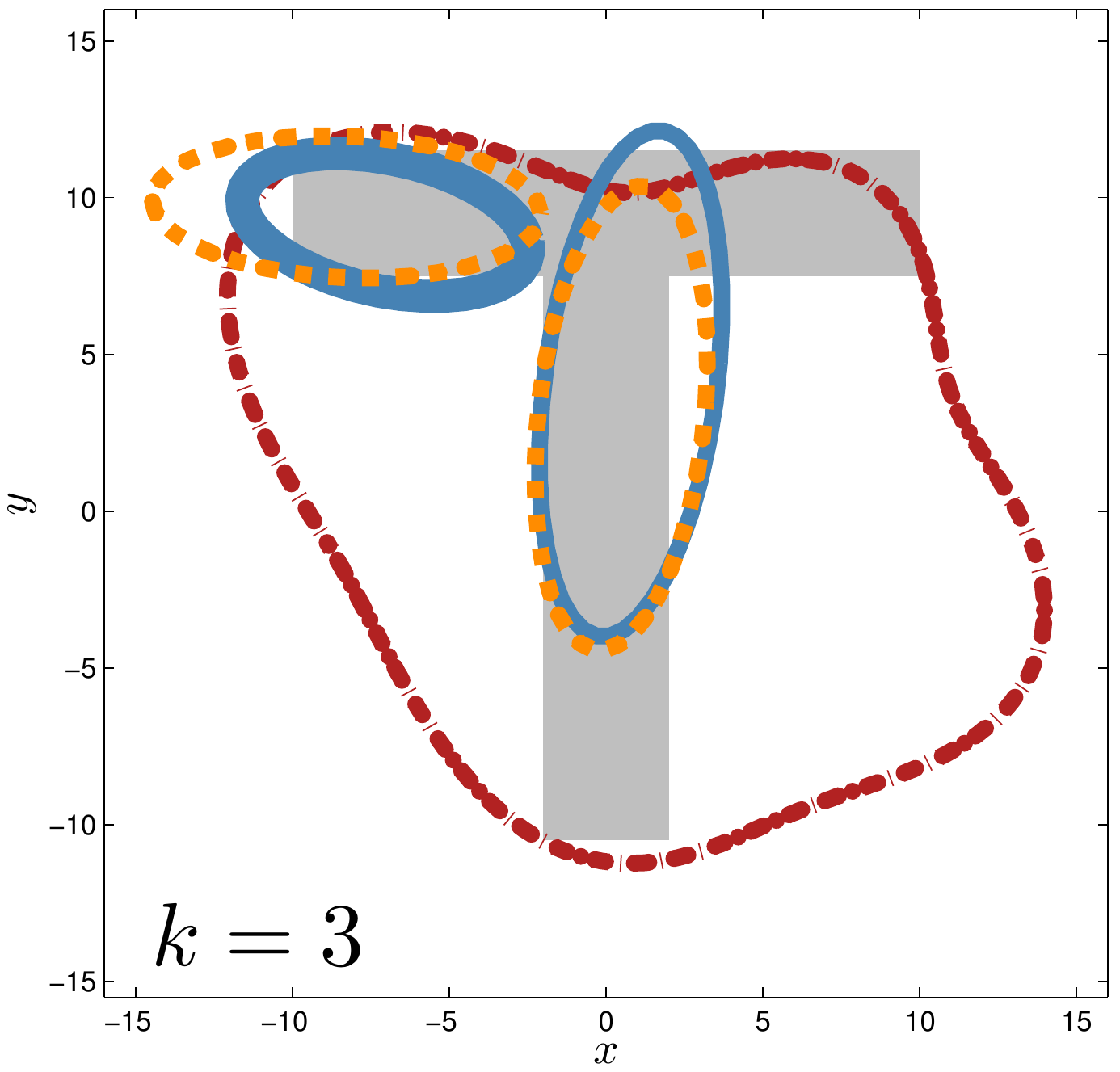} \label{fig:multi_ellipse_shape_comparison_Tshape_gam_2_ex_2}}
\hfill
\subfloat{\includegraphics[width=0.125\textwidth]{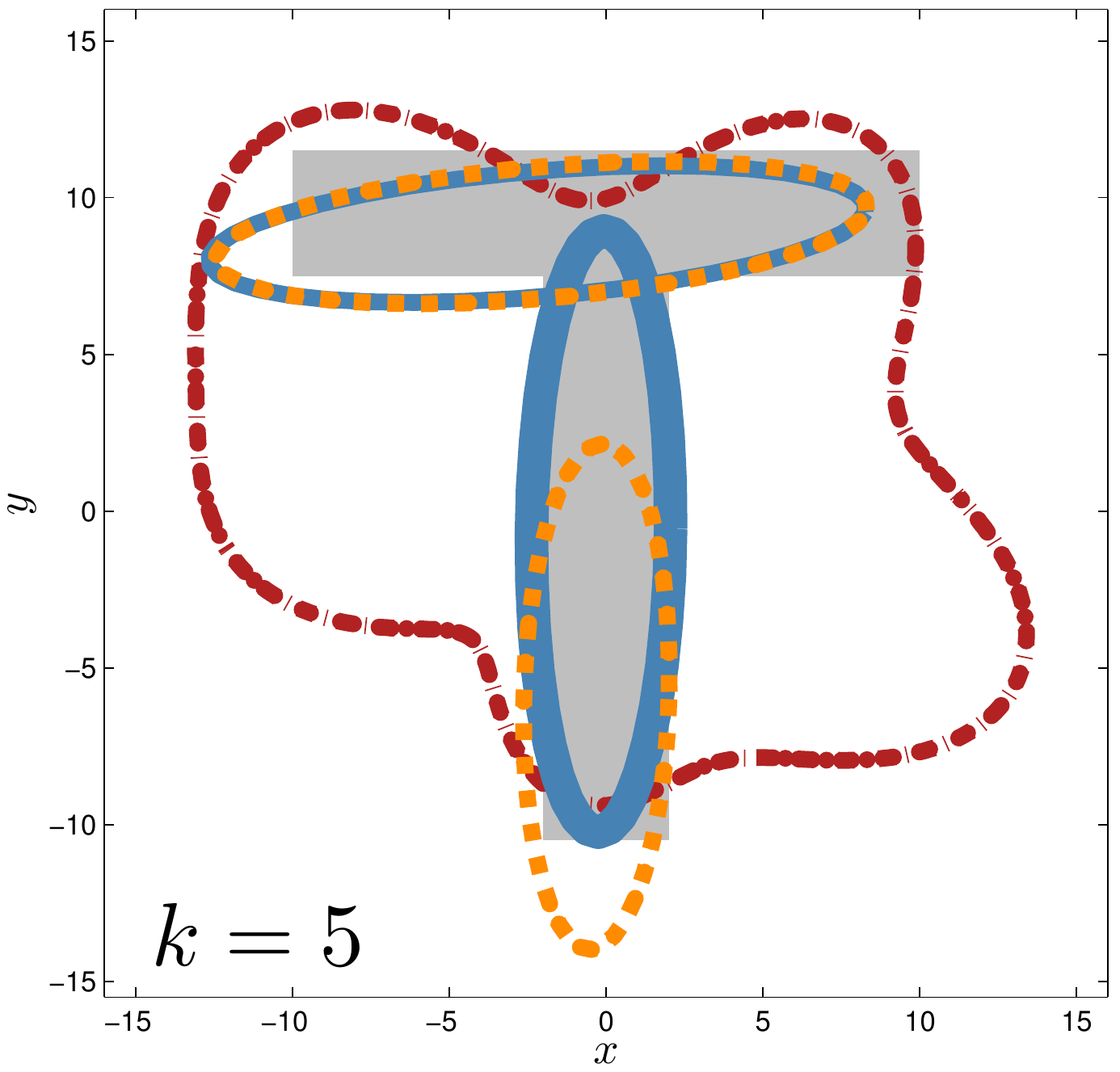} \label{fig:multi_ellipse_shape_comparison_Tshape_gam_2_ex_3}}
\hfill
\subfloat{\includegraphics[width=0.125\textwidth]{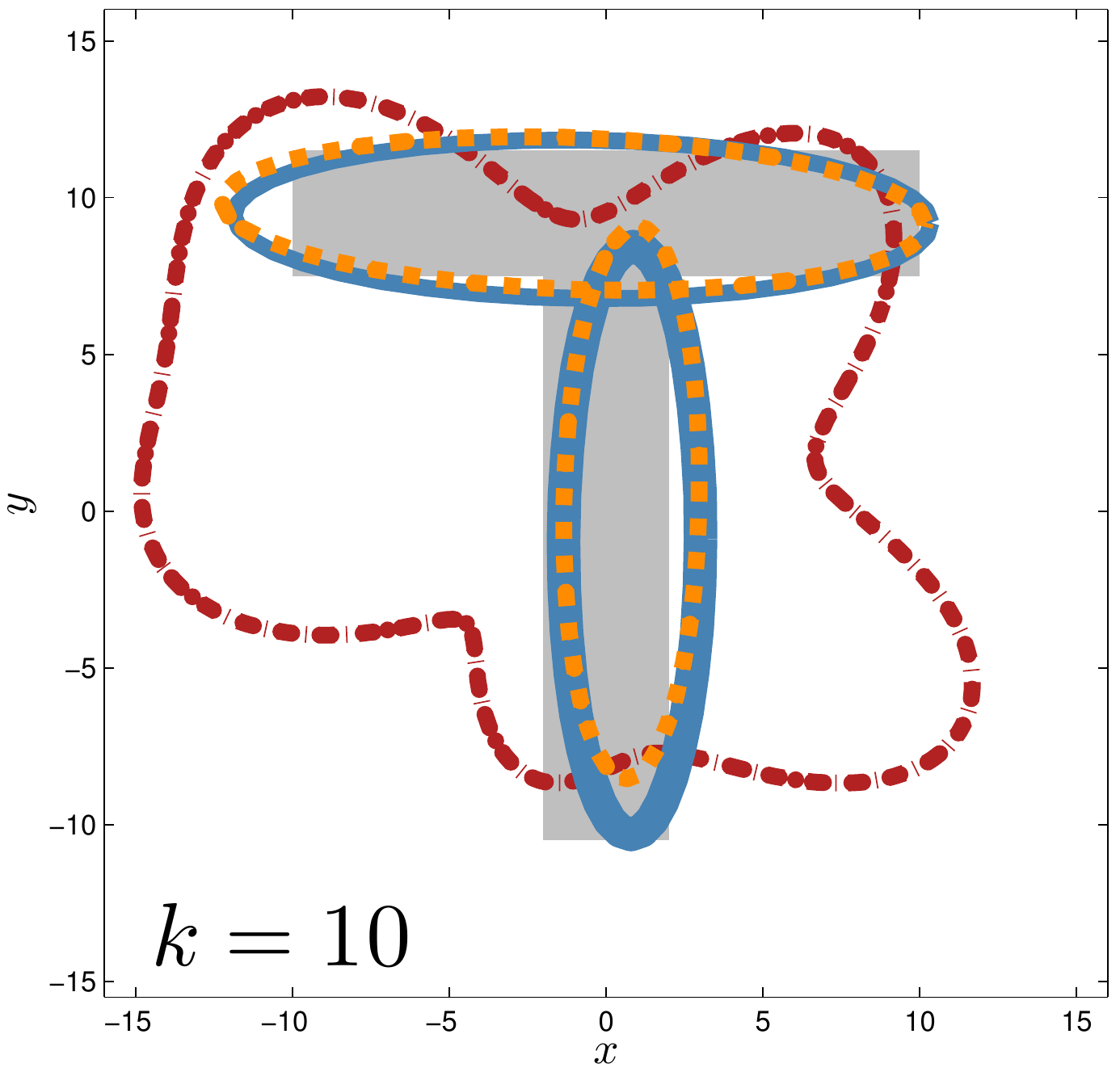} \label{fig:multi_ellipse_shape_comparison_Tshape_gam_2_ex_4}}
}
\centerline{
\subfloat{\includegraphics[width=0.125\textwidth]{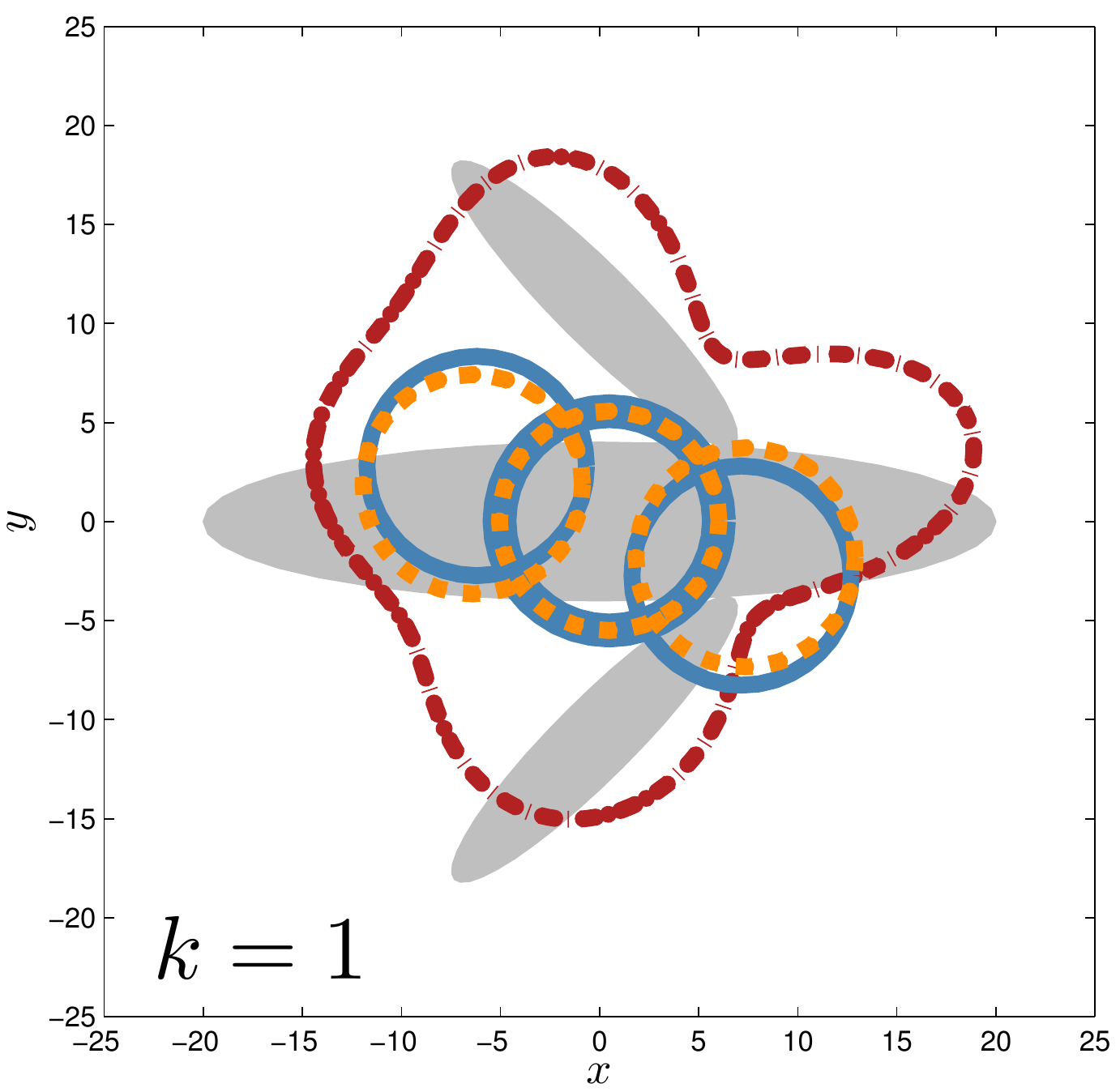} \label{fig:multi_ellipse_shape_comparison_gam_20_ex_1}}
\hfill
\subfloat{\includegraphics[width=0.125\textwidth]{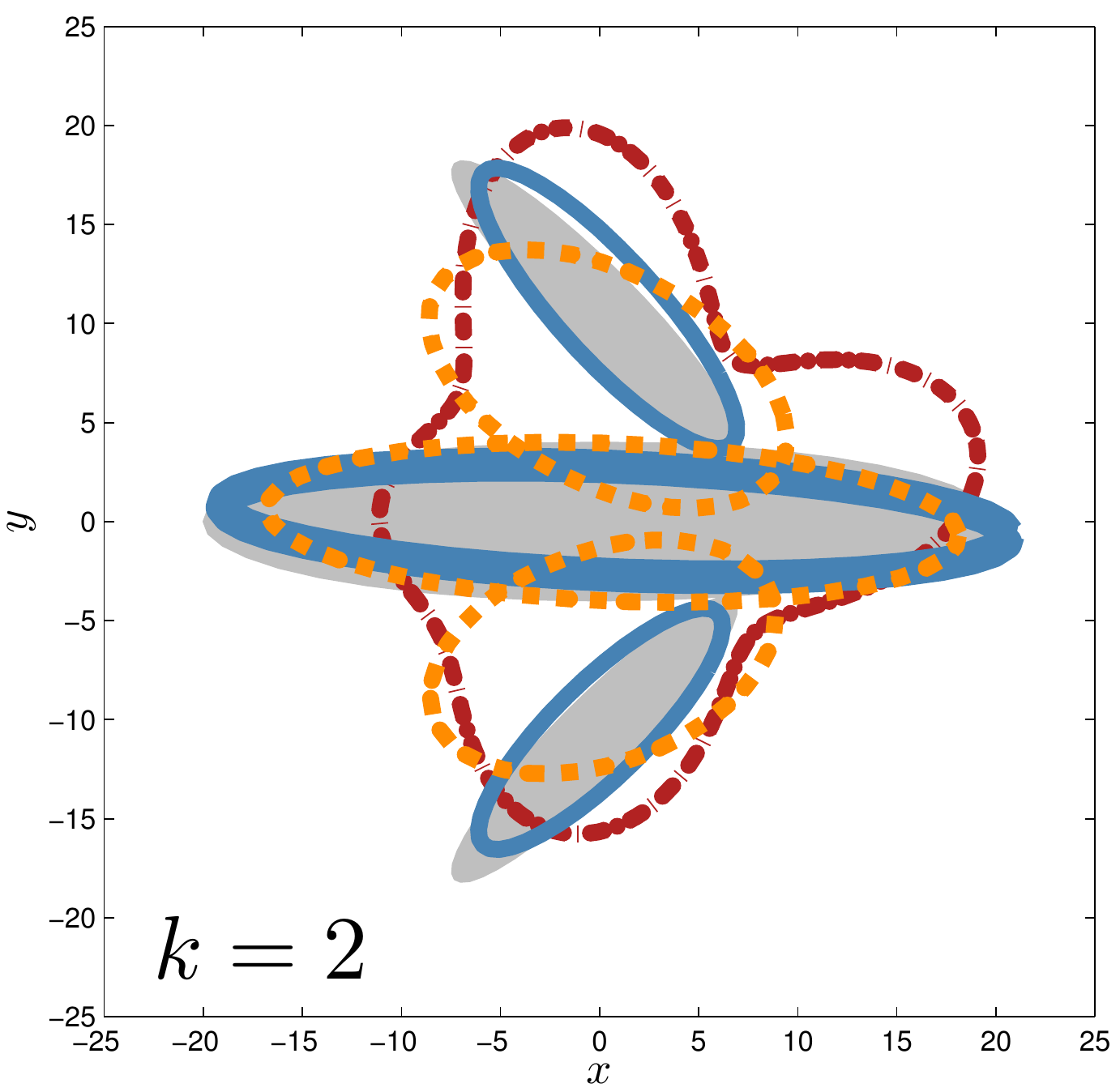} \label{fig:multi_ellipse_shape_comparison_gam_20_ex_2}}
\hfill
\subfloat{\includegraphics[width=0.125\textwidth]{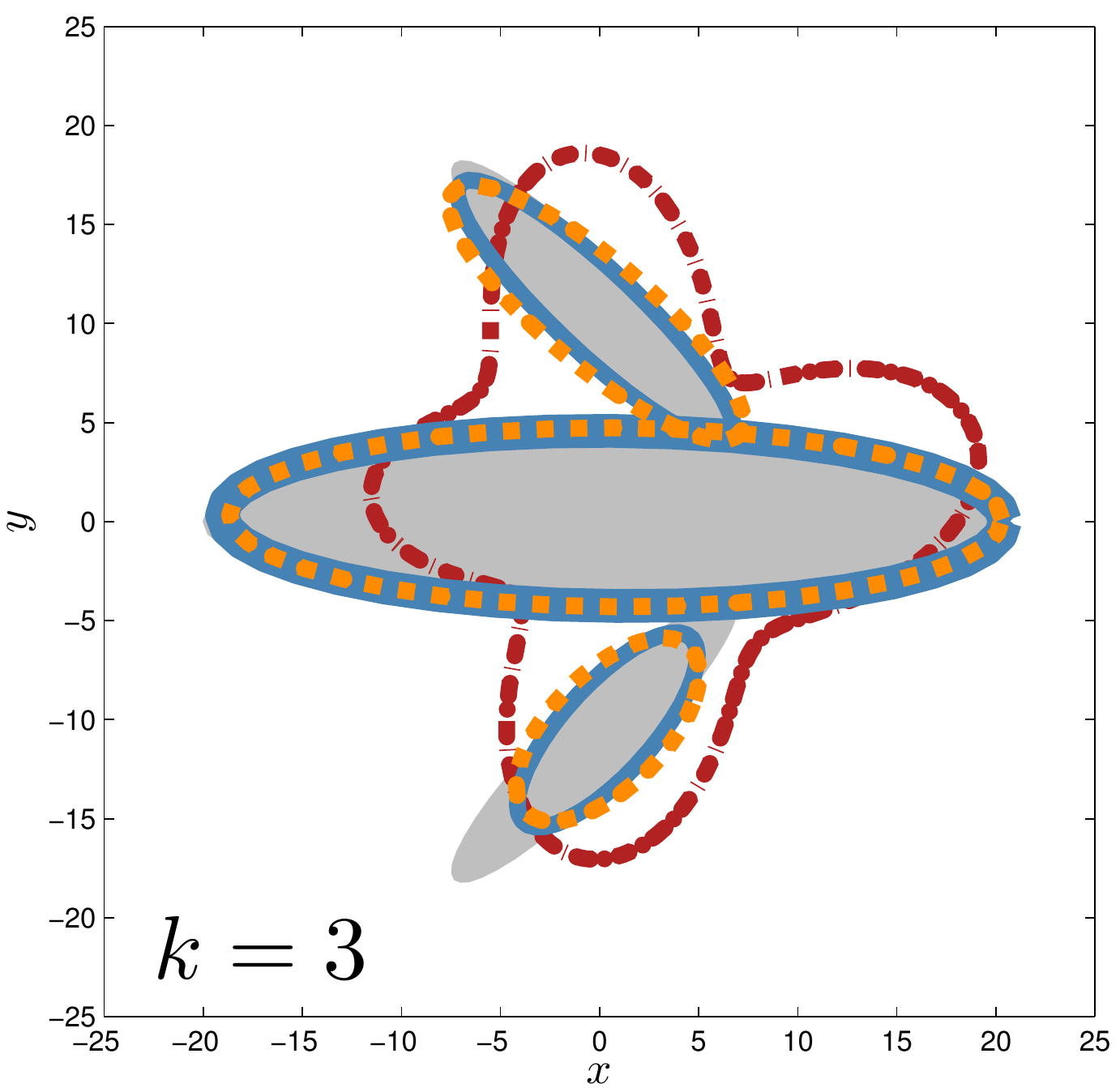} \label{fig:multi_ellipse_shape_comparison_gam_20_ex_3}}
\hfill
\subfloat{\includegraphics[width=0.125\textwidth]{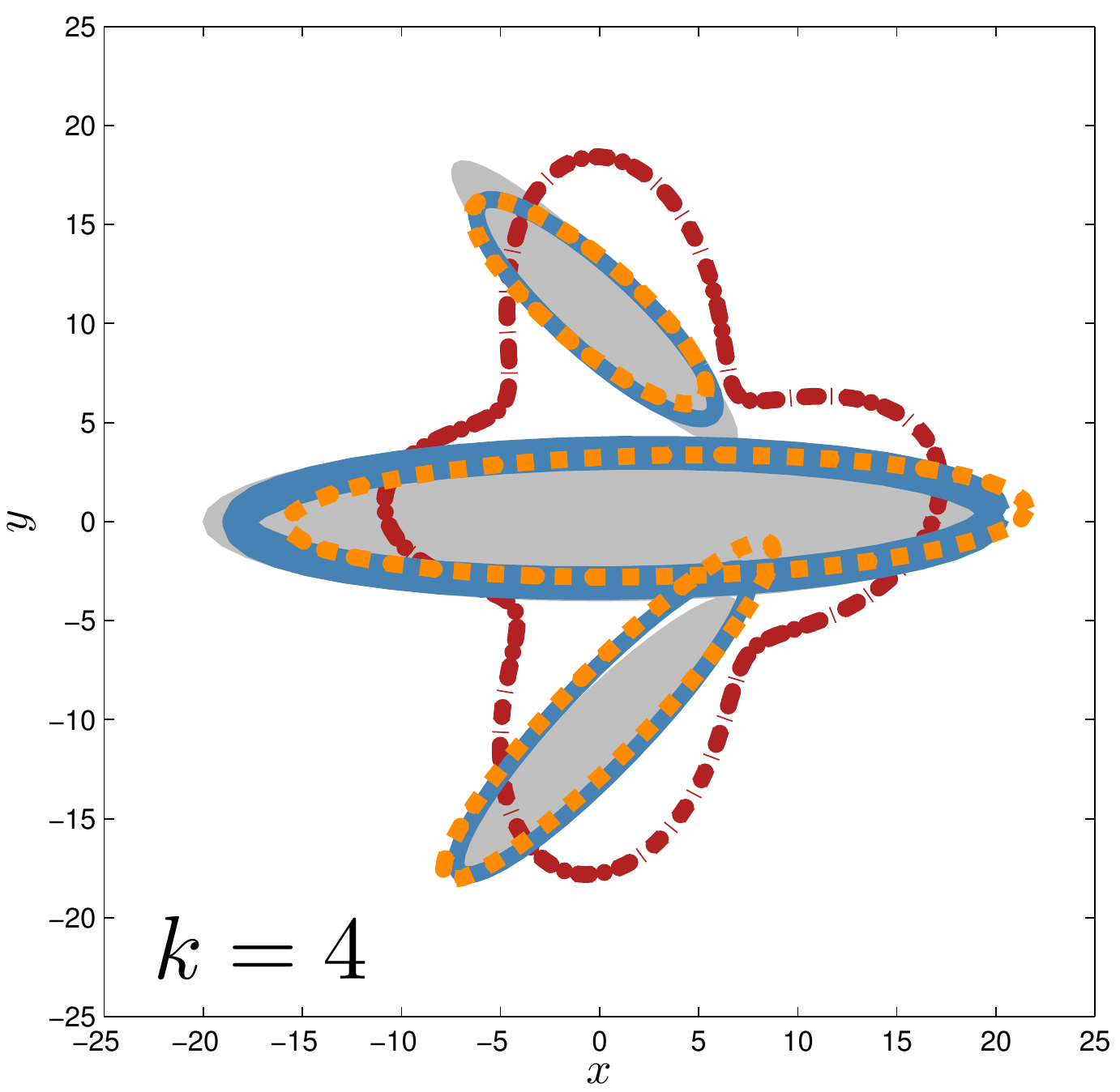} \label{fig:multi_ellipse_shape_comparison_gam_20_ex_4}}
\hfill
\subfloat{\includegraphics[width=0.125\textwidth]{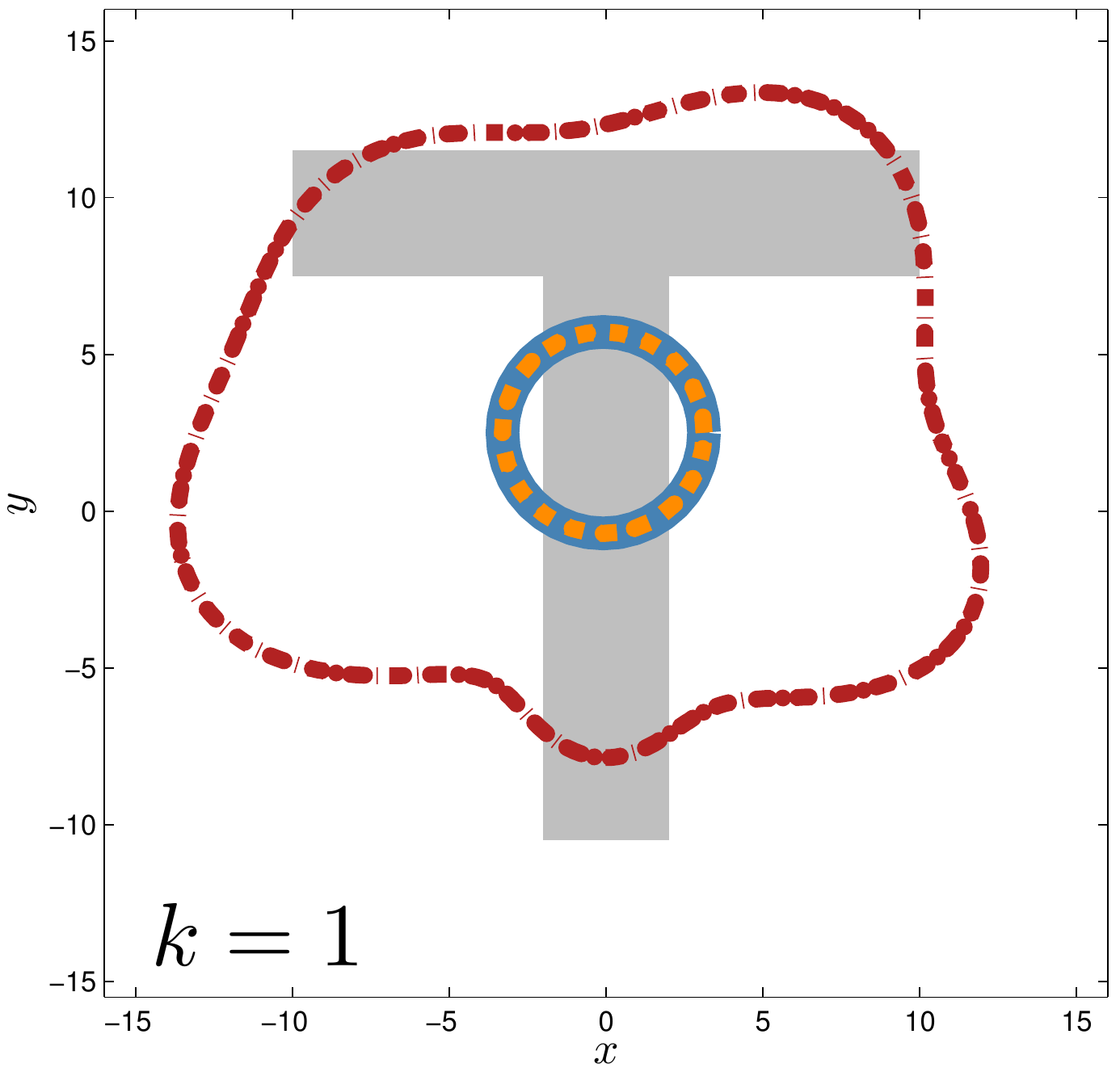} \label{fig:multi_ellipse_shape_comparison_Tshape_gam_20_ex_1}}
\hfill
\subfloat{\includegraphics[width=0.125\textwidth]{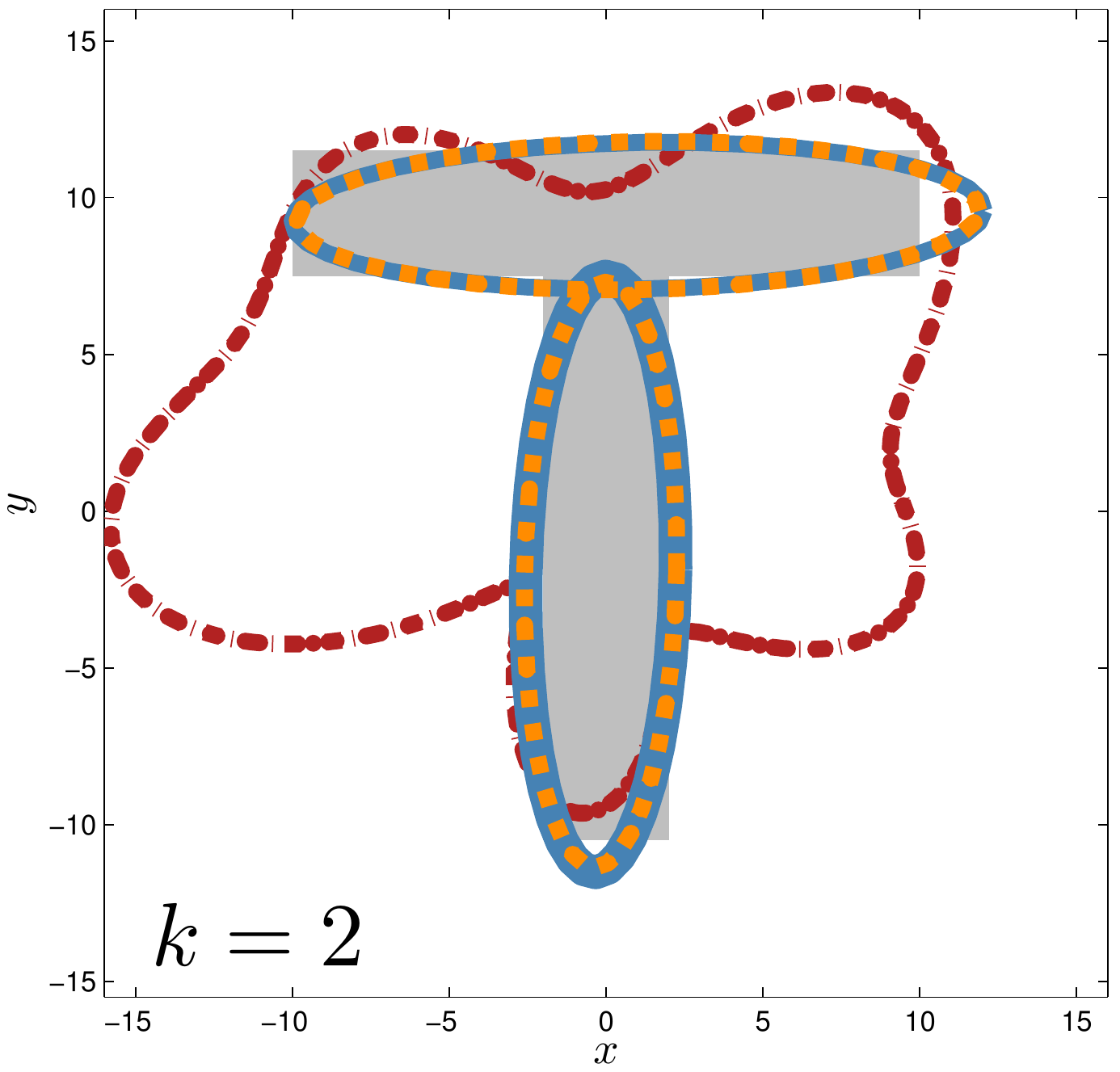} \label{fig:multi_ellipse_shape_comparison_Tshape_gam_20_ex_2}}
\hfill
\subfloat{\includegraphics[width=0.125\textwidth]{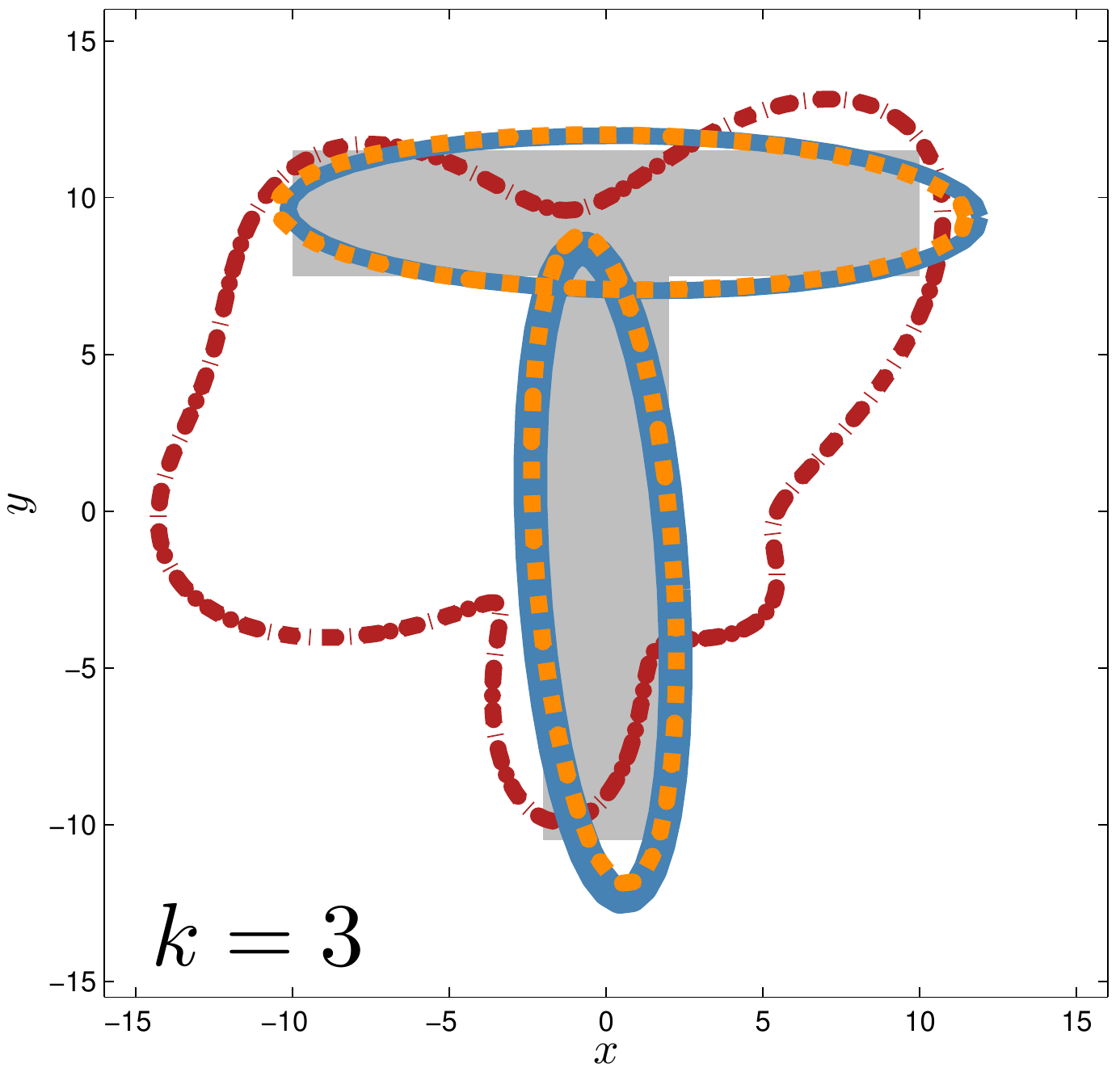} \label{fig:multi_ellipse_shape_comparison_Tshape_gam_20_ex_3}}
\hfill
\subfloat{\includegraphics[width=0.125\textwidth]{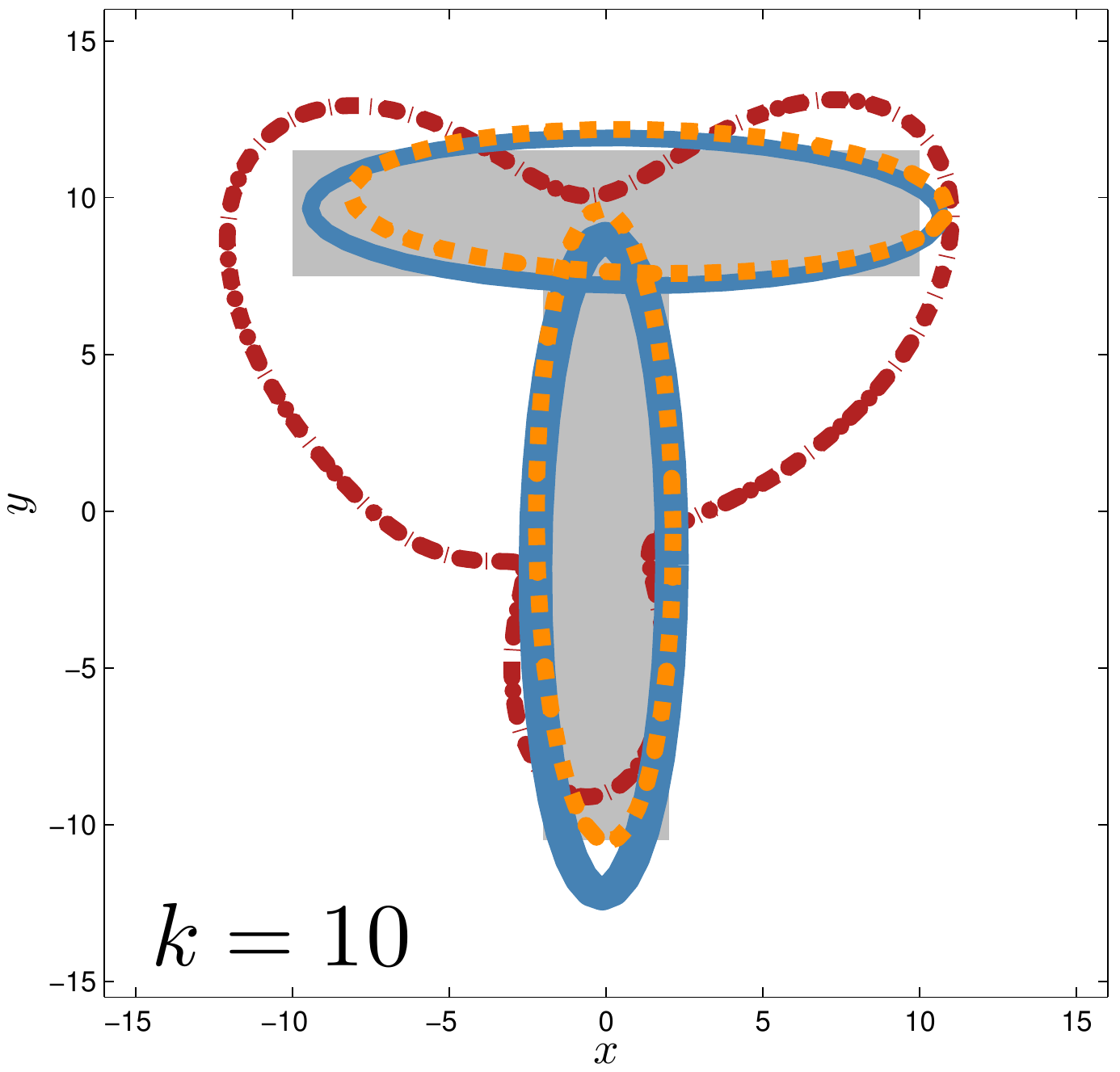} \label{fig:multi_ellipse_shape_comparison_Tshape_gam_20_ex_4}}
}
\caption{Example results for tracking of a plane-shaped and a T-shaped stationary target. Top row $\gamma_{0}=2$, bottom row $\gamma_{0}=20$. True target (gray area) compared to proposed model (solid blue line, main subobject indicated by thicker line), model M2 (dashed orange line), model M1 (dash-dotted red line). Time step $k$ written in lower left corner. As expected convergence is much faster when there are more measurements (i.e. higher $\gamma_{0}$).}
\label{fig:multi_ellipse_shape_comparison}
\end{figure*}

\begin{figure*}[!t]
\centerline{
\subfloat[$\gamma_{0}=2$]{\includegraphics[width=0.33\textwidth]{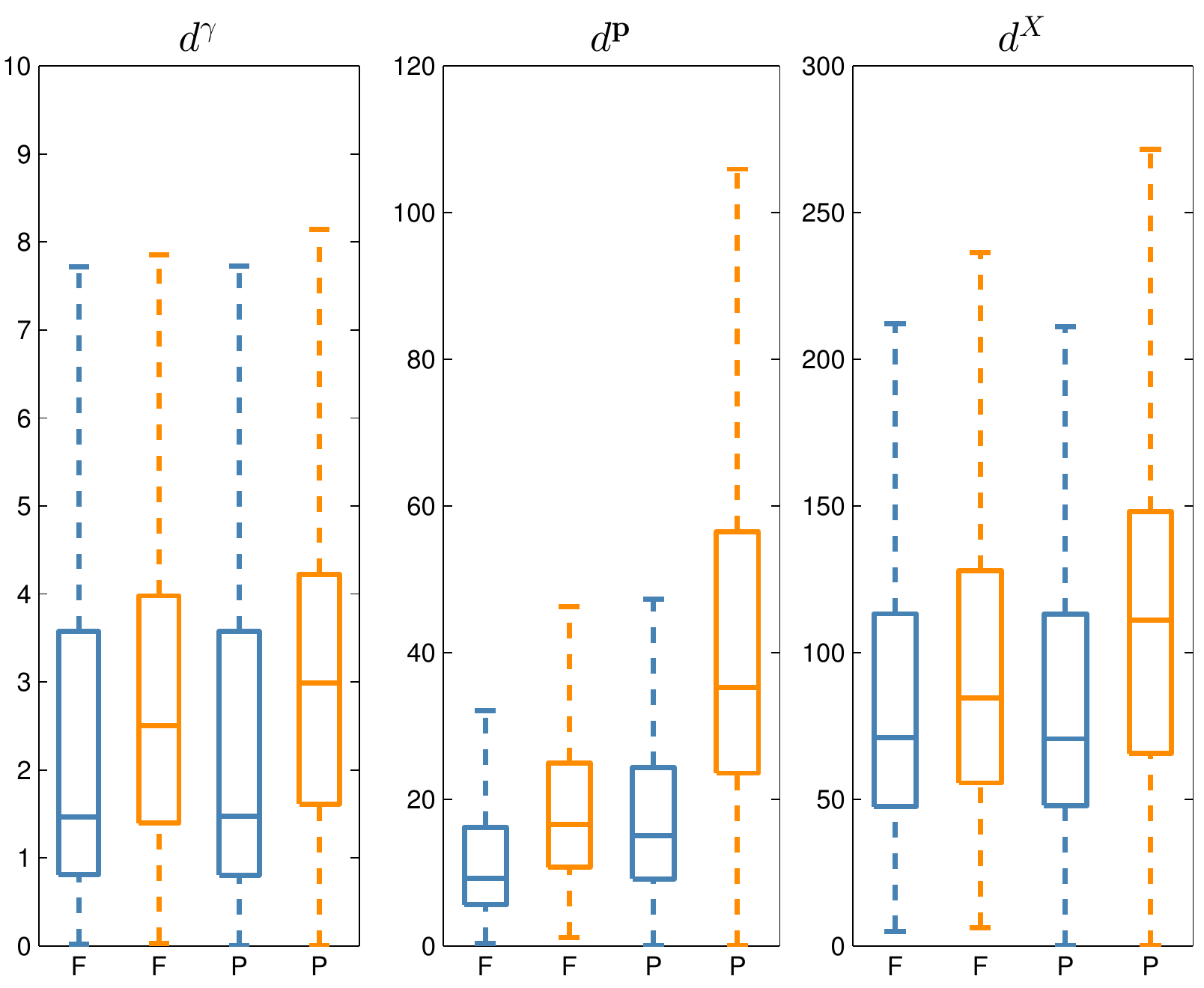} \label{fig:multi_ellipse_planeShape_MC_boxplot_gam_2}}
\vline
\subfloat[$\gamma_{0}=5$]{\includegraphics[width=0.33\textwidth]{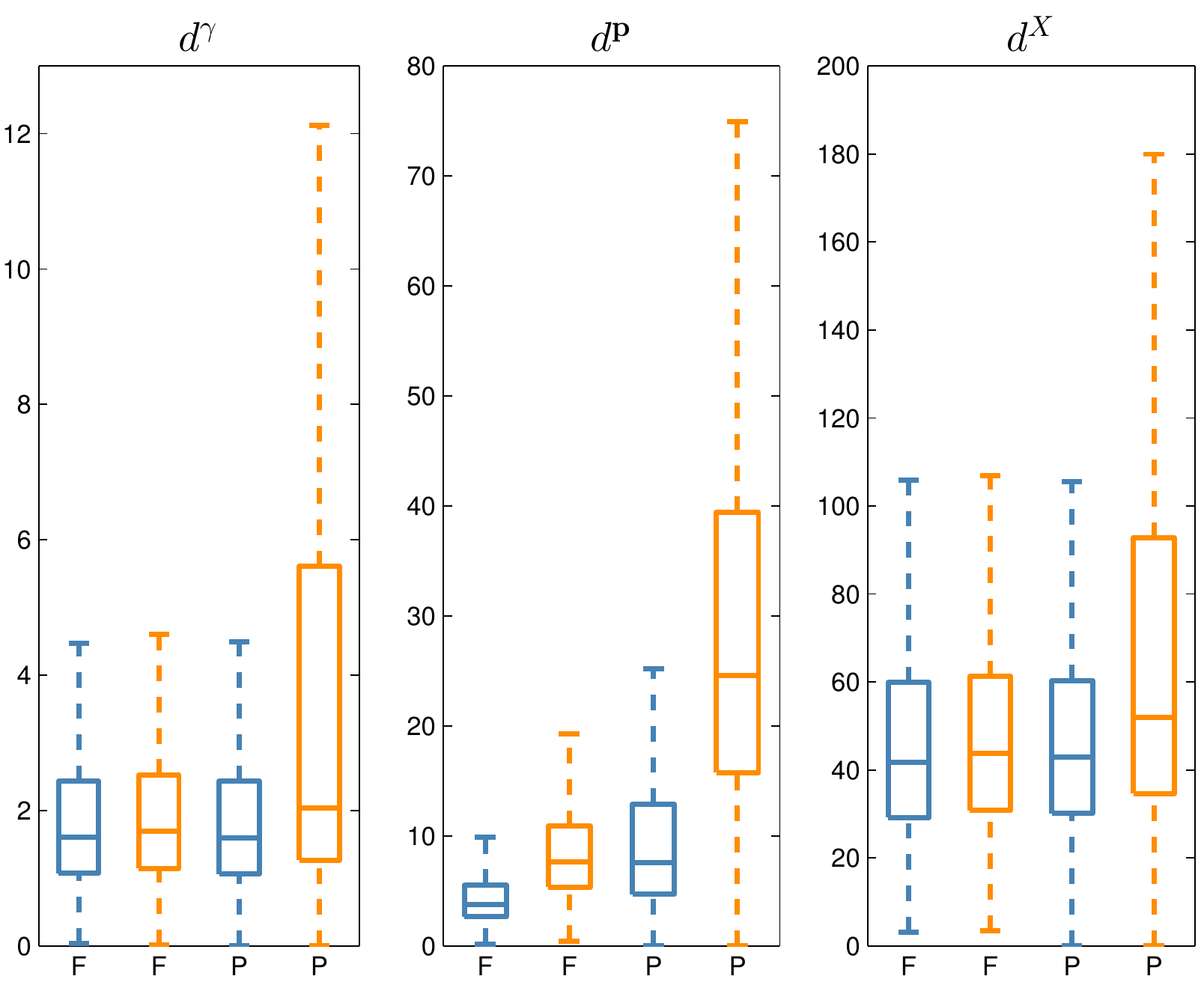} \label{fig:multi_ellipse_planeShape_MC_boxplot_gam_5}}
\vline
\subfloat[$\gamma_{0}=20$]{\includegraphics[width=0.33\textwidth]{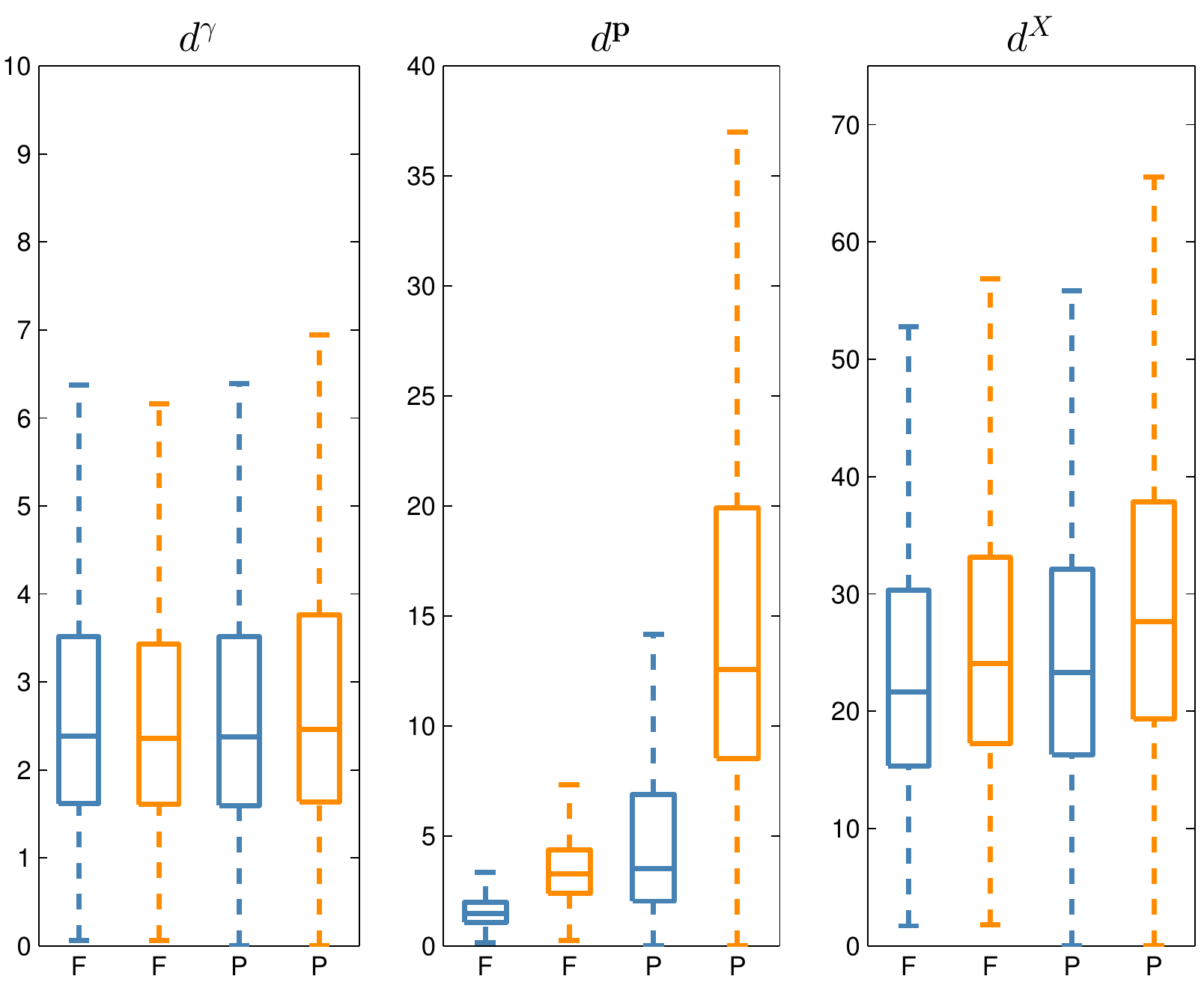} \label{fig:multi_ellipse_planeShape_MC_boxplot_gam_20}}
}
\caption{Estimation errors for plane-shaped target. Proposed model in blue, model M2 in orange. x-labels F and P denote filter errors $d_{k|k}$ and prediction errors $d_{k|k-1}$. On each box, central mark is median, edges of box are 25th and 75th percentiles, whiskers extend to most extreme datapoints the algorithm considers to be not outliers.}
\label{fig:multi_ellipse_planeShape_MC_boxplot}
\end{figure*}

The filter parameters that were used in the implementation are listed in Table~\ref{tab:FilterParameters}.

\begin{table}[htbp]
	\centering
	\caption{Parameters for proposed method}
		\begin{tabular}{lcc}
			Parameter &  & Value \\
			\hline
			Sample time & $T$ & 1 \\
			Number of initial hypotheses & $N_p$ & $2(N_{s,k}-1)$ \\
			Initial kinematics & $\mathbf{c}_{0}$ & $\mathbf{0}_{3\times 1}$ \\
			Initial covariance & $P_{0}$ & $10^{2}\mathbf{I}_{n_x}$ \\
			Measurement rate initial mean & $e$ & 15 \\
			Measurement rate initial variance & $v$ & 10 \\
			Measurement rate prediction factor & $\eta_{k}^{(m)}$ & $1.05$, $\forall m$ \\
			Prediction degrees of freedom & $n_{k+1}^{(m)}$ & $100$, $\forall m$ \\
			Pruning threshold & $\tau$ & $0.01$
		\end{tabular}
	\label{tab:FilterParameters}
\end{table}


For M1 an implementation available online was used\footnote{Thanks to M. Baum and R. Sandkuehler for providing code.\\ {\tiny\texttt{http://www.cloudrunner.eu/algorithm/12/random-hypersurface-model/version/2/}}}. Model M2 was implemented as instructed in Section VI ``Simulation studies'' in \cite{LanRL:2014}, and augmented to include estimation of the measurement rates, see \cite{GranstromO:2012c} for details. The method was parametrized with three motion models: the first corresponds to constant velocity motion; the second corresponds to a turn with turn-rate $\psi$; the third corresponds to a turn with turn-rate $-\psi$. In the simulations the parameter $\psi$ was set to $5$ degrees per second. For the lower measurement rates ($\gamma_{0}=2$ and $\gamma_{0}=5$) sometimes during the maneuvers one of the subobjects would separate from the other two subobjects. To correct this, any hypothesis $p^{(\ell)}(\cdot)$ with a subobject more than $50$ meters from the other subobjects was deleted. In case all hypotheses were deleted, the target estimate was reinitialized.

\begin{figure*}[htp]
\centerline{
\subfloat[$\gamma_{0}=2$]{\includegraphics[width=0.33\textwidth]{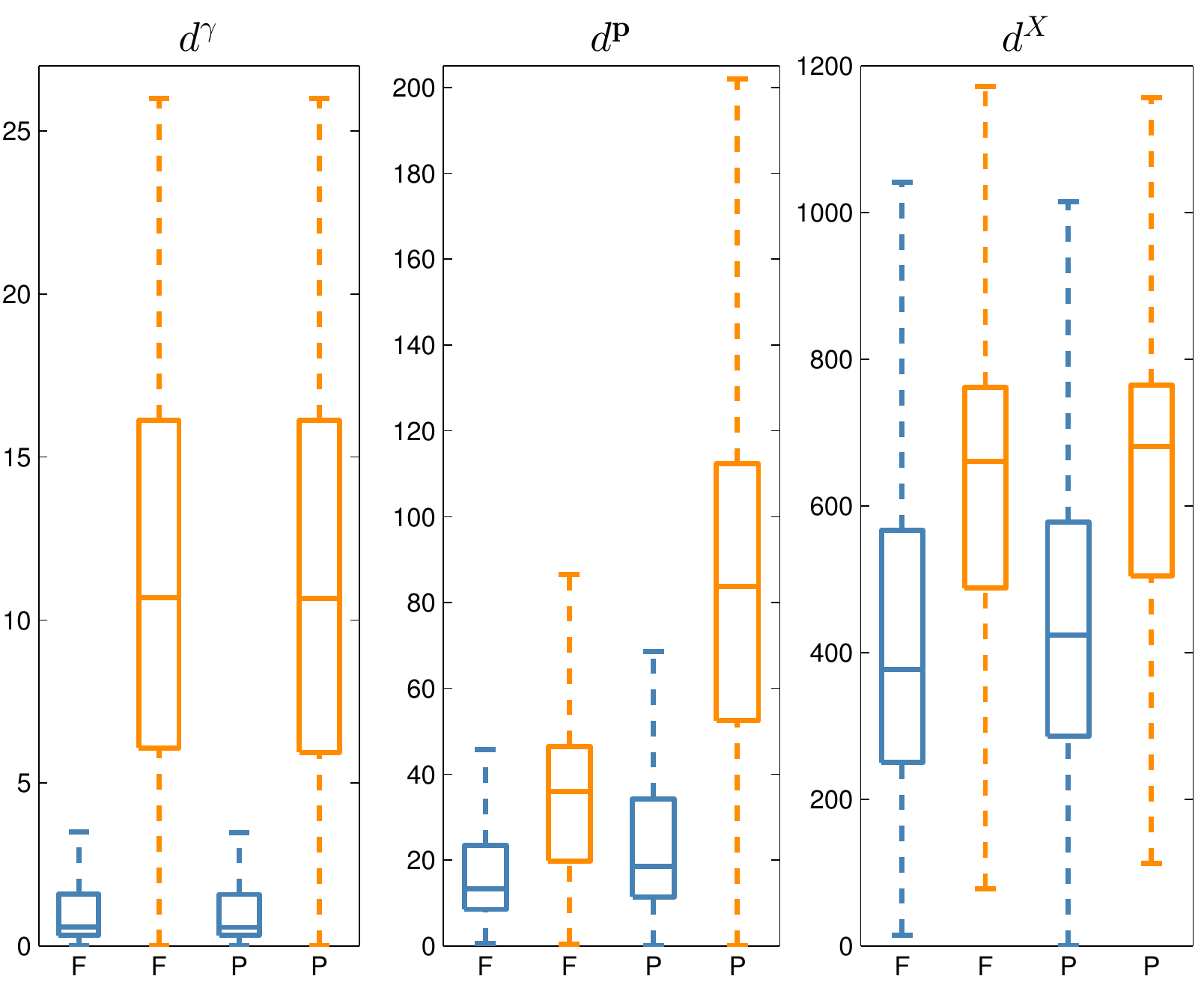} \label{fig:multi_ellipse_Vshape_MC_boxplot_gam_2}}
\vline
\subfloat[$\gamma_{0}=5$]{\includegraphics[width=0.33\textwidth]{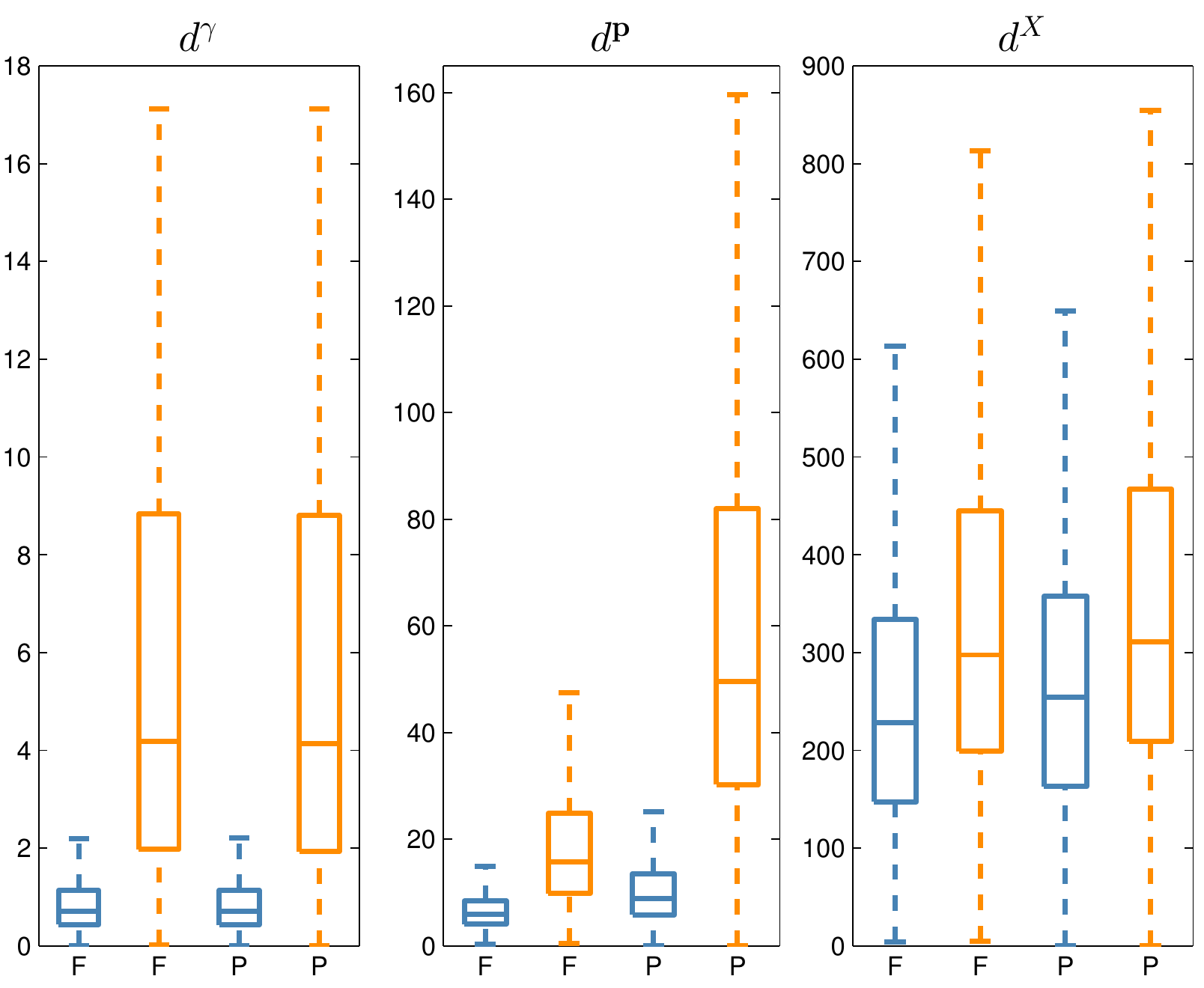} \label{fig:multi_ellipse_Vshape_MC_boxplot_gam_5}}
\vline
\subfloat[$\gamma_{0}=20$]{\includegraphics[width=0.33\textwidth]{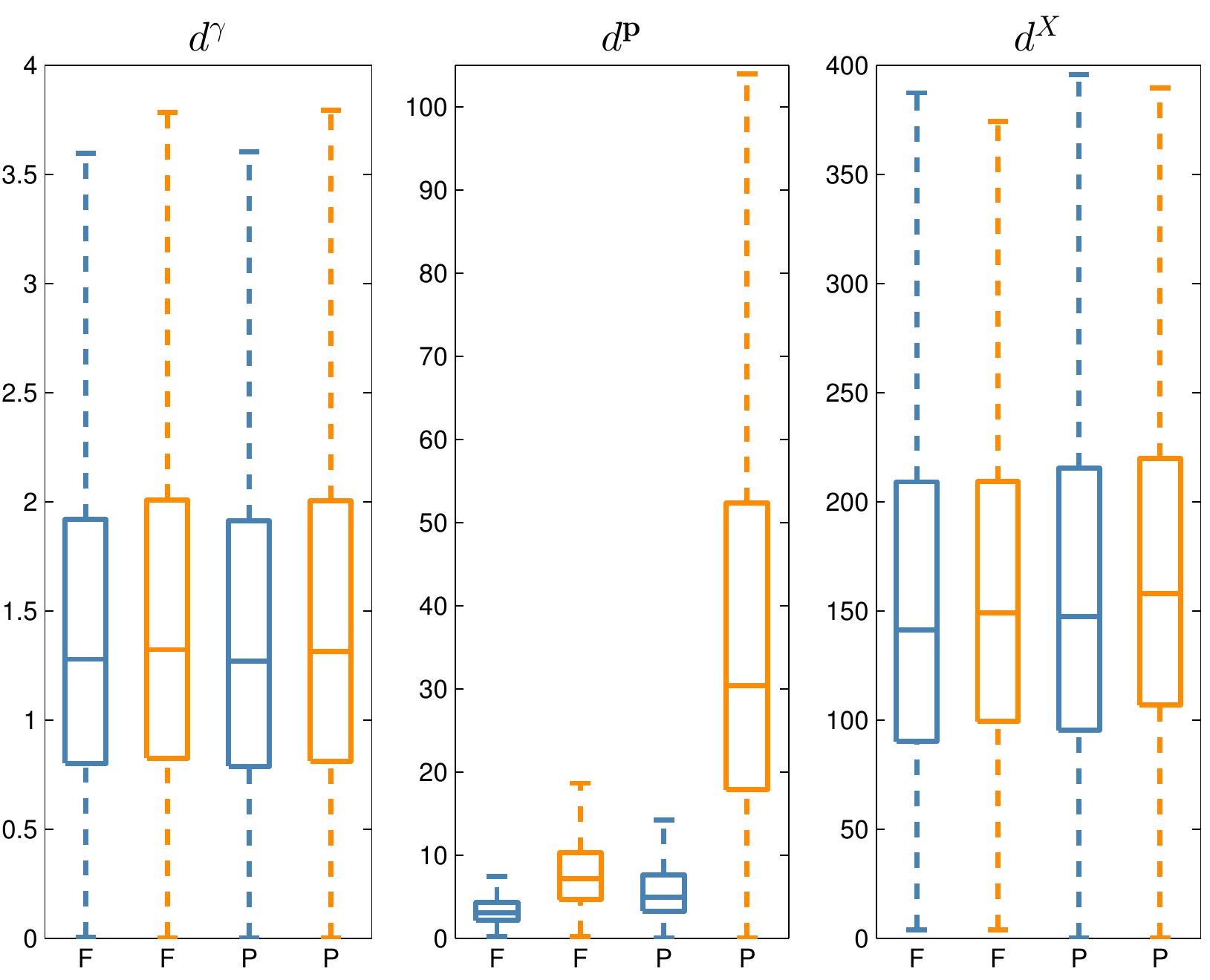} \label{fig:multi_ellipse_Vshape_MC_boxplot_gam_20}}
}
\caption{Estimation errors for V-shaped target. Proposed model in blue, model M2 in orange. x-labels F and P denote filter errors $d_{k|k}$ and prediction errors $d_{k|k-1}$. On each box, central mark is median, edges of box are 25th and 75th percentiles, whiskers extend to most extreme datapoints the algorithm considers to be not outliers.}
\label{fig:multi_ellipse_Vshape_MC_boxplot}
\end{figure*}

\subsection{Results: stationary target}
The T-shaped and the plane-shaped targets were simulated for measurement rates $\gamma_{0}=2$ and $\gamma_{0}=20$. A comparison of the proposed model and the models M1 and M2 is shown in \figurename~\ref{fig:multi_ellipse_shape_comparison}. As expected all three methods converge much faster when there are more measurements (\iep higher $\gamma_{0}$). All three methods give reasonable results, however the proposed model and the M2 are closer to the ground truth than M1. Because of this, for the moving target we only compare the proposed model and model M2.

\subsection{Results: moving target}
\begin{figure}[!t]
\centerline{
\subfloat{\includegraphics[width=0.90\columnwidth]{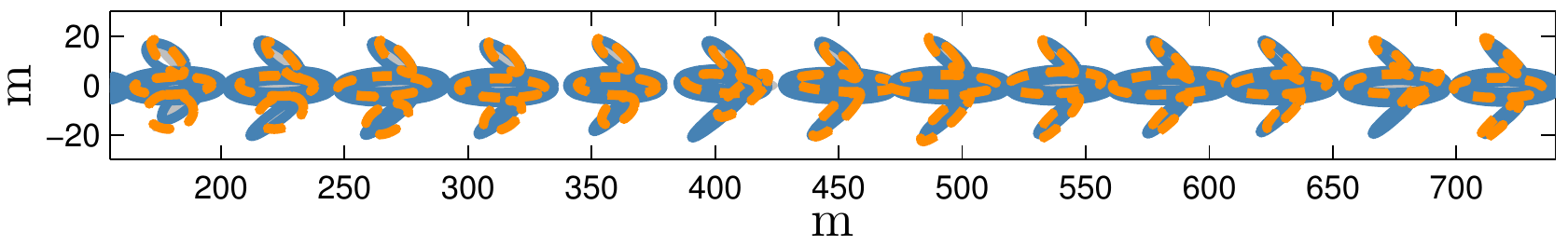} \label{fig:multi_ellipse_Example_Results_Straight_Filter}}
}
\centerline{
\subfloat{\includegraphics[width=0.90\columnwidth]{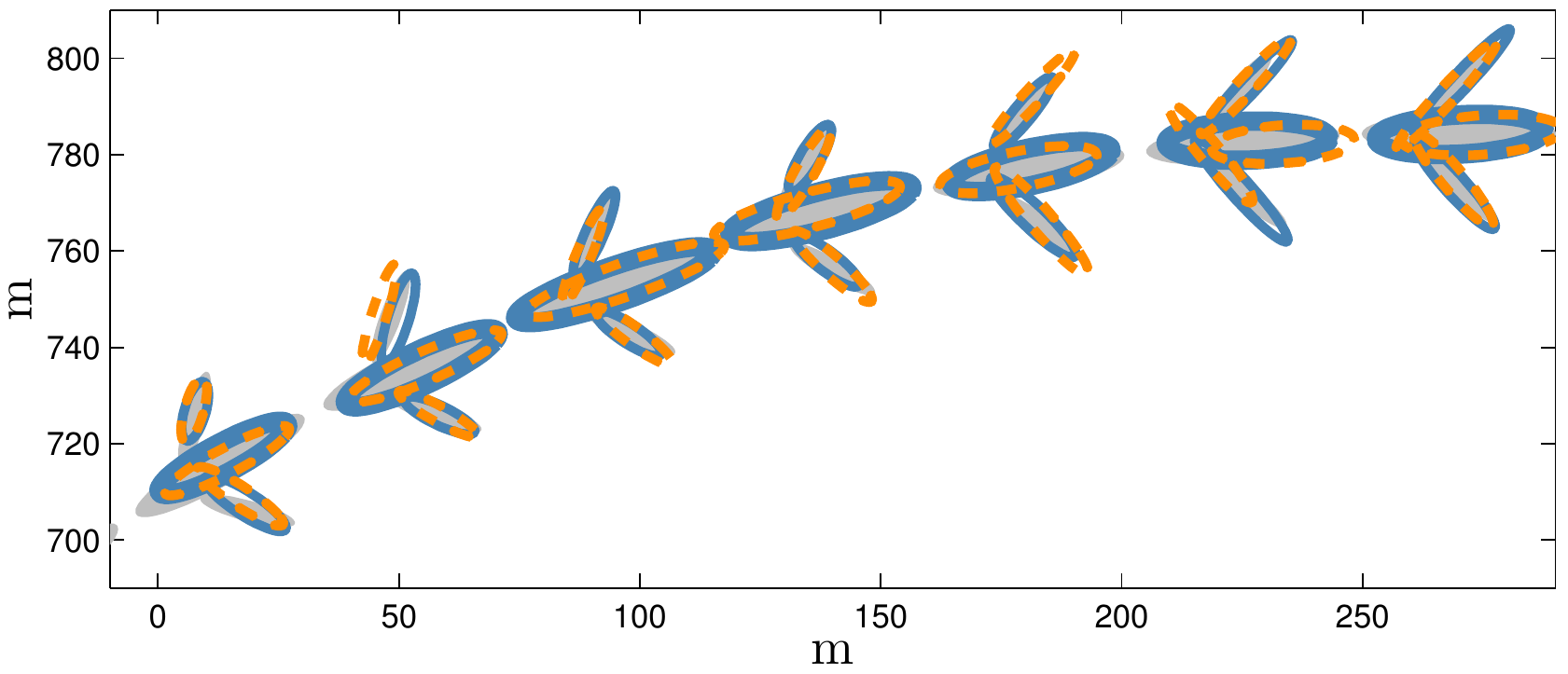} \label{fig:multi_ellipse_Example_Results_Filter}}
}
\centerline{
\subfloat{\includegraphics[width=0.90\columnwidth]{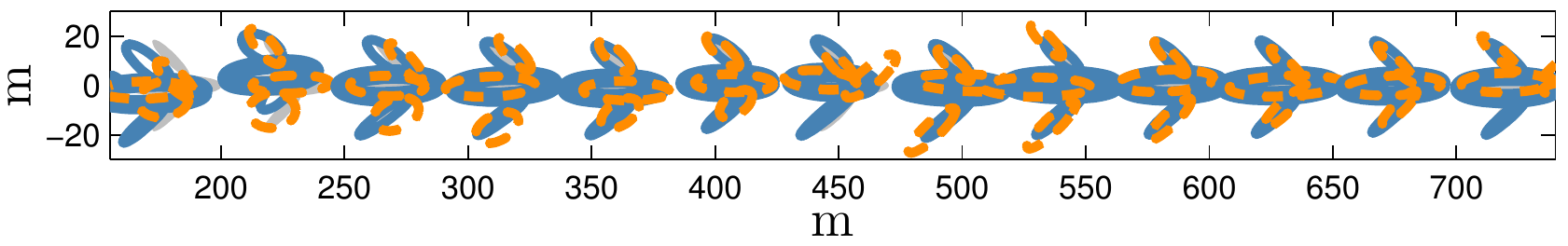} \label{fig:multi_ellipse_Example_Results_Straight_Prediction}}
}
\centerline{
\subfloat{\includegraphics[width=0.90\columnwidth]{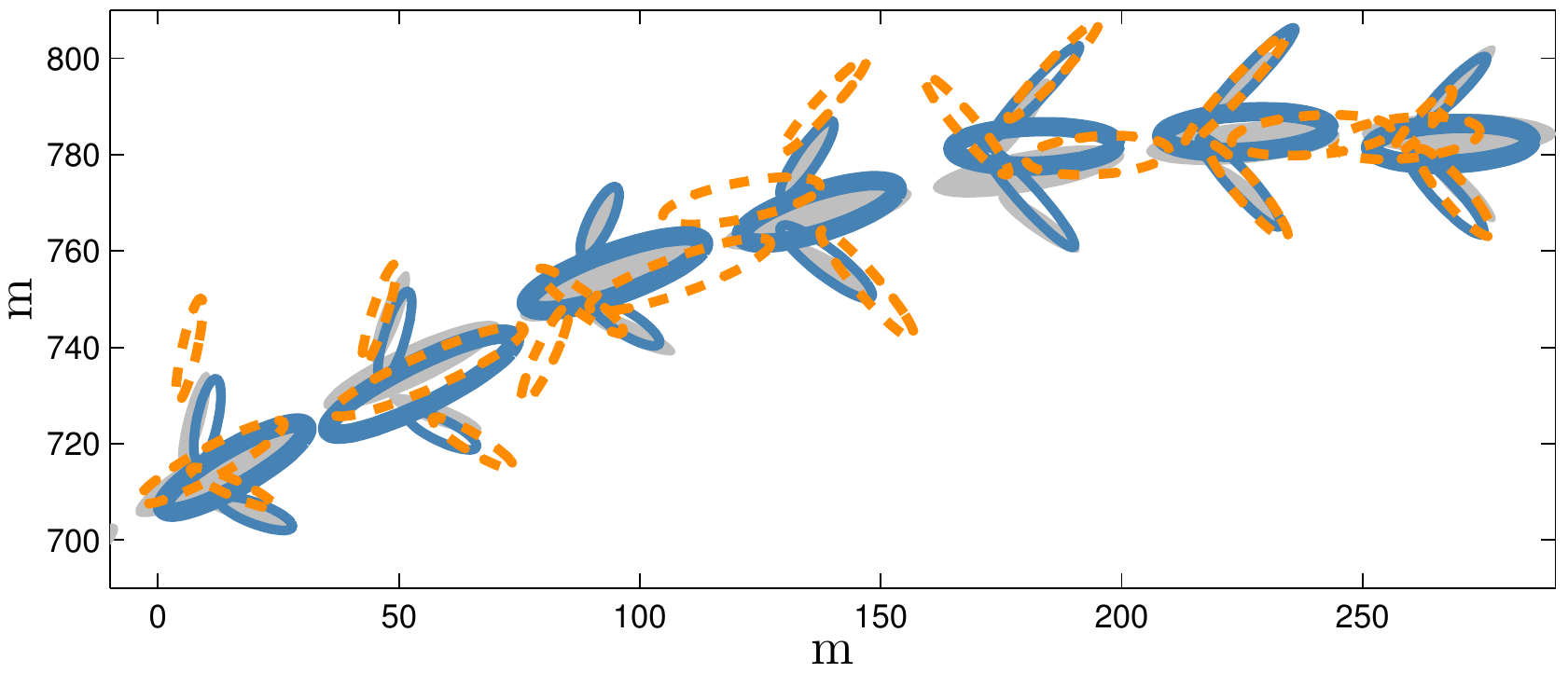} \label{fig:multi_ellipse_Example_Results_Prediction}}
}
\caption{Example results for $\gamma_{0}=5$ for the trajectory in \figurename~\ref{fig:multi_ellipse_true_tracks}. Top: filtered estimates. Bottom: predicted estimates. Ground truth (gray area), compared to model M2 (dashed orange line) and proposed method (solid blue line, main subobject drawn with thicker line).}
\label{fig:multi_ellipse_Example_Results}
\end{figure}

The plane-shaped and the V-shaped targets were simulated for $\gamma_{0}=2$, $\gamma_{0}=5$, and $\gamma_{0}=20$. For each value of the measurement rate $\gamma_{0}$ the scenarios were simulated $10^{3}$ times.
For the plane-shaped target (three subobjects) the filter errors $d_{k|k}$ and prediction errors $d_{k|k-1}$ are shown in \figurename~\ref{fig:multi_ellipse_planeShape_MC_boxplot}, for the V-shaped target (two subobjects) the results are shown in \figurename~\ref{fig:multi_ellipse_Vshape_MC_boxplot}.
Example filter and prediction outputs for the plane-shaped target for $\gamma_{0}=5$ are shown in \figurename~\ref{fig:multi_ellipse_Example_Results}. From the results the following observations can be made:
\begin{itemize}
	\item Both the prediction errors and the filter errors are smaller for the presented method for all $\gamma_{0}$.
	\item The biggest difference is for the subobject position errors, especially during maneuvers. Note that, even if the estimated random matrices have the correct size and orientation, the subobject positions are more important for the overall extended target extension estimate. The larger the subobject position errors are, the more distorted the overall shape becomes, which can be seen in \figurename~\ref{fig:multi_ellipse_Example_Results}.
\end{itemize}
The lower errors for the presented method, especially the lower position errors, are a direct effect of 
\begin{inparaenum}[\itshape a\upshape)]
	\item using a single state vector for the subobject positions and the kinematics including a full covariance matrix; and 
	\item having unified kinematics for the subobject positions.
\end{inparaenum}

\subsection{Computational complexity}
\label{sec:Computational_Complexity}
The code used in this work was implemented in \matlab and run on a $2.83$GHz Intel Core2 Quad CPU with $3.48$GB of RAM running Windows. Note that the code has not been optimized for speed.


In each time step approximately $15$ to $25$ different partitions of the set of measurements were computed. The average number of measurement-to-subobject association events are given in Table~\ref{tab:AssociationEvents}. A comparison to the number of association events if there are $\E[n_{z,k}]$ measurements shows that the set of assocation events is reduced by several orders of magnitude. It is noteworthy that for $\gamma_{0}=20$ the reduction in number of association events is by far greatest, yet the estimation errors are smaller for $\gamma_{0}=20$ than for $\gamma_{0}=2$ and $\gamma_{0}=5$. 

\begin{table}[htbp]
	\centering
	\caption{Number of association events, mean $\pm$ standard deviation}
		\begin{tabular}{ccc}
			$\gamma_{0}$ & $\left|\bar{\setTheta}\right|$ & $(N_{s,k})^{\E[n_{z,k}]}$ \\
			\hline
			$2$ & $114 \pm 29$ & $6.6 \times 10^3$ \\
			$5$ & $128 \pm 24$ & $3.5\times 10^{9}$ \\
			$20$ & $130 \pm 23$ & $1.5\times 10^{38}$
		\end{tabular}
	\label{tab:AssociationEvents}
\end{table}

The number of mixture components increase in each time step. However, in the mixture reduction step many components can be pruned, and the remaining components can be merged such that typically only $2$ to $6$ components remain.

%

The average cycle times are given in Table~\ref{tab:CycleTimes}. We see that the times for prediction and reduction are independent of $\gamma_{0}$. The time to compute clusters for data association increases when $\gamma_{0}$ increases, because with more measurements it takes more time to cluster them. The correction time decreases when $\gamma_{0}$ increases, because with more measurements the scenario is less ambiguous and the probability mixture typically has fewer components. Note that these two increases/decreases in time offset each other such that the average total cycle time is about $1.7$ seconds for all values of $\gamma_{0}$ that were tested.

The average cycle time for the method from \cite{LanRL:2014} is $2.1 \pm 1.4$ seconds for $\gamma_{0}=2$, $1.7\pm 0.7$ seconds for $\gamma_{0}=5$, and $1.6\pm 0.5$ seconds for $\gamma_{0}=20$. When comparing the cycle times, remember that neither implementation was optimized for speed.

\begin{table}[htbp]
	\centering
	\caption{Cycle times [seconds], mean $\pm$ standard deviation}
		\begin{tabular}{cccccc}
			$\gamma_{0}$ & Prediction & Clusters & Correction & Reduction & Total \\
			\hline
			$2$ & $0.4 \pm 0.2$ & $0.7 \pm 0.1$ & $0.6 \pm 0.4$ & $0.02 \pm 0.02$ & $1.7 \pm 0.6$ \\
			$5$ & $0.3 \pm 0.1$ & $0.9 \pm 0.1$ & $0.5 \pm 0.3$ & $0.01 \pm 0.02$ & $1.8 \pm 0.4$ \\
			$20$ & $0.2 \pm 0.1$ & $1.2 \pm 0.1$ & $0.3 \pm 0.2$ & $0.01 \pm 0.02$ & $1.7 \pm 0.3$
		\end{tabular}
	\label{tab:CycleTimes}
\end{table}


\section{Conclusions and future work}
\label{sec:conclusion_future_work}

The paper has presented an extended target model in which the target extension is modeled using a collection of elliptical subobjects. 
The simulation results show that the proposed model outperforms previous work.

The presented model can be reduced to the cases where either the measurement rates $\gamma_{k}^{(i)}$, the extension matrices $\ext_{k}^{(i)}$, or both, are known. The simulation study considered extended targets, however the model is applicable also to group targets. In case the targets in the group are moving relative to each other, in addition to the unified group movement, individual kinematics can be estimated along with the unified kinematics.

An important topic for future work is to include estimation of the number of subobjects. Interesting lines for future work also include integrating the proposed model into a multiple target algorithm, such as an extended target \phd- or \cphd filter, see \egp \cite{mahler_FUSION_2009_extTarg,LundquistGO:2013}. Furthermore alternative measurement models can be considered, see \egp \cite{FeldmannFK:2011,Orguner:2012,LanRL:2012} for recent work on this topic within the random matrix framework.

\appendix
The prior distribution and the measurement likelihood are
\begin{subequations}
\begin{align}
p\left(\xi\right) = & \Npdfbig{\sx}{m}{P} \nonumber \\
& \times \prod_{i=1}^{N_s} \Gammapdf{\gamma_{i}}{\alpha_{i}}{\beta_{i}} \IWishpdf{\ext_{i}}{v_{i}}{V_{i}}, \label{eq:GGIW_prior} \\
p\left(\setZ\left|\xi\right.\right) = & \prod_{i=1}^{N_s} n_{i}! \mathcal{PS}\left(n_{i};\ \gamma_{i}\right) \prod_{j=1}^{n_{i}} \Npdfbig{\sz_{ij}}{H_{i}\sx}{\ext_{i}}. \label{eq:subobject_measurement_likelihood}
\end{align}
\end{subequations}
The problem at hand is to derive the posterior distribution $p\left(\xi|\setZ\right)$ and the corresponding likelihood $\mathcal{L}$,
\begin{align}
p\left(\setZ\left|\xi\right.\right)p\left(\xi\right) = \mathcal{L} p\left(\xi|\setZ\right),
\end{align}
where the posterior $p\left(\xi|\setZ\right)$ is of the same functional form as the prior \eqref{eq:GGIW_prior}, \iep
\begin{align}
p\left(\xi|\setZ\right) = & \Npdfbig{\sx}{m^{+}}{P^{+}} \nonumber \\
& \times \prod_{i=1}^{N_s} \Gammapdf{\gamma_{i}}{\alpha_{i}^{+}}{\beta_{i}^{+}} \IWishpdf{\ext_{i}}{v_{i}^{+}}{V_{i}^{+}}, \label{eq:GGIW_posterior}
\end{align}

The product of Gaussian distributions in \eqref{eq:subobject_measurement_likelihood} can be rewritten as follows
\begin{align}
& \prod_{j=1}^{n_i} \Npdfbig{\sz_{ij}}{H_i\sx}{\ext_i} \label{eq:aux1} \\
= & \left(2\pi\right)^{-n_i d/2}|\ext_i|^{-n_i/2} \nonumber \\
& \times \etr \left(-\frac{1}{2} \left(\sum_{j=1}^{n_i}\left(\sz_{ij}-H_i\sx\right)\left(\sz_{ij}-H_i\sx\right)^{\tp}\right)\ext_i^{-1}\right), \nonumber
\end{align}
where $\etr\left(\cdot\right)=\exp\left(\tr\left(\cdot\right)\right)$ is exponential trace. Define the centroid measurements as
\begin{equation}
\bar{\sz}_i \triangleq \frac{1}{n_i}\sum_{j=1}^{n}\sz_{ij}
\end{equation}
and the scatter matrices as 
\begin{equation}
Z_i\triangleq\sum_{j=1}^{n_i} \left(\sz_{ij} - \bar{\sz}_i\right)\left(\sz_{ij} - \bar{\sz}_i\right)^{\tp},
\end{equation}
and rewrite the summation as
\begin{align}
& \sum_{j=1}^{n_i}\left(\sz_{ij}-{H}_i\sx\right)\left(\sz_{ij}-{H}_i\sx\right)^{\tp} \nonumber \\
= & Z_i+ n_i \left(\bar{\sz}_i - {H}_i\sx\right)\left(\bar{\sz}_i - \tilde{H}_i\sx\right)^{\tp}.
\label{eq:aux2}
\end{align}
Inserting \eqref{eq:aux2} into \eqref{eq:aux1} gives
\begin{subequations}
\begin{align}
&\prod_{j=1}^{n_i} \Npdfbig{\sz_{ij}}{H_i\sx}{\ext_i}\\
=& \left(2\pi\right)^{-\frac{n_i d}{2}}|\ext_i|^{-\frac{n_i}{2}}{ \etr \left(-\frac{1}{2} Z_i\ext_i^{-1}\right)} \nonumber \\
& \times \etr \left(-\frac{1}{2} \left(\bar{\sz}_i - {H}_i\sx\right)\left(\bar{\sz}_i - {H}_i\sx\right)^{\tp}\left(\frac{\ext_i}{n_i}\right)^{-1}\right) \\
=& \left(2\pi\right)^{-\frac{(n_i-1)d}{2}}|\ext_i|^{-\frac{n_i-1}{2}}{n_i}^{-\frac{d}{2}} \nonumber \\
& \times { \etr \left(-\frac{1}{2} Z_i\ext_i^{-1}\right)}\Npdfbig{\bar{\sz}_i}{{H}_i\sx}{\frac{\ext_i}{n_i}} \\
=& \mathcal{A}_{i} \Npdfbig{\bar{\sz}_i}{{H}_i\sx}{\frac{\ext_i}{n_i}}.
\end{align}
\label{eq:aux3}
\end{subequations}
The product of Gaussian distributions in \eqref{eq:subobject_measurement_likelihood} is thus rewritten as
\begin{subequations}
\begin{align}
\prod_{j=1}^{n_i} & \Npdfbig{\sz_{ij}}{H_{i}\sx}{\ext_{i}} = \mathcal{A}_{i} \Npdfbig{\bar{\sz}_{i}}{H_{i}\sx}{\frac{\ext_{i}}{n_{i}}},\\
& \mathcal{A}_{i} = \frac{\left|\ext_{i}\right|^{-\frac{n_i-1}{2}}\etr\left(-\frac{1}{2}Z_{i}\ext_{i}^{-1}\right)}{\left(2\pi\right)^{\frac{(n_i-1)d}{2}} n_i^{\frac{d}{2}} }.
\end{align}%
\end{subequations}%
The product of the likelihood and the prior distribution is
\begin{align}
& p\left(\setZ\left|\xi\right.\right)  p\left(\xi\right) \nonumber \\
= & \Npdfbig{\sx}{m}{P} \left(\prod_{i=1}^{N_s} \mathcal{A}_{i} \Npdfbig{\bar{\sz}_{i}}{H_{i}\sx}{\frac{\ext_{i}}{n_{i}}} \IWishpdf{\ext_{i}}{v_{i}}{V_{i}} \right) \nonumber \\
&\times \left( \prod_{i=1}^{N_s} n_{i}! \mathcal{PS}\left(n_{i};\ \gamma_{i}\right) \Gammapdf{\gamma_{i}}{\alpha_{i}}{\beta_{i}} \right).
\end{align}
For the measurement rates we have the following
\begin{subequations}
\begin{align}
& n_i! \mathcal{PS}\left(n_{i};\ \gamma_{i}\right) \Gammapdf{\gamma_{i}}{\alpha_{i}}{\beta_{i}} \nonumber \\
= & \frac{\beta_{i}^{\alpha_{i}}\gamma_{i}^{\alpha_{i}+n_{i}-1}e^{-(\beta_{i}+1)\gamma_{i}}}{\Gamma\left(\alpha_{i}\right)}  \\
= & \Gammapdf{\gamma_{i}}{\alpha_{i}+n_{i}}{\beta_{i}+1} \frac{\Gamma\left(\alpha_{i}+n_{i}\right)\beta_{i}^{\alpha_{i}}}{\Gamma\left(\alpha_{i}\right)\left(\beta_{i}+1\right)^{\alpha_{i}+n_{i}}} \\
= & \mathcal{L}_{i}^{\gamma} \Gammapdf{\gamma_i}{\alpha_{i}^{+}}{\beta_{i}^{+}}, \label{eq:gamma_meas_update}
\end{align}%
\label{eq:updated_measurement_rates}
\end{subequations}%
where
\begin{subequations}
\begin{align}
\alpha_{i}^{+} &=  \alpha_{i} + n_{i}, \\
\beta_{i}^{+} &= \beta_{i} + 1, \\
\mathcal{L}_{i}^{\gamma} &= \frac{\Gamma\left(\alpha_{i}^{+}\right)\beta_{i}^{\alpha_{i}}}{\Gamma\left(\alpha_{i}\right)\left(\beta_{i}^{+}\right)^{\alpha_{i}^{+}}},
\end{align}
\label{eq:corr_meas_rate}%
\end{subequations}%
The likelihood $\mathcal{L}_{i}^{\gamma}$ is proportional to a negative binomial distribution, see \egp \cite{GelmanCSR:2004}. For the kinematic vector and the random matrices we have
\begin{subequations}
\begin{align}
& \Npdfbig{\sx}{m}{P} \prod_{i=1}^{N_s} \mathcal{A}_{i} \Npdfbig{\bar{\sz}_{i}}{H_{i}\sx}{\frac{\ext_{i}}{n_{i}}} \IWishpdf{\ext_{i}}{v_{i}}{V_{i}} \nonumber \\
= & \Npdfbig{\sx}{m}{P} \Npdfbig{\bar{\sz}}{\mathbb{H}\sx}{\mathbb{X}} \prod_{i=1}^{N_s} \mathcal{A}_{i} \IWishpdf{\ext_{i}}{v_{i}}{V_{i}},
\end{align}
where
\begin{align}
& \bar{\sz} =  \begin{bmatrix} \bar{\sz}_{1}^{\tp} & \cdots & \bar{\sz}_{i}^{\tp} & \cdots & \bar{\sz}_{N_s}^{\tp} \end{bmatrix}^{\tp} \\
& \mathbb{H} =  \begin{bmatrix} H_{1}^{\tp} & \cdots & H_{i}^{\tp} & \cdots & H_{N_s}^{\tp} \end{bmatrix}^{\tp} \\
& \mathbb{X} =  \blkdiag\left( \frac{\ext_{1}}{n_{1}} , \ \ldots , \ \frac{\ext_{i}}{n_{i}} , \ \ldots , \ \frac{\ext_{N_s}}{n_{N_{s}}} \right).
\end{align}%
\end{subequations}%
Using the Kalman filter \cite{Kalman:1960} measurement update we get
\begin{subequations}
\begin{align}
& \Npdfbig{\sx}{m}{P} \Npdfbig{\bar{\sz}}{\mathbb{H}\sx}{\mathbb{X}} \prod_{i=1}^{N_s} \mathcal{A}_{i} \IWishpdf{\ext_{i}}{v_{i}}{V_{i}} \\
= & \Npdfbig{\sx}{\tilde{m}^{+}}{\tilde{P}^{+}} \Npdfbig{\bar{\sz}}{\mathbb{H}m}{\mathbb{H}P\mathbb{H}^{\tp}+\mathbb{X}} \nonumber \\
& \times \prod_{i=1}^{N_s} \mathcal{A}_{i} \IWishpdf{\ext_{i}}{v_{i}}{V_{i}} \label{eq:approximation_equation}
\end{align}
where
\begin{align}
\tilde{m}^{+} = & m+\tilde{K}\left(\bar{\sz}-\mathbb{H}m\right), \\
\tilde{P}^{+} = &  P-\tilde{K}\mathbb{H}P, \\
\tilde{K} = & P\mathbb{H}^{\tp}\tilde{S}^{-1}, \\
\tilde{S} = &  \mathbb{H}P\mathbb{H}^{\tp} + \mathbb{X}. \label{eq:kinematic_correction_innovation_covariance}
\end{align}%
\end{subequations}%
At this point we make two approximations.

\begin{approximation}
In $\tilde{S}$ in \eqref{eq:kinematic_correction_innovation_covariance} the random variable $\mathbb{X}$ is approximated by its expected value
\begin{subequations}
\begin{align}
\hat{\mathbb{X}} = & \E\left[ \mathbb{X} \right] = \blkdiag\left( \frac{\hat{\ext}_{1}}{n_{1}} , \ldots , \frac{\hat{\ext}_{i}}{n_{i}} , \ldots , \frac{\hat{\ext}_{N_s}}{n_{N_{s}}} \right),\\
\hat{\ext}_{i} = & \E\left[\ext_{i}\right] = \frac{V_{i}}{v_{i}-2d-2}
\end{align}%
\end{subequations}%
\hfill$\square$
\end{approximation}
\begin{remark}
This approximation is analogous to an approximation made by Feldmann \etal, see \cite[Equations 31 and 33]{FeldmannFK:2011}.\hfill$\square$
\end{remark}
\begin{approximation}
In $\Npdfbig{\bar{\sz}}{\mathbb{H}m}{\mathbb{H}P\mathbb{H}^{\tp}+\mathbb{X}}$ in \eqref{eq:approximation_equation} the matrix $\mathbb{H}P\mathbb{H}^{\tp}$ is approximated by the block-diagonal matrix
\begin{align}
\blkdiag\left( H_{1}PH_{1}^{\tp} , \ \ldots , \ H_{i}PH_{i}^{\tp} , \ \ldots , \ H_{N_s}PH_{N_s}^{\tp} \right).
\end{align}
\hfill$\square$
\end{approximation}
\begin{remark}
This approximation is necessary to obtain a posterior distribution that is of the same functional form as the prior distribution.
\hfill$\square$
\end{remark}%
Under these approximations instead of \eqref{eq:approximation_equation} we have
\begin{align}
& \Npdfbig{\sx}{m^{+}}{P^{+}} \label{eq:updated_kinematic_vector} \\
\times & \prod_{i=1}^{N_s} \mathcal{A}_{i} \Npdfbig{\bar{\sz}_{i}}{H_{i}m}{H_{i}PH_{i}^{\tp}+\frac{\ext_{i}}{n_{i}}} \IWishpdf{\ext_{i}}{v_{i}}{V_{i}} \nonumber
\end{align}
where
\begin{subequations}
\begin{align}
{m}^{+} = &  m+{K}\left(\bar{\sz}-\mathbb{H}m\right), \\
{P}^{+} = &  P-{K}\mathbb{H}P, \\
{K} = &  P\mathbb{H}^{\tp}{S}^{-1}, \\
{S} = &  \mathbb{H}P\mathbb{H}^{\tp} + \hat{\mathbb{X}}.
\end{align}%
\label{eq:corr_kinematics}
\end{subequations}%
For the factors in the product in \eqref{eq:updated_kinematic_vector}, the Gaussian covariances $H_{i}P H_{i}^{\tp}+\frac{\ext_{i}}{n_{i}}$ can be expanded (\egp using Cholesky Factorization) as
\begin{align}
\left(H_{i}PH_{i}^{\tp}+\frac{\ext_{i}}{n_{i}}\right)^{\frac{1}{2}}\ext_{i}^{-\frac{1}{2}}\ext_{i}\ext_{i}^{-\frac{\tp}{2}}\left(H_{i}PH_{i}^{\tp}+\frac{\ext_{i}}{n_{i}}\right)^{\frac{\tp}{2}}. \label{eq:expanded_Gaussian_covariance}
\end{align}
A third approximation is now made.
\begin{approximation}
Equation \eqref{eq:expanded_Gaussian_covariance} is approximated by
\begin{align}
S_{i}^{\frac{1}{2}} \hat{\ext}_{i}^{-\frac{1}{2}} \ext_{i} \hat{\ext}_{i}^{-\frac{\tp}{2}} S_{i}^{\frac{\tp}{2}}, \qquad S_{i} = H_{i}PH_{i}^{\tp}+\frac{\hat{\ext}_{i}}{n_{i}}.
\end{align}
\hfill$\square$
\end{approximation}
\begin{remark}
This approximation is analogous to an approximation made by Feldmann \etal, see \cite[Equations 38 and 39]{FeldmannFK:2011}.\hfill$\square$
\end{remark}
Under this approximation the Gaussians in \eqref{eq:updated_kinematic_vector} can be rewritten as
\begin{subequations}
\begin{align}
& \Npdfbig{\bar{\sz}_{i}}{H_{i}m}{ S_{i}^{\frac{1}{2}} \hat{\ext}_{i}^{-\frac{1}{2}} \ext_{i} \hat{\ext}_{i}^{-\frac{\tp}{2}} S_{i}^{\frac{\tp}{2}} } \\
= & \left(2\pi\right)^{-\frac{d}{2}} \left| S_{i}^{\frac{1}{2}} \hat{\ext}_{i}^{-\frac{1}{2}} \ext_{i} \hat{\ext}_{i}^{-\frac{\tp}{2}} S_{i}^{\frac{\tp}{2}} \right|^{-\frac{1}{2}} \nonumber \\
& \times \etr\left(-\frac{1}{2}\left(\bar{\sz}_{i}-H_{i}m\right)^{\tp} \left(S_{i}^{\frac{1}{2}} \hat{\ext}_{i}^{-\frac{1}{2}} \ext_{i} \hat{\ext}_{i}^{-\frac{\tp}{2}} S_{i}^{\frac{\tp}{2}}\right)^{-1} \left(\bar{\sz}_{i}-H_{i}m\right) \right) \\
= & \left(2\pi\right)^{-\frac{d}{2}} \left| S_{i}^{\frac{1}{2}} \hat{\ext}_{i}^{-\frac{1}{2}} \right|^{-\frac{1}{2}} \left| \ext_{i} \right|^{-\frac{1}{2}} \left| \hat{\ext}_{i}^{-\frac{\tp}{2}} S_{i}^{\frac{\tp}{2}} \right|^{-\frac{1}{2}} \nonumber \\
& \times \etr\left(-\frac{1}{2}\left(\bar{\sz}_{i}-H_{i}m\right)^{\tp} S_{i}^{-\frac{\tp}{2}} \hat{\ext}_{i}^{\frac{\tp}{2}} \ext_{i}^{-1} \hat{\ext}_{i}^{\frac{1}{2}} S_{i}^{\frac{\-1}{2}} \left(\bar{\sz}_{i}-H_{i}m\right) \right) \\
= & \left(2\pi\right)^{-\frac{d}{2}} \left| \hat{\ext}_{i}^{-\frac{\tp}{2}} S_{i} \hat{\ext}_{i}^{-\frac{1}{2}} \right|^{-\frac{1}{2}} \left| \ext_{i} \right|^{-\frac{1}{2}} \nonumber \\
& \times \etr\left(-\frac{1}{2}\hat{\ext}_{i}^{\frac{1}{2}} S_{i}^{\frac{\-1}{2}} \left(\bar{\sz}_{i}-H_{i}m\right)\left(\bar{\sz}_{i}-H_{i}m\right)^{\tp} S_{i}^{-\frac{\tp}{2}} \hat{\ext}_{i}^{\frac{\tp}{2}} \ext_{i}^{-1} \right) \\
= & \left(2\pi\right)^{-\frac{d}{2}} \left| \hat{\ext}_{i}^{-\frac{\tp}{2}} S_{i} \hat{\ext}_{i}^{-\frac{1}{2}} \right|^{-\frac{1}{2}} \left| \ext_{i} \right|^{-\frac{1}{2}} 
\etr\left(-\frac{1}{2}N_{i} \ext_{i}^{-1} \right)
\end{align}
where
\begin{align}
N_{i} & = \hat{\ext}_{i}^{\frac{1}{2}} S_{i}^{-\frac{1}{2}} \left(\bar{\sz}_{i}-H_{i}m\right)\left(\bar{\sz}_{i}-H_{i}m\right)^{\tp} S_{i}^{-\frac{\tp}{2}} \hat{\ext}_{i}^{\frac{\tp}{2}}
\end{align}
\end{subequations}
The factors in \eqref{eq:updated_kinematic_vector} can now be rewritten as
\begin{subequations}
\begin{align}
\mathcal{A}_{i} & \Npdfbig{\bar{\sz}_{i}}{H_{i}m}{ S_{i}^{\frac{1}{2}} \hat{\ext}_{i}^{-\frac{1}{2}} \ext_{i} \hat{\ext}_{i}^{-\frac{\tp}{2}} S_{i}^{\frac{\tp}{2}} } \IWishpdf{\ext_{i}}{v_{i}}{V_{i}} \\
= & \frac{\left|\ext_{i}\right|^{-\frac{n_i-1}{2}}\etr\left(-\frac{1}{2}Z_{i}\ext_{i}^{-1}\right)}{\left(2\pi\right)^{\frac{(n_i-1)d}{2}} n_i^{\frac{d}{2}} } \nonumber \\
& \times \left(2\pi\right)^{-\frac{d}{2}} \left| \hat{\ext}_{i}^{-\frac{\tp}{2}} S_{i} \hat{\ext}_{i}^{-\frac{1}{2}} \right|^{-\frac{1}{2}} \left| \ext_{i} \right|^{-\frac{1}{2}} 
\etr\left(-\frac{1}{2}N_{i} \ext_{i}^{-1} \right) \nonumber \\
& \times \frac{2^{-\frac{v_{i}-d-1}{2}}\left|V_{i}\right|^{\frac{v_{i}-d-1}{2}}}{\Gamma_{d}\left(\frac{v_{i}-d-1}{2}\right)\left|\ext_{i}\right|^{\frac{v_{i}}{2}}}\etr\left(-\frac{1}{2}\ext_{i}^{-1}V_{i}\right) \\
= & n_i^{-\frac{d}{2}}\left(2 \pi\right)^{-\frac{n_i d}{2}} \left| \hat{\ext}_{i}^{-\frac{\tp}{2}} S_{i} \hat{\ext}_{i}^{-\frac{1}{2}} \right|^{-\frac{1}{2}} \frac{2^{-\frac{v_{i}-d-1}{2}}\left|V_{i}\right|^{\frac{v_{i}-d-1}{2}}}{\Gamma_{d}\left(\frac{v_{i}-d-1}{2}\right)\left|\ext_{i}\right|^{\frac{v_{i}+n_{i}}{2}}} \nonumber \\
& \times  \etr\left(-\frac{1}{2}\left(V_i+Z_i+N_i\right)\ext_{i}^{-1}\right) \\
= & \frac{ n_i^{-\frac{d}{2}}\left(2 \pi\right)^{-\frac{n_i d}{2}}}{ \left| \hat{\ext}_{i}^{-\frac{\tp}{2}} S_{i} \hat{\ext}_{i}^{-\frac{1}{2}} \right|^{\frac{1}{2}}}  \frac{2^{-\frac{v_{i}-d-1}{2}}}{2^{-\frac{v_{i}+n_{i}-d-1}{2}}} \nonumber \\
& \times \frac{\Gamma_{d}\left(\frac{v_{i}+n_i-d-1}{2}\right)}{\Gamma_{d}\left(\frac{v_{i}-d-1}{2}\right)}  \frac{\left|V_{i}\right|^{\frac{v_{i}-d-1}{2}}}{\left|V_{i}+Z_{i}+N_{i}\right|^{\frac{v_{i}+n_i-d-1}{2}}} \nonumber \\
& \times \frac{2^{-\frac{v_{i}+n_{i}-d-1}{2}}\left|V_{i}+Z_{i}+N_{i}\right|^{\frac{v_{i}+n_i-d-1}{2}}}{\Gamma_{d}\left(\frac{v_{i}+n_i-d-1}{2}\right)\left|\ext_{i}\right|^{\frac{v_{i}+n_{i}}{2}}} \nonumber \\
& \times \etr\left(-\frac{1}{2}\left(V_i+Z_i+N_i\right)\ext_{i}^{-1}\right) \\
= & \mathcal{L}^{\sx,\ext}_{i} \IWishpdf{\ext_{i}}{v_{i}^{+}}{V_{i}^{+}}
\end{align}
\label{eq:updated_random_matrices}%
\end{subequations}%
where
\begin{subequations}
\begin{align}
v_{i}^{+} & = v_{i} + n_{i}, \\
V_{i}^{+} & = V_{i} + Z_{i} + N_{i} \\
\mathcal{L}^{\sx,\ext}_{i} & = \frac{ \left(n_i\pi^{n_i}\right)^{-\frac{d}{2}}2^{-\frac{n_i (d-1)}{2}}}{ \left| \hat{\ext}_{i}^{-\frac{\tp}{2}} S_{i} \hat{\ext}_{i}^{-\frac{1}{2}} \right|^{\frac{1}{2}}} \frac{\Gamma_{d}\left(\frac{v_{i}^{+}-d-1}{2}\right)}{\Gamma_{d}\left(\frac{v_{i}-d-1}{2}\right)}  \frac{\left|V_{i}\right|^{\frac{v_{i}-d-1}{2}}}{\left|V_{i}^{+}\right|^{\frac{v_{i}^{+}-d-1}{2}}} \label{eq:kin_ext_likelihood}
\end{align}%
\label{eq:corr_iWish}
\end{subequations}%
The likelihood $\mathcal{L}^{\sx,\ext}_{i}$ is proportional to a generalized matrix variate beta type two distribution, see \egp \cite{GuptaN:2000}.

By the combination of \eqref{eq:updated_measurement_rates}, \eqref{eq:updated_kinematic_vector} and \eqref{eq:updated_random_matrices}, under the three approximations given above, the parameters of the posterior \eqref{eq:GGIW_posterior} are given by \eqref{eq:corr_meas_rate}, \eqref{eq:corr_kinematics}, and \eqref{eq:corr_iWish}, and the likelihood is $\mathcal{L} = \prod_{i=1}^{N_s} \mathcal{L}^{\gamma}_{i} \mathcal{L}^{\sx,\ext}_{i}$, where $\mathcal{L}^{\gamma}_{i}$ is given in \eqref{eq:gamma_meas_update} and $\mathcal{L}^{\sx,\ext}_{i}$ is given in \eqref{eq:kin_ext_likelihood}.

\ifCLASSOPTIONcaptionsoff
  \newpage
\fi

\bibliographystyle{IEEEtran}
\bibliography{PHD_ext_targ_track}

\end{document}